\documentclass[a4paper]{article}

\usepackage[english]{babel}
\usepackage[utf8]{inputenc}
\usepackage{amssymb}
\usepackage{amsmath}
\usepackage{array}
\usepackage{graphicx}
\usepackage{epstopdf}
\usepackage{dcolumn}
\usepackage{bm}
\usepackage{color}
\usepackage{hyperref}
\usepackage{algpseudocode}
\usepackage{algorithm}
\usepackage{pbox}
\usepackage{caption}
\usepackage{subcaption}
\usepackage{amsthm}
\usepackage{authblk}
\usepackage{cleveref}
\usepackage{tikz}
\usepackage{enumerate}
\usepackage{epstopdf}
\usepackage{kbordermatrix}
\usepackage{pdfpages}
\usetikzlibrary{arrows}
\usetikzlibrary{positioning}

\theoremstyle{plain}

\newlength{\actualtopmargin}
\newlength{\actualsidemargin}
\title{Efficient estimation of perturbative error with cellular automata}

\author{Yudong Cao\footnote{Department of Computer Science, Purdue University. West Lafayette, IN 47906, USA. Email: cao23@purdue.edu}, Sabre Kais\footnote{Department of Chemistry, Physics and Computer Science, Purdue University. West Lafayette, IN 47906, USA; Qatar Energy and Environment Research Institute, HBKU,  Doha, Qatar and  Santa Fe Institute, 1399 Hyde Park Rd., Santa Fe, NM 87501, USA. Email: kais@purdue.edu}}

\date{}

\begin{document}
\maketitle

\begin{abstract}
From celestial mechanics to quantum theory of atoms and molecules, perturbation theory has played a central role in natural sciences. Particularly in quantum mechanics, the amount of information needed for specifying the state of a many-body system commonly scales exponentially as the system size. This poses a fundamental difficulty in using perturbation theory at arbitrary order. As one computes the terms in the perturbation series at increasingly higher orders, it is often important to determine whether the series converges and if so, what is an accurate estimation of the total error that comes from the next order of perturbation up to infinity. Here we present a set of efficient algorithms that compute tight upper bounds to perturbation terms at arbitrary order. We argue that these tight bounds often take the form of symmetric polynomials on the parameter of the quantum system. We then use cellular automata as our basic model of computation to compute the symmetric polynomials that account for all of the virtual transitions at any given order. At any fixed order, the computational cost of our algorithm scales \emph{polynomially} as a function of the system size. We present a non-trivial example which shows that our error estimation is nearly tight with respect to exact calculation.
\end{abstract}

An overwhelming majority of problems in quantum physics and quantum chemistry do not admit exact, analytical solutions. Therefore one has to resort to approximation methods based on for instance series expansions \cite{BO99,HJO00,SO96,P63,HAG93,Kais}. Often these expansions are truncated to a finite order $r$ as an approximation of the true solution an the remaining terms from the $(r+1)$-th order on are errors. It is then important to estimate the magnitude of errors at arbitrary order as a gauge of how the series performs as an approximate solution. The main challenge in this task is that exact calculation of the perturbative terms commonly scales exponentially as the size of the system under consideration, making it hard to pinpoint the regime where perturbation theory yields acceptable accuracy \cite{HJO00}.

\begin{figure}
\begin{center}
\includegraphics[scale=1]{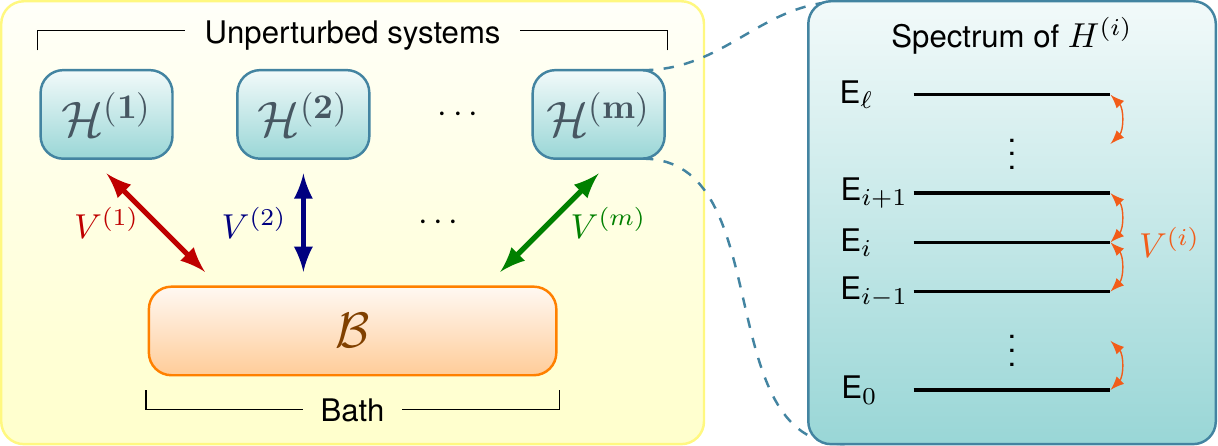}
\end{center}
\caption{General setting of the perturbation theory.}
\label{fig:gen_setting}
\end{figure}

Here we present an efficient method for deriving tight upper bounds for the norm of perturbative expansion terms at arbitrary order. The use of perturbation theory starts with identifying a physical system $\tilde{H}$ as a sum of an unperturbed Hamiltonian $H$ that acts on a Hilbert space $\mathcal{H}$ and a perturbation $V$. As shown in Figure \ref{fig:gen_setting}a, we assume that $H=H^{(1)}+H^{(2)}+\cdots+H^{(m)}$ consists of $m$ identical and non-interacting unperturbed subsystems with Hilbert space $\mathcal{H}^{(i)}$, $i=1,\cdots,m$. Each subsystem interacts with a ``bath'' $\mathcal{B}$ through perturbation $V$ that is presumably small. We further assume that for each subsystem $H^{(i)}$, $V$ can only cause transitions in neighboring energy levels (Figure \ref{fig:gen_setting}b). This form of physical setting is typical in for example spin systems with perturbation on individual spins via local fields \cite{MBSZ+13,ISCK+13}, or in Hartree approximation where $m$ identical particles interact with a mean field \cite{SO96}. Here $V$ does not necessarily act identically on each $\mathcal{H}^{(i)}\otimes\mathcal{B}$ for every $i$. For a given $V$, one could determine an upper bound $\lambda_i$ for each subsystem $i$ such that $|\langle\phi|V|\phi'\rangle|\le\lambda_i$ for any $|\phi\rangle$, $|\phi'\rangle$ being eigenstates of $H^{(i)}$. We could also determine an upper bound $\omega$ such that for any $|\phi\rangle$ that is an eigenstate of $H$, $|\langle\phi|V|\phi\rangle|\le\omega$. With the spectrum of each $H^{(i)}$ fully known, one could also determine for each energy level $s$ and $t$ the maximum number of possible ways for an eigenstate at energy level $s$ to make a transition to a state of energy level $t$ via the perturbation $V$. We let this number be $M_{st}$ for all $H^{(i)}$, since their spectra are identical.

In many cases we are only concerned about the property of the effective Hamiltonian below certain cutoff energy $E_*$. Assume that the ground state energy of every $H^{(i)}$ is 0 and $E_*=\Delta/2$ where $\Delta=E_1$ is the spectral gap between the ground and the first excited state. For $\|V\|$ small enough compared to $\Delta$ we could extract this information using the operator valued resolvent $G(z)=(zI-H)^{-1}$ with a small expansion parameter $z$ and construct the self-energy
\begin{equation}\label{eq:self_energy}
\Sigma_-(z)=H_{--}+V_{--}+V_{-+}G_{++}V_{+-}+V_{-+}G_{++}V_{++}G_{++}V_{+-}+\cdots
\end{equation}
where we partition $\mathcal{H}$ into subspaces $\mathcal{L}_-$ and $\mathcal{L}_+$, with $\mathcal{L}_-$ being the subspace of $\mathcal{H}$ spanned by $H$ eigenstates with energy below $E_*$ and $\mathcal{L}_+$ being the complement of $\mathcal{L}_-$ in $\mathcal{H}$, and let $O_{\pm\pm}=\Pi_\pm O\Pi_\pm$ be projections of any operator $O$ onto the $\mathcal{L}_\pm$ subspaces. $\Pi_-$ and $\Pi_+$ are projectors onto $\mathcal{L}_-$ and $\mathcal{L}_+$ respectively. To compute an approximation to the low-energy effective Hamiltonian of $\tilde{H}$, one simply truncates Equation \ref{eq:self_energy} at low orders to obtain an effective Hamiltonian $H_\text{eff}$ and discard the remaining terms which constitutes the error of the perturbation series. Here we are only restricted to convergent series. For divergent series one may resort to resummation techniques such as Pad\'{e} approximation \cite{BO99}. If we denote the $r$-th order term in the self energy expansion \eqref{eq:self_energy} as $T_r=V_{-+}(G_{++}V_{++})^{r-2}G_{++}V_{+-}$ for $r\ge 2$ and $T_1=V_{--}$, then our effective Hamiltonian $H_\text{eff}=T_1+T_2+\cdots+T_R$ for some $R$ and the remaining terms $T_{R+1}+T_{R+2}+\cdots$ are error. The connection between the magnitude of the error $\|\Sigma_-(z)-H_\text{eff}\|_2$ and the spectral difference between $\tilde{H}$ and $H_\text{eff}$ is well established. If for a suitable range of $z$, $\|\Sigma_-(z)-H_\text{eff}\|_2$ is no greater than $\epsilon$, then the energies of $H_\text{eff}$ are at most $\epsilon$ apart from their counterparts in the low energy spectrum of $\tilde{H}$ (see \cite{KKR06,OT06}). Our goal is precisely to find tight upper bounds for the magnitude of the error terms $\|\Sigma_-(z)-H_\text{eff}\|_2$. 

\begin{figure}
\begin{center}
\makebox[4.5in]{\includegraphics[scale=0.7]{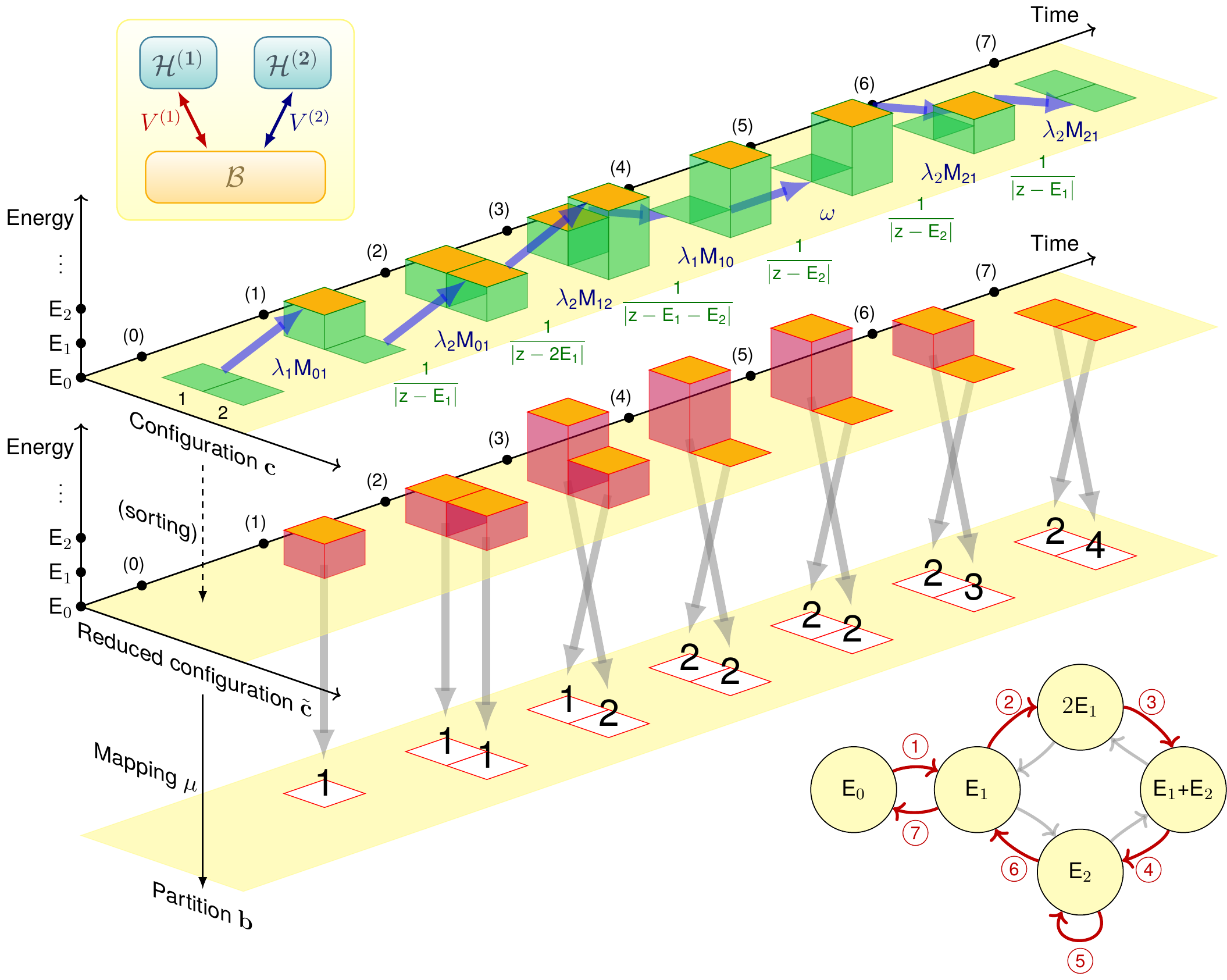}}
\end{center}
\caption{An example of a walk arising at 7th order perturbation theory $T_7=V_{-+}(G_+V_+)^5G_+V_{+-}$. Top left: the specific physical setting concerned, where the number of subsystems is $m=2$. Top layer: the relationship between the 7-step walk in the space of energy configurations ${\bf c}$ and an upper bound associated with it. Each transition due to $V$ is associated with a factor of either $\lambda_iM_{st}$ or $\omega$. Each intermediate step with energy $E^{(i)}$ contributes a term ${1}/{|z-E^{(i)}|}$ due to $G_+$. Middle layer: the corresponding walk in $\tilde{\bf c}$, where at each step $\tilde{\bf c}^{(i)}$ is obtained by sorting ${\bf c}$ in descending order. Bottom layer: the corresponding change in the partition $\bf b$ and the mapping $\mu:\tilde{\bf c}\mapsto{\bf b}$ maintained throughout. By convention, the partition ${\bf b}$ is always of non-decreasing order. Bottom right: the walk in the space of energy combination ${\bf n}$ corresponding to the walk in $\tilde{\bf c}$. This walk in ${\bf n}$ is what the cellular automaton algorithm essentially implements.}
\label{fig:walk_steps}
\end{figure}

For convergent series it suffices to be able to find tight estimates for the $\infty$-norm of the $r$-th order term $\|T_r\|_\infty$ for any $r\ge 2$.  The $\infty$-norm of a matrix $A\in\mathbb{C}^{m\times n}$ is defined as $\max_{i=1,\cdots,m}\sum_{j=1}^n|a_{ij}|$. We could bound $\|T_r\|_\infty$ from above by a function of $\lambda_i$, $M_{st}$ and $\omega$. Because $T_r$ is essentially a matrix product, one could think of the matrix element $\langle\phi|T_r|\phi'\rangle$ as a sum of $r$-step walks on the eigenstates of $H$, which can be written as $|\phi\rangle\rightarrow|\phi^{(1)}\rangle\rightarrow\cdots\rightarrow|\phi^{(r-1)}\rangle\rightarrow|\phi'\rangle$, with each $|\phi^{(i)}\rangle$ being an eigenstate of $H$ and each step of the walk contributing a factor and the total weight of the walk is the product of all the factors. Using the scalar quantities $\lambda_i$, $M_{st}$ and $\omega$ symbols we could derive an upper bound to $|\langle\phi|T_r|\phi'\rangle|$ by noting that 
\begin{equation}\label{eq:walk_Hstate}
\begin{array}{ccl}
|\langle\phi|T_r|\phi'\rangle| & \le & \displaystyle\sum_{\{|\phi^{(i)}\rangle\}}|\langle\phi|V|\phi^{(1)}\rangle|\cdot|\langle\phi^{(1)}|G|\phi^{(1)}\rangle|\cdot|\langle\phi^{(1)}|V|\phi^{(2)}\rangle| \\[0.1in]
& & \cdots|\langle\phi^{(r-2)}|V|\phi^{(r-2)}\rangle|\cdot|\langle\phi^{(r-1)}|G|\phi^{(r-1)}\rangle|\cdot|\langle\phi^{(r-1)}|V|\phi'\rangle|
\end{array}
\end{equation}
where the summation is over all possible $r$-step walks on the eigenstates of $H$ that starts at $|\phi\rangle$ and ends at $|\phi'\rangle$.
The factors $|\langle\phi^{(i)}|G|\phi^{(i)}\rangle|=1/|z-E^{(i)}|$, where $E^{(i)}=\langle\phi^{(i)}|H|\phi^{(i)}\rangle$, can be computed easily since the spectrum of $H$ is known. Suppose $V$ transforms an $H$ eigenstate $|\phi^{(i)}\rangle$ into $V|\phi^{(i)}\rangle=|\phi^{(i+1)}\rangle$ by changing the energy level of one of the subsystems (say $H^{(i)}$) from $s$ to $t$. Then $|\langle\phi^{(i)}|V|\phi^{(i+1)}\rangle|\le\lambda_iM_{st}$. However, if $|\phi^{(i)}\rangle=|\phi^{(i+1)}\rangle$, then we have $|\langle\phi^{(i)}|V|\phi^{(i+1)}\rangle|\le\omega$. For each walk on the eigenstates of $H$ we could then assemble an upper bound that looks like for example (Figure \ref{fig:walk_steps} top layer)
\begin{equation}\label{eq:walk_bound_example}
\lambda_iM_{st}\cdot\frac{1}{|z-E^{(1)}|}\cdot\lambda_jM_{pq}\cdot\frac{1}{|z-E^{(2)}|}\cdot\omega\cdots.
\end{equation}
At the second order we could use this technique to bound $\|T_2\|_\infty$ from above as
\begin{equation}\label{eq:walk_bound_example_T2}
\begin{array}{ccl}
\|T_2\|_\infty & \le & \displaystyle \lambda_1M_{01}\cdot\frac{1}{|z-E_1|}\cdot\lambda_1M_{10}+\lambda_2M_{01}\cdot\frac{1}{|z-E_1|}\cdot\lambda_2M_{10}+\cdots \\[0.1in]
& & \displaystyle \cdots+\lambda_mM_{01}\cdot\frac{1}{|z-E_1|}\cdot\lambda_mM_{10}.
\end{array}
\end{equation}
where we recall that $E_1$ is the first excited state energy of any subsystem $H^{(i)}$ (Figure \ref{fig:gen_setting}b). Each term in Equation \ref{eq:walk_bound_example_T2} with $\lambda_j$ corresponds to a 2-step walk where the $j$-th subsystem is excited from the ground state ($0$-th energy level) into the first excited state and then transitions back to the ground state energy subspace.

The expressions for the upper bounds to $\|T_r\|_\infty$ such as on the right hand side of Equation \ref{eq:walk_bound_example_T2} looks simple for $r=2$. At higher order, however, the situation quickly becomes more complicated. Intuitively this is because each unperturbed system has $\ell$ possible energy levels, and $m$ such subsystems could manifest $\ell^m$ possible ways in which the energies of each subsystems are assigned. Therefore any matrix element of $T_r$ should be a sum of roughly at most $O(\ell^{mr})$ walks, yielding an exponential complexity with respect to the total system size $m$. However, we note that such exponential complexity could be reduced to merely poly$(m)$ by exploiting the inherent permutation symmetry of upper bounds such as Equation \ref{eq:walk_bound_example_T2}. The essential observation is that these upper bounds are invariant with respect to permutation of the subsystems. This implies that they are \emph{symmetric functions} over the $\lambda_i$ variables. In particular, these upper bounds to $\|T_r\|_\infty$ are linear combinations of \emph{monomial symmetric polynomials}, which can be written in form of \cite{M98} \[m_{\bf b}({\boldsymbol\lambda})=\sum_{\pi\in S_k}\lambda_{\pi(1)}^{b_1}\lambda_{\pi(2)}^{b_2}\cdots\lambda_{\pi(k)}^{b_k}\]where ${\bf b\in\mathbb{N}^k}$ is a vector which we call \emph{partition}, $\boldsymbol\lambda=(\lambda_1,\cdots,\lambda_m)$ and the summation is over a permutation group $S_k$, where any permutation $\pi$ chooses $k$ elements from $m$ elements and permutes them. For example, $m_{(1,2)}(\lambda_1,\lambda_2,\lambda_3)=\lambda_1\lambda_2^2+\lambda_1\lambda_3^2+\lambda_2\lambda_3^2+\lambda_2\lambda_1^2+\lambda_3\lambda_2^2+\lambda_3\lambda_1^2$ is a monomial symmetric polynomial. Equation \ref{eq:walk_bound_example_T2} could be compactly represented as $\|T_2\|_\infty\le\frac{1}{|z-E_1|}M_{01}M_{10}m_{(2)}$. At $4$-th order we could show that
\begin{equation}\label{eq:walk_bound_example_T4}
\|T_4\|_\infty \le \displaystyle \frac{M_{01}M_{10}\omega^2m_{(2)}}{|(z-E_1)^3|}+\frac{2M_{01}^2M_{10}^2m_{(2,2)}}{|(z-E_1)^2(z-2E_1)|} +\frac{M_{01}M_{12}M_{21}M_{10}m_{(4)}}{|(z-E_1)^2(z-E_2)|}.
\end{equation}

By respecting the matrix product structure of $T_r$, the symmetric polynomial upper bounds such as those in Equations \ref{eq:walk_bound_example_T2} and \ref{eq:walk_bound_example_T4} turn out to be a much more accurate estimation of the true magnitude of $\|T_r\|_\infty$ than crude bounds using geometric series such as $\|T_r\|_2\le \|V\|_2\cdot\|G_{++}\|_2\cdot\|V\|_2\cdots\|G_{++}\|_2\cdot\|V\|_2$. In later discussions we will demonstrate this point using numerical examples.

The question then becomes how we may assemble expressions such as \eqref{eq:walk_bound_example_T2} and \eqref{eq:walk_bound_example_T4} in an algorithmic fashion. We accomplish this efficiently by using \emph{cellular automata} as the basic data structure. In a nutshell, a cellular automaton is a computational model consisting of a network of basic units called \emph{cells} that are connected by directed edges. Each cell stores some data which represent its current \emph{state}. All the cells are assigned an initial state and the computation proceeds by evolving each cell using an identical rule for updating its state. The new state of each cell is only dependent on the previous states of the same cell and its neighbors. The study of cellular automata dates back to the 1940s \cite{vN48}, followed by interesting constructions \cite{WR46,G70,S91} and formal, systematic study over the past decades \cite{S58,W83}. Though computationally rich, the structure of cellular automata considered in these contexts are commonly rather simple, with cells that have discrete sets of possible states and are connected by simple network geometries (such as a 2D grid). In our case, as we will discuss later, the cells in cellular automata store more complex data structures and are connected with often non-planar network geometries. The update rules designed specifically so that the coordination of cells as a whole computes the symmetric polynomial upper bound for $\|T_r\|_\infty$.

The connection between cellular automata and perturbation theory seems unusual at first glance. However, the connection between cellular automata and random walks is well documented \cite{TM87,EN92,E93}. Such connection, combined with our earlier discussion on how the symmetric polynomial upper bounds could arise from summing over walks on the set of $H$ eigenstates, suggests that one may also be able to use cellular automata for the summation over these walks. One could further think of our task of computing a symmetric polynomial upper bound to $\|T_r\|_\infty$ as summing over walks in a space of \emph{energy configurations} ${\bf c}$, which are $m$-dimensional vectors of indices ranging from $0$ to $\ell-1$ indicating the energy level of each subsystem in a particular $H$ eigenstate. In other words, ${\bf c}=(c_1,\cdots,c_m)\in\{0,1,\cdots,\ell-1\}^m$ and $\langle\phi|H^{(i)}|\phi\rangle=E_{c_i}$ for any particular $H$ eigenstate $|\phi\rangle$. Therefore each $r$-step walk in the space of $H$ eigenstates corresponds to a walk in the space of energy configuration ${\bf c}$, which is of size $O(\ell^m)$. We could reduce the size of this space by taking every energy configuration ${\bf c}$ and sort its elements to produce a new vector $\tilde{\bf c}$, which we call \emph{reduced energy configuration}. Like the number of energy levels in $H$, the set of $\tilde{\bf c}$ is also of size $O(m^\ell)$, which is polynomial is $m$ assuming $\ell$ is a constant and intensive property of each subsystem (for instance a spin-1/2 particle has $\ell=2$ if we are only concerned with the spin degree of freedom). Each energy level of $H$ is a sum of the energies of the subsystems: $\langle\phi|H|\phi\rangle=\sum_{i=0}^{\ell-1}n_iE_i=E({\bf n})$ where $E_i$ is one of the $\ell$ possible energy levels of a subsystem. We could write each energy level of $H$ as an $\ell$-dimension vector ${\bf n}=(n_0,n_1,\cdots,n_{\ell-1})$ which we call \emph{energy combination} (Figure \ref{fig:walk_steps} middle layer). 

With the discussion so far we have reduced the problem of summing over walks on the set of $H$ eigenstates, whose number scales exponentially with respect to system size parameter $m$, to one that concerns only with walks on the set of ${\bf n}$, which is of only polynomial size in $m$. In accomplishing this reduction, we introduced the notion of energy configuration ${\bf c}$ and reduced energy configuration $\tilde{\bf c}$. Going from walks in ${\bf c}$ to $\tilde{\bf c}$ is a major step that takes advantage of the permutation symmetry with respect to the $m$ subsystems in the $r$-th order from $T_r$. We capture this symmetry with the use of symmetric polynomials $m_{\bf b}(\boldsymbol\lambda)$. We illustrate this concept in Figure \ref{fig:walk_steps}. We note that the partition ${\bf b}$ does not contain all of the information associated with a walk in $\tilde{\bf c}$. Consider a particular walk on the set of $H$ eigenstates and its associated weight whose functional form is shown in Equation \ref{eq:walk_bound_example}, ${\bf b}$ only records the number of times that some subsystem is acted on by $V$, without the information about the order and the energies of the subsystem before and after the action (Figure \ref{fig:walk_steps} bottom layer). For example the partition $(1,2)$ means ``one of the subsystems is acted on by $V$ once and another is acted on by $V$ twice''. The expression $m_{(1,2)}(\boldsymbol\lambda)$ sums over the weights of walks that fits that description. But there are more than one possible walks, be it on the set of $H$ eigenstates or ${\bf c}$ or $\tilde{\bf c}$, that fits the description. Therefore in order for a symmetric polynomial to accurately represent an upper bound to the contributions to $\langle\phi|T_r|\phi'\rangle$ from all walks in $\tilde{\bf c}$, a mapping must be maintained between ${\bf b}$ and $\tilde{\bf c}$ to indicate which subsystem is being acted on at the current step. Figure \ref{fig:walk_steps} shows an example that illustrates the connection between $\tilde{\bf c}$, ${\bf b}$, and $\mu$ to a walk in the configuration space ${\bf c}$.

In our construction cellular automata that executes the summation over walks in $\tilde{\bf c}$, each cell corresponds to an energy level of $H$. Hence there are in total $O(m^\ell)$ cells. We use the energy combinations ${\bf n}$ to uniquely label each cell. Then the cells are connected with directed edges such that cell ${\bf n}$ will only be connected to cell ${\bf n}'$ if there are eigenstates $|\phi\rangle$, $|\phi'\rangle$ of $H$ with energy combinations ${\bf n}$ and ${\bf n}'$ respectively such that $|\langle\phi|V|\phi'\rangle|\neq 0$. In our algorithm each monomial symmetric polynomial $\xi m_{\bf b}(\boldsymbol\lambda)$ is represented with a 4-tuple $(\tilde{\bf c},{\bf b},\xi,\mu)$ where $\xi$ is a scalar quantity indicating the weight of $m_{\bf b}(\boldsymbol\lambda)$ in the overall symmetric polynomial upper bound. $\tilde{\bf c}$ and ${\bf b}$ are respectively the reduced energy configuration and partition at the current step of the walk. $\mu:\tilde{\bf c}\mapsto{\bf b}$ is a bijective mapping between $\tilde{\bf c}$ and ${\bf b}$, as justified in previous discussion.  

Each cell of the automaton stores a list of 4-tuples $(\tilde{\bf c},{\bf b},\xi,\mu)$ as its state. As shown in Figure \ref{fig:multi_cubewalk}, at each iteration the state of each cell is updated in a two-phase process. In phase I (Figure \ref{fig:multi_cubewalk}a), the list of 4-tuples stored in $\mathcal{S}_{\bf n}$ is first merged with thosed stored in all of the incident edges to $\mathcal{S}_{\bf n}$ and then the coefficients of all the 4-tuples in $\mathcal{S}_{\bf n}$ are multiplied by a factor $1/|z-E({\bf n})|$. The intuition is that each 4-tuple corresponds to a particular walk such as the one shown in Figure \ref{fig:walk_steps}. The multiplication by $1/|z-E({\bf n})|$ essentially accounts for the contribution from $G_+$ in $T_r$. In phase II, we account for the contribution from $V$ terms in $T_r$ by first computing new 4-tuples with $\tilde{\bf c}$ that can be generated from the current 4-tuples in $\mathcal{S}_n$ with one application of $V$, and then distributing the new 4-tuples among the outgoing edges $\mathcal{S}_{{\bf n},{\bf n}''}$, as shown in Figure \ref{fig:multi_cubewalk}b.

As the cells evolve, the 4-tuples are updated and passed along between the cells so that at the end of $r$ iterations, we could glean the symmetric polynomial upper bound from the states of the cells. The update rules for each cell are designed to maintain the property that at any iteration, each cell ${\bf n}$ contains a list of 4-tuples $(\tilde{\bf c},{\bf b},\xi,\mu)$ each of which corresponds to the set of all walks in $\tilde{\bf c}$ that leads up to a state with energy combination ${\bf n}$, and $\xi m_{\bf b}(\boldsymbol\lambda)$ is an upper bound to the total contribution of the walks on the set of $H$ eigenstates that share the same corresponding walk in $\tilde{\bf c}$. In other words, $\xi m_{\bf b}(\boldsymbol\lambda)$ is a sum of expressions such as Equation \ref{eq:walk_bound_example} for these walks on the set of $H$ eigenstates. We are able to rigorously show that with suitable initialization, after $r$ iterations the cellular automaton is indeed able to find a symmetric polynomial upper bound for $\|T_r\|_\infty$ similar to that of $\|T_4\|_\infty$ in Equation \ref{eq:walk_bound_example_T4}. 

We stress that the overall time complexity of our algorithm scales \emph{polynomially} as the system size grows. The degree of the polynomial, however, depends on the order of perturbation theory. For convergent series, the exponential dependence on the order $r$ of perturbation theory could be handled in practice by for instance setting a threshold $\eta$ such that when the symmetric polynomial upper bound computed by the cellular automaton is below $\eta$ at some order $r_c$ of perturbation, we bound the remaining terms up to infinity by a geometric series. For different problems and choices of $\eta$, the value of $r_c$ may vary. But the overall polynomial scaling with respect to the system size $m$ should not be affected.

$\quad$\\
$\quad$

In the mathematical developments of physical theories one is often concerned with the \emph{representation} of the solution to a problem. For very few problems are we able to find a close-form, explicit formula as a representation of the solution. Series expansions are then introduced to largely enhance our ability to solve difficult problems far beyond analytical solution, as they allow for representation of a much wider class of mathematical objects. If we think of these representations as efficient procedures that allow us to construct our solution, then in greater generality we could argue that the outputs of efficient algorithms are also valid representations of our solution. Our scheme based on cellular automata essentially produces this type of representation: the symmetric polynomial upper bound to $\|T_r\|_\infty$ that we have devised is most conveniently expressed in form of an algorithmic output, rather its explicit self as a sum of monomials. A similar example to this situation is perhaps the development of tensor networks as representations for quantum ground states \cite{VCM08,CV09,ACL12}. As is the case for our algorithmic development, tensor networks are also intended to cope with the exponential size of Hilbert space as the physical system grows. Using innovative data structures based on tensors, one obtains a polynomial size approximation to the true ground state. The resulting ground state is then most conveniently represented in form of a tensor network rather than its exponential-size self as a linear combination of basis states. Our cellular automaton algorithm could also be thought of as producing an approximation to $\|T_r\|_\infty$, in the sense that we replace the action of $V$ on the  unperturbed eigenstates $|\phi\rangle$, $|\phi'\rangle$ of each subsystem $i$ by scalar quantities $\lambda_i$ and $\omega$, and we use the integers $M_{st}$ to obtain a sketch of the structure of $V$. Such approximations may seem crude at first sight, but they preserve the combinatorial structure of $T_r$ as a matrix product, and allow for compact description using symmetric polynomials. We use iteration of cell evolution as a natural means to compute these symmetric polynomials. As a result, the output of our cellular automaton algorithm is the most natural representation for the upper bound to $\|T_r\|_\infty$ that we have devised.

One of the areas where our algorithm could find direct application is quantum computation. Though perturbation theory has been pervasively used for calculating properties of quantum systems, the lack of efficient and effective methods for estimating the error even for convergent series has cast a wide shadow of uncertainty on these calculations. Such problem becomes ever more imminent when one tries to engineer quantum systems that are intended to meet specific application requirements such as quantum computing \cite{LT15,CRBK14,CN14}. As the implementations of quantum devices scale up and perturbation theory finds its inevitable use in analyzing these devices, it is imperative to have a scalable method for estimating the error in the perturbative expansion. 

\begin{figure}
\makebox[-1.5cm]{}
\vspace{-2cm}
\begin{center}
\includegraphics[scale=1]{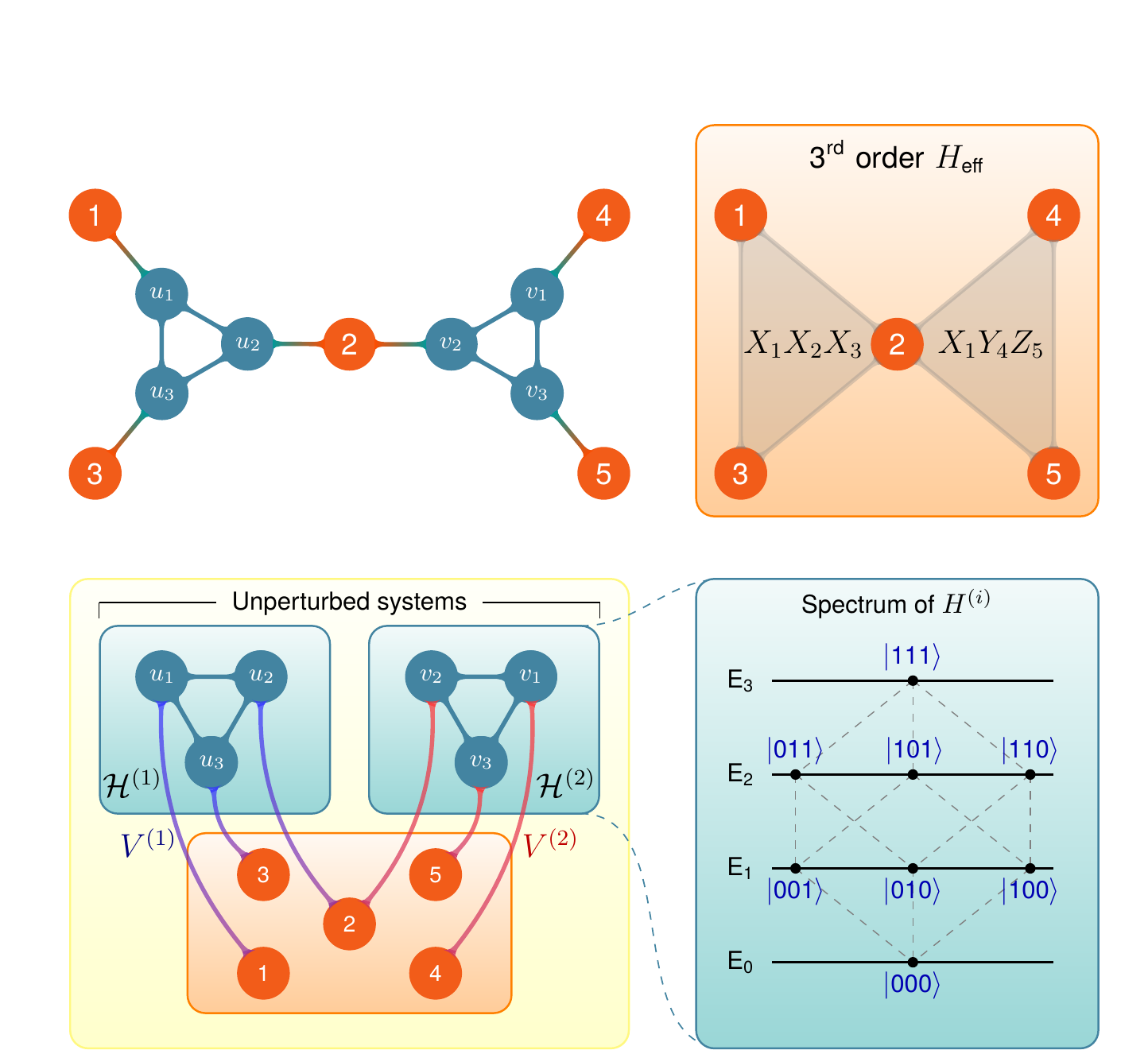}
\end{center}
\caption{A numerical example for demonstrating our algorithm estimating the perturbative error. (a) The 11-spin system constructed for testing. Each node corresponds to a spin-1/2 particle and each edge represents an interaction term in the Hamiltonian between two spins. (b) Effective Hamiltonian truncating at 3rd order perturbation theory. Here each triangle represents a 3-body interaction term. Using the perturbative expansion in Equation \ref{eq:self_energy} we could show that the low-energy effective Hamiltonian truncated at 3rd order is $H_\text{eff}=\alpha_1X_1X_2X_3+\alpha_2X_2Y_4Z_5$ up to a constant energy shift. (c) Rearranging and partitioning the system in (a) according to the setting of perturbation theory used. Here each unperturbed system $H^{(i)}$ consists of three ferromagnetically interacting spins (details in the long version). (d) Spectrum of each subsystem $H^{(i)}$ in (a), $i\in\{1,2\}$. Here each node represents an eigenstate of $H^{(i)}$. Nodes on a same horizontal dashed line belong to the same energy subspace $\mathcal{P}_j$. There is an edge $(u,v)$ iff $\|\langle u|V|v\rangle\|\neq 0$. For example, if we consider this diagram as representing $H^{(1)}$, since $V^{(1)}|001\rangle_{u_1u_2u_3}\propto(|101\rangle+|011\rangle+|000\rangle)_{u_1u_2u_3}$ we connect the $|001\rangle$ with the nodes representing $|101\rangle$, $|011\rangle$ and $|000\rangle$.}
\label{fig:num_ex_big}
\end{figure}

\begin{figure}
\makebox[-3cm]{}
\begin{center}
\includegraphics[scale=1.3]{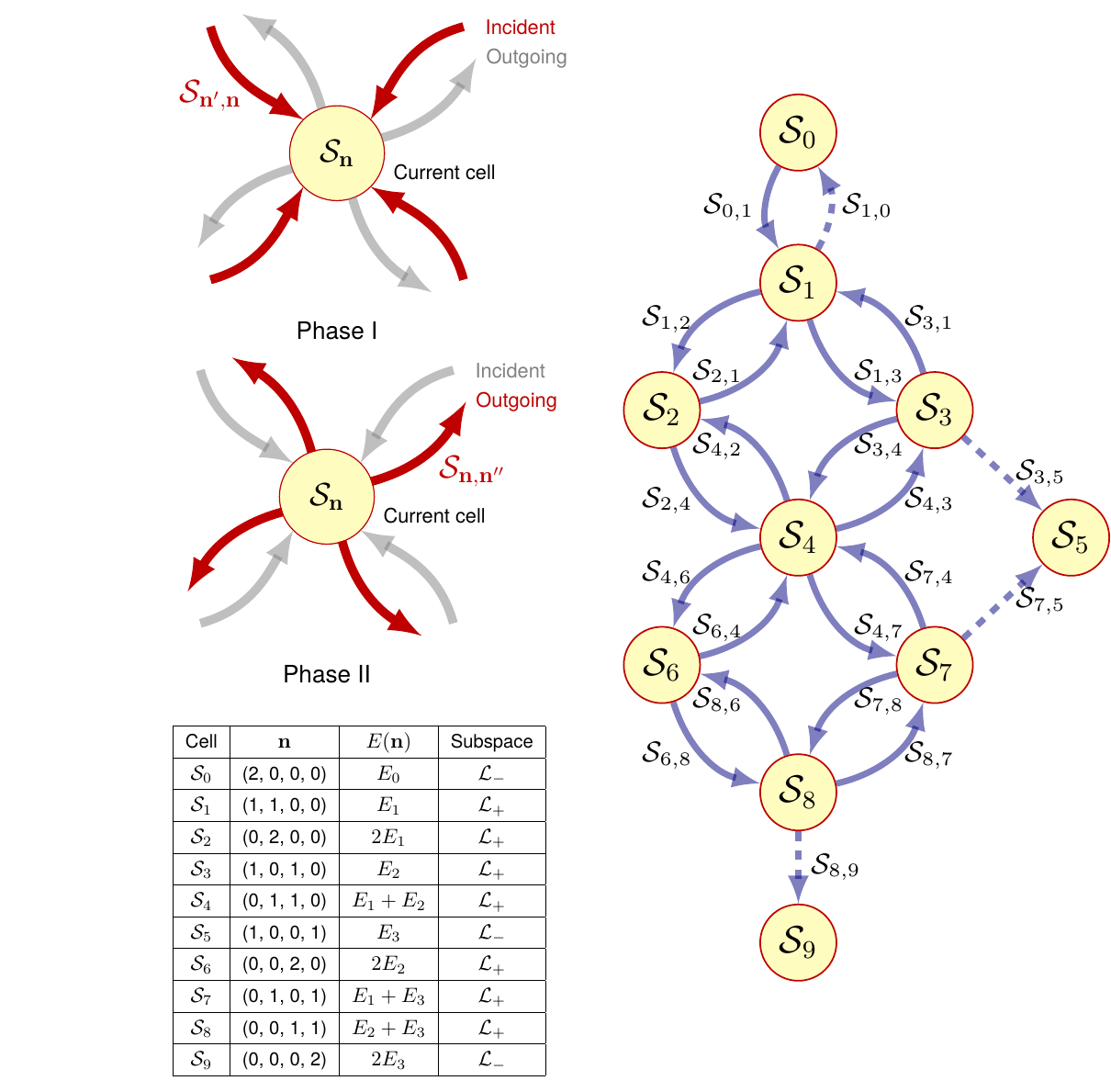}
\end{center}
\caption{The cellular automaton generated for the example considered in Figure \ref{fig:num_ex_big}. Here each cell corresponds to an energy level of the unperturbed system $H=H^{(1)}+H^{(2)}$. The sets of 4-tuples $\mathcal{S}_i$ and  $\mathcal{S}_{i,j}$ at each cell and each directed edge store lists of 4-tuples $(\tilde{\bf c},{\bf b},\xi,\mu)$. For details, refer to the long version. (a) and (b): Schematic diagrams for illustrating the two sequential steps executed when updating the state of each cell during an iteration. (c) A table listing the energy combinations ${\bf n}$, energy $E({\bf n})$ and the subspace (low energy $\mathcal{L}_-$ or high energy $\mathcal{L}_+$) associated with each cell. (d) The cellular automaton constructed for the example considered in Figure \ref{fig:num_ex_big} and Equation \ref{eq:bitflip}. Here the dashed lines corresponds to edges that go from a node in $\mathcal{L}_+$ to one in $\mathcal{L}_-$, which is only present in the automaton during the final step.}
\label{fig:multi_cubewalk}
\end{figure}

\begin{figure}
\hspace*{-0.8cm}
\begin{center}
\includegraphics[scale=1]{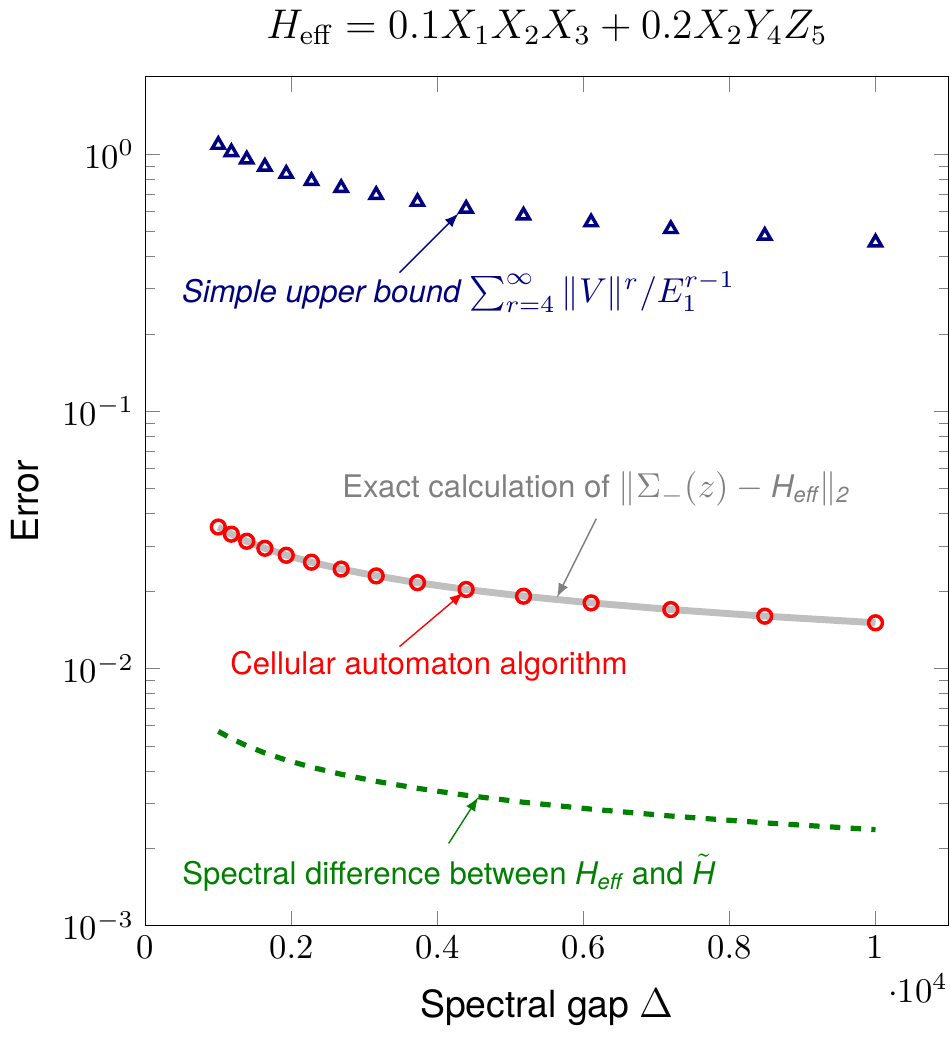}
\end{center}
\caption{Comparison between the upper bounds computed using the cellular automaton algorithm and the norm computed using (inefficient) explicit method. The ``actual spectral error'' in this plot shows the maximum difference between the eigenvalues of $H_\text{eff}$ and their counterparts in $\tilde{H}$, which are the energies of its $2^N$ lowest eigenstates with $N=5$ being the number of particles that $H_\text{eff}$ acts on (Figure \ref{fig:num_ex_big}b). The actual spectral error is always lower than the error computed based on $\|\Sigma_-(z)-H_\text{eff}\|_2$ because $\|\Sigma_-(z)-H_\text{eff}\|_2\le\epsilon$ is only a \emph{sufficient} condition that guarantees the spectral difference between $\tilde{H}$ and $H_\text{eff}$ being within $\epsilon$ (see Theorem 1 of the long version).}
\label{fig:num_plot}
\end{figure}

For example, in quantum simulation one often wishes to construct a two-body physical system $\tilde{H}$ whose low energy effective interactions $H_\text{eff}$ are many-body \cite{KKR06,OT06,JF08}. The most general construction of $\tilde{H}$ to date that could generate arbitrary many-body dynamics in $H_\text{eff}$ is based on perturbation theory. Here in Figure \ref{fig:num_ex_big} we show one example of such construction with $H_\text{eff}=\alpha_1X_1X_2X_3+\alpha_2X_2Y_4Z_5$ being three-body while $\tilde{H}=H+V$ is entirely two-body \cite{JF08}: 
\begin{equation}\label{eq:bitflip}
\begin{array}{ll}
H = H^{(1)} + H^{(2)}, & \displaystyle\qquad\qquad H^{(1)}=\frac{\Delta}{4}(Z_{u_1}Z_{u_2}+Z_{u_2}Z_{u_3}+Z_{u_1}Z_{u_3}) \\[0.1in]
& \displaystyle\qquad\qquad H^{(2)}=\frac{\Delta}{4}(Z_{v_1}Z_{v_2}+Z_{v_2}Z_{v_3}+Z_{v_1}Z_{v_3}) \\[0.1in]
V = V^{(1)} + V^{(2)}, & \qquad\qquad V^{(1)}=\mu_1(X_1X_{u_1}+X_2X_{u_2}+X_3X_{u_3}) \\[0.05in]
& \qquad\qquad V^{(2)}=\mu_2(Y_4X_{v_1}+X_2X_{v_2}+Z_5X_{v_3})
\end{array}
\end{equation}
where spins with $u_i$ and $v_i$ labels belong to the two unperturbed subsystems. Here we let $\Delta$ be orders of magnitude larger than $\mu_1$ and $\mu_2$ and keep the coefficients $\mu_1$ and $\mu_2$ as
$\mu_1 = ({\alpha_1\Delta^2}/{6})^{1/3}, \mu_2 = ({\alpha_2\Delta^2}/{6})^{1/3}$.
Perturbative calculation on $\tilde{H}$ show that the leading three orders $T_1+T_2+T_3=H_\text{eff}\otimes\Pi$ for some projector $\Pi$ acting on a Hilbert space separate from that of $H_\text{eff}$. The simulator Hamiltonian $\tilde{H}$ is constructed such that the perturbative series converges. In our example $\tilde{H}$ consists of only two-body spin interactions and parameters $\omega=0$, $\lambda_1=\mu_1$, $\lambda_2=\mu_2$ and $M_{st}$ can be computed from Figure \ref{fig:num_ex_big}d. The cellular automaton in this case is set up as in Figure \ref{fig:multi_cubewalk}. We then proceed to evolve the cellular automaton, gathering outputs from the cells corresponding to the low energy subspace. As shown in Figure \ref{fig:num_plot}, even with the convergence, simple geometric series upper bounds fail to capture the true magnitude of $\|T_r\|_\infty$ while the output of our cellular automaton algorithm is essentially tight with respect to the true value. Note that the true value takes an exponential amount of computational effort in $m$ while our cellular automaton algorithm costs only polynomial in $m$, as discussed before. This implies that we could obtain efficient and accurate estimations for the error of our quantum simulation that are not previously available.

Beyond quantum computing, our algorithm should retain its effectiveness for general spin systems and find its application in greater areas of condensed matter physics. 
For example, dimensional scaling method, pioneered by Herschbach \cite{HAG93}, uses  the inverse space dimensionality  as a  perturbation free parameter to solve complex many-body problems
 by taking the large-dimensional limit as the zeroth order approximation.  At this limit many problems admit a simple solution, as in the electronic structure calculations of atoms and molecules. 
Moreover, the second-order term also can be calculated but the higher order terms are cumbersome and hard to estimate \cite{HAG93}.   
This new proposed algorithm might be useful to estimate the perturbation error in dimensional scaling method which will lead to a very powerful and efficient approach to solve complex many-body problems.  
 Like tensor networks, which triggered an entirely new direction of research, it would be exciting to see what deeper truths of our quantum world could be unveiled by innovative proposals of algorithms and data structures.

\bibliographystyle{unsrt}
\bibliography{ref}

\newpage


\section*{Supplementary material}

# Supplementary material for "Efficient estimation of perturbation error with cellular automaton"

Yudong Cao[*][1] and Sabre Kais[†][2,3,4]

[1]Department of Computer Science, Purdue University, West Lafayette, IN 47906, USA
[2]Department of Chemistry, Purdue University, West Lafayette, IN 47906, USA
[3]Qatar Energy and Environment Research Institute, HBKU, Doha, Qatar
[4]Santa Fe Institute, 1399 Hyde Park Rd, Santa Fe, NM 87501, USA

## 1 Summary of main ideas

The framework of perturbation theory starts from identifying a physical system $\tilde{H}$ as a combination of an unperturbed system $H$ and a perturbation $V$. In this work we consider a general setting where the unperturbed system consists of $m$ identical constituents $H^{(i)}$ we call *subsystems* for which the spectrum is fully known. The perturbation $V$ couples each of the unperturbed subsystems to a common "bath". The couplings between the bath and each of the subsystems need are not necessarily identical. For computing a (for example) low-energy effective Hamiltonian of $\tilde{H}$ one usually truncates the perturbation expansion to some fixed order $r$. The terms from $(r+1)$-th order and onward are considered as the perturbation error. Our goal here is to find tight upper bounds for the total magnitude of these error terms without incurring computational cost that scales exponentially in $m$, which is the cost of explicitly computing the error terms at any order due to the exponential size of the Hilbert space. We will describe an algorithm with cost scaling $O(rm^r)$, which is polynomial for fixed $r$. We use a numerical example to demonstrate that in some cases our algorithm is able to find the magnitudes of error terms almost exactly.

Drawing on intuitions from linear algebra regarding matrix products, we observe that the $r$-th order perturbation is associated with specific types of $r$-step walks among the eigenstates of $H$. Let $\ell$ be the total number of energy levels of each subsystem. Then there are in total $O(\ell^{mr})$ such walks, each of which contributes a term at the $r$-th order perturbation. Since each subsystem has the freedom to be any of the $\ell$ energy levels, this clearly yields exponential complexity with respect to $m$.

Our strategy for dealing with the exponential size of $\tilde{\mathcal{H}}$ is to take each possible *energy configuration* $\mathbf{c} \in \{0, 1, \cdots, \ell-1\}^m$, which is an assignment of energy levels to each subsystem $1, 2, \cdots, m$, and permute the subsystems such that the energy levels are non-decreasing. We call this new configuration the *reduced energy configuration* $\tilde{\mathbf{c}}$. This sorting operation substantially reduces the space of configurations that need to be concerned because for each reduced energy configuration $\tilde{\mathbf{c}}$ there could be up to $m!$ energy configurations $\mathbf{c}$ that are *consistent* with $\tilde{\mathbf{c}}$ *i.e.* sorting the elements of $\mathbf{c}$ in non-descending order generates $\tilde{\mathbf{c}}$. We show that the total number of such possible reduced energy configuration is $O(m^\ell)$, which is polynomial assuming $\ell$ is a constant that is only dependent on each subsystem.

The transformation from the space of energy configuration $\mathbf{c}$ to the space of reduced energy configuration $\tilde{\mathbf{c}}$ can be regarded as partitioning the set of vectors $\mathbf{c}$, which is of size $O(\ell^m)$, into $O(m^\ell)$ subsets each of which contains $\mathbf{c}$ vectors that are consistent with the same reduced energy configuration $\tilde{\mathbf{c}}$. The task of summing over $r$-step walks among the $H$ eigenstates is first translated into a summation over $r$-step walks in the space of energy configuration $\mathbf{c}$. Then we use the mapping from $\mathbf{c}$ to $\tilde{\mathbf{c}}$ to reduce the summation over walks in $\mathbf{c}$ to walks in $\tilde{\mathbf{c}}$, which is manageable with $O(m^{\ell r})$ computation. We connect a single walk in $\tilde{\mathbf{c}}$, which we denote as $\mathcal{W}^{\tilde{\mathbf{c}}}$, with the set of all walks in $\mathbf{c}$ that is consistent with $\mathcal{W}^{\tilde{\mathbf{c}}}$ using *monomial symmetric polynomials*. Key to the efficiency of our approach is the property that any symmetric polynomial of $m$ variables could contain up to $m!$ monomial terms but can be evaluated in poly$(m)$ time. With the specific setup of our algorithms, these symmetric polynomials turn out to accurate reflect the permutation symmetry involved in the relationship between $\tilde{\mathbf{c}}$ and the set of energy

---

[*]cao23@purdue.edu
[†]kais@purdue.edu

configurations **c** that are consistent with **c̃**. This enables us to sum over walks in an exponentially large set with only polynomial computational cost.

The summation over walks in the space of reduced energy configurations **c̃** is coordinated by algorithms that use *cellular automaton* as their data structure. In a nutshell, a cellular automaton is a network of *cells* where each cell stores some data as its *state*. In our case both the cells and the edges that connect the cells store data. With specific initial assignments of states, the states of the cells undergo a process of evolution which is essentially repeated updates of the cell states. At each iteration, the state of a cell is updated according to a fixed rule and its new state is only dependent on its neighboring cells *i.e.* cells that are connected to the current cell. We present algorithms that construct a cellular automaton given $\tilde{H}$, and performs summation over the $r$-step walks in **c̃** by setting appropriate initial states and evolve the automaton $r$ times. Finally we glean the computed upper bound for the $r$-th order term from the final states of the cells. Assuming $\ell$ and $r$ are fixed, the entire algorithm requires computational cost that is polynomial in $m$, which is related to the size of the system.

In order to demonstrate the use of our algorithms, we consider a concrete physical system of 11 spins and find its low-energy effective Hamiltonian by truncating to 3rd order perturbation theory. The terms that are 4th order and higher are considered error terms. We apply our algorithm on this example and numerically show that our method is able to almost exactly estimate the magnitude of the error terms. This provides evidence that tighter bounds than what our algorithm produces are likely hard to achieve.

### Organization of the Supplementary Material

Section 2 lays the mathematical foundations for presenting the algorithm. Section 2.1 introduces the assumed physical setting. Section 2.2 introduces the perturbation theory formalism that we use. Section 2.3 expands on the intuition about viewing matrix products as walking on a graph and introduces its connection to infinity norm, which will become useful in later developments. Section 2.4 introduces symmetric polynomials, which serve as the bedrock of our algorithms. Section 2.5 discusses cellular automaton from the perspective of existing literature and the differences and similarities between our construction and existing ones.

Section 3 further elaborates the content of Section 2 in the context of perturbation theory and derives an upper bound for the magnitude of $r$-th order term as a sum of walks in the space of reduced energy configurations. Section 3.1 builds on Section 2.1 to elaborate on the structure of $V$. Section 3.2 builds on the perturbation theory outlined in Section 2.2 by applying the notions introduced in Section 3.1. Section 3.3 builds on the linear algebraic intuition described in Section 2.3 by incorporating it into the perturbation theory in Section 3.2. Section 3.4 carries the notion of walking among $H$ eigenstates, which is introduced in Section 3.3, into the domain of energy configurations **c**. Section 3.5 describes how to transform the sum over walks in energy configurations **c** to a sum over walks in *reduced* energy configurations **c̃** by using the symmetric polynomial defined in 2.4, see Lemma 6.

Section 4 is the main section introducing our algorithms for computing the upper bounds established in Section 3. Section 4.1 describes the algorithm used for constructing the cellular automaton given the physical setting. Section 4.2 describes the update rules for the cells. Section 4.3 shows the final algorithm for computing upper bounds of perturbative terms at arbitrary order $r$.

Section 5 shows a concrete example of a physical system and we conclude with Section 6, where we discuss the potential uses of our technique in a broader context of physical theories that require perturbative treatment. **Due to a large amount of symbols and notations introduced in this Supplementary Material, we provide a glossary for these symbols in alphabetical order in Appendix A.**

## 2 Preliminaries

### 2.1 Basic setting

We consider the most general setting of perturbation theory, where we have an *unperturbed Hamiltonian* $H$ with an energy gap $\Delta$ between its ground state subspace $\mathcal{L}_-$ and the rest of its spectrum which we denote as $\mathcal{L}_+$. Naturally in the eigenbasis of $H$ we could write down $H$ as a block diagonal operator:

$$H = \begin{pmatrix} H_+ & \\ & H_- \end{pmatrix}. \tag{1}$$

Then we add a perturbation $V$ to the unperturbed Hamiltonian. Here we assume $\|V\|_2 < \Delta/2$. Here $\|\cdot\|_2$ is the 2-norm defined as $\|A\|_2 = \max_{\||\psi\rangle\|=1} \|A|\psi\rangle\|$. In the same basis we could write $V$ as a block matrix

$$V = \begin{pmatrix} V_+ & V_{+-} \\ V_{-+} & V_- \end{pmatrix}. \tag{2}$$

For a parameter $z$ such that $|z| \ll \Delta$, define *operator valued resolvent* $G(z) = (zI - H)^{-1}$. Then like $H$, $G$ is also block diagonal in the eigenbasis of $H$.

Suppose we are most concerned with the low-energy subspace of the perturbed Hamiltonian $\tilde{H} = H + V$, which is spanned by all the eigenvectors of $\tilde{H}$ with eigenvalues that are less than $\Delta/2$. However, we do not require that the ground state of $H$ be necessarily non-degenerate.

The unperturbed Hamiltonian $H$ should correspond to some finite physical system with $\ell$ energy levels $E_0$, $E_1$, $E_2$, $\cdots$, $E_{\ell-1}$ with the corresponding eigenspaces which we denote as $\mathcal{P}_0$, $\mathcal{P}_1$, $\mathcal{P}_2$, $\cdots$, $\mathcal{P}_{\ell-1}$ and the respective projectors as $P_0, P_1, P_2, \cdots, P_{\ell-1}$.

Without loss of generality assume $E_0$, the ground state energy of $H$, is zero. The energy values $E_i$ do not have to be distinct or monotonically increasing but they should be separable into two subsets with one corresponding to the low-energy subspace $\mathcal{L}_- = \text{span}\{|E_j\rangle | E_j < \Delta/2\}$ and the other one corresponding to the rest of the spectrum $\mathcal{L}_+ = \text{span}\{|E_j\rangle | E_j > \Delta/2\}$.

Now let us consider a setting with $m$ identical copies of such systems described by $H$, each of which we call a *subsystem*. In this case all of the $m$ subsystems are mutually non-interacting. The possible total energy values of the this $m$-copy system are thus simply linear combinations of energy levels of each subsystem. In essence, the spectrum of the $m$-copy system can be described by the set

$$\left\{ E = \sum_{i=0}^{\ell} n_i E_i \ \Big| \ \sum_{i=0}^{\ell} n_i = m, \quad n_i \in \mathbb{Z}, \quad 0 \le n_i \le m \right\}. \tag{3}$$

Let $\mathcal{H}$ be the Hilbert space where $H$ dwells. As a notation we use $\mathcal{H}^{(i)}$, $i = 1, \cdots, m$, to denote the Hilbert space associated with the $i$-th subsystem. Let $H^{(i)}$ be the Hamiltonian of the $i$-th subsystem. Correspondingly we introduce the notations for eigenvalues $E_j^{(i)}$, eigenspaces $\mathcal{P}_j^{(i)}$ spanned by eigenvectors $|\psi_{j,p}^{(i)}\rangle$ with $p$ ranging from 1 to $\dim(\mathcal{P}_j^{(i)})$ and their projectors $P_j^{(i)}$ defined as

$$P_j^{(i)} = \sum_{p=1}^{\dim\left(\mathcal{P}_j^{(i)}\right)} |\psi_{j,p}^{(i)}\rangle\langle\psi_{j,p}^{(i)}|. \tag{4}$$

where $|\psi_{j,p}^{(i)}\rangle$ represents the $p$-th degenerate eigenstate of $H^{(i)}$ with energy $E_j$.

Now we further introduce perturbation $V$ for each subsystem, by letting each of the subsystems interact with a common "bath" with Hilbert space $\mathcal{B}$, as illustrated in Figure 1a of the main text. $V$ contains a sum of terms $V^{(i)}$ that couples the eigenspace $\mathcal{H}^{(i)}$ of the $i$-th unperturbed subsystem with the Hilbert space of the "bath" $\mathcal{B}$ by acting non-trivially on the joint space $\mathcal{H}^{(i)} \otimes \mathcal{B}$.

The "bath" by itself has its own internal dynamics governed by some Hamiltonian we write as $H_\mathcal{B}$. This Hamiltonian describes interactions in $\mathcal{B}$ that are independent of each subspace $\mathcal{H}^{(i)}$. We point out that both the $H^{(i)}$'s and $V^{(i)}$'s act on the total Hilbert space $\tilde{\mathcal{H}} = \mathcal{H}^{(1)} \otimes \mathcal{H}^{(2)} \otimes \cdots \otimes \mathcal{H}^{(m)} \otimes \mathcal{B}$ but only non-trivially on $\mathcal{H}^{(i)}$ for the $H^{(i)}$'s and $\mathcal{H}^{(i)} \otimes \mathcal{B}$ for the $V^{(i)}$'s. Like before we could also partition each of the local subspace $\mathcal{H}^{(i)}$ into low and high energy subspaces $\mathcal{L}_-^{(i)}$ and $\mathcal{L}_+^{(i)}$ such that $\mathcal{H}^{(i)} = \mathcal{L}_-^{(i)} \oplus \mathcal{L}_+^{(i)}$. Then the total Hilbert space can be written as $\tilde{\mathcal{H}} = (\mathcal{L}_- \oplus \mathcal{L}_+) \otimes \mathcal{B}$ where $\mathcal{L}_- = \mathcal{L}_-^{(1)} \otimes \mathcal{L}_-^{(2)} \otimes \cdots \otimes \mathcal{L}_-^{(m)}$ and $\mathcal{L}_+$ is the complement of $\mathcal{L}_-$ in the subspace $\mathcal{H}^{(1)} \otimes \mathcal{H}^{(2)} \otimes \cdots \otimes \mathcal{H}^{(m)}$.

With definitions of subspaces in place, we define the unperturbed Hamiltonian $H$ and the perturbation $V$ as

$$H = \sum_{i=1}^{m} H^{(i)}, \qquad V = H_\mathcal{B} + \sum_{i=1}^{m} V^{(i)}. \tag{5}$$

For each subsystem $i$, we assume that the perturbation $V$ induces only transitions between $\mathcal{P}_j^{(i)}$ and $\mathcal{P}_k^{(i)}$ such that $j$ and $k$ differ by at most one. In other words, for any $i = 1, 2 \cdots, m$, we assume that the perturbation $V$ be block tridiagonalizable in the eigenbasis of $H$:

$$V^{(i)} = \mathbf{1}_{\mathcal{H}^{(1)}} \otimes \mathbf{1}_{\mathcal{H}^{(2)}} \otimes \cdots \mathbf{1}_{\mathcal{H}^{(i-1)}} \otimes \begin{pmatrix} O_{00}^{(i)} & O_{01}^{(i)} & & & \\ O_{10}^{(i)} & O_{11}^{(i)} & O_{12}^{(i)} & & \\ & O_{21}^{(i)} & \ddots & \ddots & \\ & & \ddots & O_{\ell-2,\ell-2}^{(i)} & O_{\ell-2,\ell-1}^{(i)} \\ & & & O_{\ell-1,\ell-2}^{(i)} & O_{\ell-1,\ell-1}^{(i)} \end{pmatrix} \otimes \mathbf{1}_{\mathcal{H}^{(i+1)}} \otimes \cdots \otimes \mathbf{1}_{\mathcal{H}^{(m)}}. \quad (6)$$

Here each block $O_{jk}^{(i)}$ represents the transition driven by the perturbation $V$ from states in the eigenspace $\mathcal{P}_j^{(i)}$ to those in $\mathcal{P}_k^{(i)}$. With further block permutation by grouping the blocks $O_{jk}^{(i)}$ according to whether indices $j$ and $k$ correspond to $+$ or $-$ subspace, we could rewrite $\tilde{H}$ in the block form consistent with 1 and 2.

## 2.2 Perturbation theory

Let $\Pi_-$ and $\Pi_+$ be projectors onto the subspaces $\mathcal{L}_-$ and $\mathcal{L}_+$ respectively. Then the block form of Equations 1 and 2 still holds for the definitions of $H$ and $V$ in Equation 5. More generally, any operator $\mathcal{O}$ can be written as the block form

$$\begin{pmatrix} \mathcal{O}_+ & \mathcal{O}_{+-} \\ \mathcal{O}_{-+} & \mathcal{O}_- \end{pmatrix}. \quad (7)$$

Our goal is to find a series expansion that approximates the low-energy effective Hamiltonian of the perturbed system $\tilde{H}_-$. In Section 2.1 we defined the operator-valued resolvent $G(z) = (zI - H)^{-1}$. We could similarly define operator-valued resolvent $\tilde{G}(z) = (zI - \tilde{H})^{-1}$ for $z \ll \Delta$ where $I$ is the identity acting on $\tilde{\mathcal{H}}$. We could relate $\tilde{G}$ with $G$ by $\tilde{G} = (G^{-1} - V)^{-1}$, which gives rise to a Taylor expansion

$$\tilde{G} = G(I - VG)^{-1} = G + GVG + GVGVG + \cdots \quad (8)$$

We could then introduce the central object of our concern, namely the *self-energy expansion* $\Sigma_-(z) = zI - (\tilde{G}_-(z))^{-1}$. Applying 7 and 8 on $\Sigma_-(z)$ leads to

$$\Sigma_-(z) = H_- + V_- + V_{-+}G_+V_{+-} + V_{-+}G_+V_+G_+V_{+-} + \cdots = H_- + V_- + \sum_{r=2}^{\infty} T_r. \quad (9)$$

The self-energy $\Sigma_-(z)$ is important for approximating the low-energy effective Hamiltonian $\tilde{H}_-$. The following theorem makes this intuition precise.

**Theorem 1** ([6], [7]). *Given a Hamiltonian $\tilde{H} = H + V$. Suppose $\|V\|_2 \leq \Delta/2$ with $\Delta$ being the spectral gap between the ground and the first excited state of $H$. If there exists a Hamiltonian $H_{eff}$ whose energies are contained in the interval $[a, b]$ and some real constant $\epsilon > 0$ such that $a < b < \Delta/2 - \epsilon$ and for any $z \in [a - \epsilon, b + \epsilon]$,*

$$\|\Sigma_-(z) - H_{eff}\|_2 \leq \epsilon, \quad (10)$$

*then the $j$-th eigenvalue $\tilde{\lambda}_j$ of $\tilde{H}_-$ and the corresponding $j$-th eigenvalue of $H_{eff}$ differ by at most $\epsilon$, for any appropriate range of $j$ values.*

Most uses of perturbation theory involve truncating the perturbative expansion 9 to a specific order to obtain an effective Hamiltonian $H_{\text{eff}}$ that approximates the exact solution. Theorem 1 is valuable in the sense that it establishes a connection between the magnitude of the error term $\|\Sigma_-(z) - H_{\text{eff}}\|_2$ and the quality of $H_{\text{eff}}$ as an approximation to $\tilde{H}$, modulo certain conditions that are clearly satisfied by our assumed physical setting described in Section 2.1. The task of evaluating the quality of perturbative approximation is then reduced to the task of estimating the perturbative error $\|\Sigma_-(z) - H_{\text{eff}}\|_2$. More specifically, our goal is to find a tight yet efficiently computable upper bound for the norm of the $r$-th order term $T_r$ which is

$$T_r \equiv V_{-+}(G_+V_+)^{r-2}G_+V_{+-}, \quad (11)$$

Obviously one can obtain a crude bound by triangle inequality and submultiplicativity of operator norm (namely $\|AB\|_2 \leq \|A\|_2 \cdot \|B\|_2$)

$$\|T_r\|_2 \leq \|V\|_2^r \cdot \|G_+\|_2^{r-1}. \quad (12)$$

However, as we will demonstrate with a concrete example in Section 5, this does not serve as a bound tight enough to capture the true magnitude of $\|T_r\|_2$. In order to find a tighter bound for $\|T_r\|_2$, an extreme would be to explicitly form $T_r$ and compute $\|T_r\|_2$ directly. But the computation cost is evidently exponential in the size of the system. For the remainder of the Supplementary Material we present a middle-ground possibility where a tighter bound than $\|V\|_2^r \cdot \|G_+\|_2^{r-1}$ can be obtained by efficient computation. We show that in certain cases the bound obtained is even equal to the value of $\|T_r\|_2$, providing evidence that significant improvement over our approach for general settings is likely difficult.

## 2.3 Matrix product, walks on graphs and the infinity norm

In this section we note a few intuitions concerning matrices that will be instrumental to our later discussions. We start by pointing out the connection between matrix products and walks on graphs. An $N \times N$ matrix $A = \sum_{i,j} a_{ij}|i\rangle\langle j|$ could be considered as a weighted directed graph on $N$ nodes with the edge from $i$ to $j$ having weight $a_{ij}$. In other words, each element $a_{ij}$ signifies the "weight" of a walk $i \to j$. If we consider the product between $A$ and another $N \times N$ matrix $B = \sum_{i,j} b_{ij}|i\rangle\langle j|$, the $(i,j)$ element of the product $AB$ is $(AB)_{ij} = \sum_k a_{ik}|i\rangle\langle k| \cdot b_{kj}|k\rangle\langle j|$, which is a 2-step walk $i \to j \to k$. One could think of our central object $T_r$ defined in Equation 11 as a collection of $r$-step walks in the space of $H$ eigenstates. We will make this notion precise later.

Much of our arguments in our proofs of correctness for the algorithms will be based on $\infty$-norm, instead of 2-norm, of matrices. As a simple reminder, the $\infty$-norm of an $m \times n$ matrix $A$ is defined as $\|A\|_\infty = \max_{1 \leq i \leq m} \sum_{j=1}^n |a_{ij}|$, which is simply the maximum absolute row sum of the matrix. We will be using the following properties of the infinity norm of matrices:

1. For any matrices $A$ and $B$ of compatible dimensions, $\|A + B\|_\infty \leq \|A\|_\infty + \|B\|_\infty$;

2. For any matrices $A$ and $B$ of compatible dimensions, $\|AB\|_\infty \leq \|A\|_\infty \cdot \|B\|_\infty$;

3. $\|A \otimes \mathbf{1}\|_\infty = \|A\|_\infty$ where $\mathbf{1}$ is an identity matrix of any finite dimension. Similarly $\|\mathbf{1} \otimes A\|_\infty = \|A\|_\infty$;

4. If $A$ is a block matrix and let $A_{ij}$ be the $(i,j)$ block[1], then $\|A\|_\infty \leq \max_i \sum_j \|A_{ij}\|_\infty$;

5. For a Hermitian matrix $A$, we have $\|A\|_2 \leq \|A\|_\infty$. This follows from $\|A\|_2^2 \leq \|A\|_1 \cdot \|A\|_\infty$ and $\|A\|_1 = \|A\|_\infty$ for Hermitian matrices.

Here Property 5 is useful because it ties directly to 2-norm, which has a natural connection to the spectrum of the matrix and is more commonly used for characterizing the magnitude of perturbative error $T_r$ at any order $r$. Our algorithms, on the other hand are intended for computing upper bounds to $\|T_r\|_\infty$. Property 5 thus guarantees that the upper bounds computed for $\|T_r\|_\infty$ also serve as upper bounds to $\|T_r\|_2$.

We prefer to use infinity norm in the context of this work because of its natural connection to the element-wise or block-wise structure of a matrix. Drawing on the connection mentioned in the opening paragraph, consider the powers of a block matrix $A$, namely $A^n$. Following the notation in Property 4, let $A_{ij}$ be the $(i,j)$ block. Assume $A$ is an $k \times k$ block matrix. If we think of the matrix $A$ as a directed weighted graph on $k$ nodes where each edge going from node $i$ to $j$ is associated with "weight" $A_{ij}$, then the $(i,j)$ block of $A^n$ essentially is a sum over contributions from all $n$-step walks $i_0 \to i_1 \to i_2 \to \cdots \to i_n$ on the graph of $A$ that starts from $i_0 = i$ and ends at $i_n = j$. Each one of such $n$-step walk contributes a term $A_{i_0 i_1} A_{i_1 i_2} \cdots A_{i_{n-1} i_n}$ to the $(i,j)$ block of $A^n$. Hence if use $(A^n)_{ij}$ to denote the $(i,j)$ block of $A^n$,

$$(A^n)_{ij} = \sum_{i_1, i_2, \cdots, i_{n-1}} A_{ii_1} A_{i_1 i_2} \cdots A_{i_{n-2} i_{n-1}} A_{i_{n-1} j}. \tag{13}$$

Using Property 1, 2, 4 and 5 of infinity norm on Equation 13 we could find an upper bound

$$\|A^n\|_2 \leq \max_{i=1,\cdots,k} \sum_{j=1}^k \sum_{i_1, i_2, \cdots, i_{n-1}} \|A_{ii_1}\|_\infty \cdot \|A_{i_1 i_2}\|_\infty \cdots \|A_{i_{n-2} i_{n-1}}\|_\infty \cdot \|A_{i_{n-1} j}\|_\infty. \tag{14}$$

Equation 14 underlies the basic intuition of our approach in finding a tight upper bound to $\|T_r\|_2$. Similar to Equation 14, $T_r = V_{-+}(G_+ V_+)^{r-2} G_+ V_{+-}$ also contains a basic structure of powering the matrix $G_+ V_+$. As later discussion would reveal, in the context of bounding $\|T_r\|_\infty$ the walks over which the right hand side of Equation 14

---

[1] In our notation $(i,j)$ block means the block on the $i$-th row and $j$-th column.

5

sums over correspond to sequences of transitions among eigenstates of the unperturbed Hamiltonian $H$. However, note that the sum over $i_1, i_2, \cdots, i_{n-1}$ in Equation 14 contains an exponential number of terms in $n$ due to the permutation of indices, which means any naive algorithm that computes the right hand side of Equation 14 will likely be inefficient. We introduce a mathematical tool in the next section to help with this inefficiency due to combinatorics.

## 2.4 Symmetric polynomials

*Symmetric polynomials* are used in our algorithms as a fundamental data structure to address the combinatorics of arbitrary-order virtual transitions in the perturbative expansion. We start with a few definitions. Any *monomial* in $n$ variables $x_1, x_2, \cdots, x_n$ can be written as $x_1^{a_1} \cdots x_n^{a_n}$ where the exponents $\alpha_i \in \{0, 1, 2, \cdots\}$. Writing $\mathbf{a} = (a_1, a_2, \cdots, a_n)$ and $\mathbf{x} = (x_1, x_2, \cdots, x_n)$ gives the abbreviated notation $\mathbf{x}^{\mathbf{a}} = x_1^{a_1} \cdots x_n^{a_n}$.

**Definition 1** (Monomial symmetric polynomial). *The* monomial symmetric polynomial $m_{\mathbf{a}}(\mathbf{x})$ *is defined as the sum of all monomials $\mathbf{x}^{\mathbf{a}}$ where $\mathbf{a}$ ranges over all distinct permutations of elements in $\mathbf{a} = (a_1, a_2, \cdots, a_n)$. Here $\mathbf{a}$ can be thought of as a partition of an integer $K = \sum_{i=1}^{n} a_i$ and we say $\mathbf{a}$ is the* partition *of $m_{\mathbf{a}}(\mathbf{x})$.*

Note that by definition, a monomial symmetric polynomial is invariant with respect to the ordering of elements in the partition. For convenience we impose the following restrictions to the representations of partitions, which we call *reduced partition*. From here on we will only use the reduced partition to uniquely describe a monomial symmetric polynomial.

**Definition 2** (Reduced partition). *For an $n$-variable monomial symmetric polynomial $m_{\mathbf{a}}(\mathbf{x})$, let $k$ be the number of nonzero elements in $\mathbf{a}$. Then we define the* reduced partition $\mathbf{b}$ *of $m_{\mathbf{a}}(\mathbf{x})$ to be a $k$-dimensional vector formed by taking all the $k$ nonzero elements of $\mathbf{a}$ and order them in non-descending order* i.e. $b_1 \leq b_2 \leq \cdots \leq b_k$.

There is a certain combinatorial intuition associated with monomial symmetric polynomials which is important in the context of later discussions. For instance consider $m_{(1,2,3)}(a,b,c) = ab^2c^3 + ba^2c^3 + ac^2b^3 + ca^2b^3 + bc^2a^3 + cb^2a^3$. As an analogy, we could think of each variable $a, b, c$ as a bucket of coins and each term in $m_{(1,2,3)}(a,b,c)$ as a result of flipping the coins in the three buckets one at a time such that in the end one bucket gets 1 coin flips, one gets 2 coin flips and the other gets 3. Each coin flip does not have to be on different coins. For example the first term, $ab^2c^3$, corresponds to the case where we administer 1 coin flip in bucket $a$, 2 coin flips in bucket $b$ and 3 in $c$.

Another feature of monomial symmetric polynomial that we use is its compactness in representation. For $\mathbf{b}$ such that $\sum_{i=1}^{|\mathbf{b}|} = r$, $m_{\mathbf{b}}(x_1, \cdots, x_n)$ contains $O(n^r)$ terms, while all the information for generating these terms can be condensed to $\mathbf{b}$, a $k$-element vector. As is shown in [1], for a fixed partition $\mathbf{b}$, evaluating $m_{\mathbf{b}}(x_1, x_2, \cdots, x_n)$ takes $O(r!n)$ time. In our context $r$ is the order of perturbation, which is assumed to be fixed. Hence the cost of evaluating symmetric polynomials scales *linearly* as the number of variables (or in our context the number of unperturbed subsystems).

## 2.5 Cellular automata

A *cellular automaton* (CA) is typically defined as a collection of finite-state machines called *cells* that are positioned on a grid of any finite dimension. Each cell in the grid also has a defined set of other cells as its *neighborhood*. The initial configuration of the automaton is specified by assigning states to each cell in the grid. The cells evolve together in discrete time steps, each time updating the state of each cell by a rule that is identical for each cell and does not change over time. During each time step, the rule determines the new state of each cell in terms of the current state of the cell and the states of the cells in its neighborhood.

While the initially proposed CA constructions adhere strictly to the definitions above, CA constructions that deviate from the above definitions abound. This has significantly added flexibility in the use of the terminology. For example,

- The states of cells need not be discrete; continuous-valued CAs in two-dimensions have been explored [8];
- The grid that joins the cells could be more than two-dimensional [5];
- More generally, the states of the cells do not necessarily have to be single numbers, but could also be data structures [8].

In this work we construct CAs that admit all three variations, namely CAs with cells connected in form of a (possibly high dimensional) grid and cell states that consist of data structures designed to specifically suit our purpose. However, our construction retains some typical features of cellular automata:

- The update rules are *local* in the sense that the states of the cells are only dependent on their neighbors;
- The update rules are *homogeneous* in that they are identical and time independent for all cells;
- The states of the cells are updated *in parallel* to produce a new generation.

An important problem concerning the theory of cellular automata is "What higher-level descriptions of information processing in cellular automata can be given?"[9]. There have been prior works [4] on CA constructions that are strongly based on analogues with conventional serial-processing computers. However, information processing in cellular automata occurs in a fundamentally distributed and parallel fashion. In this sense, the CAs constructed in this work perform computations in ways that departs from conventional serial computer models: to obtain an upper bound to the norm of $m$-th order perturbative term, we evolve the CA for $m$ evolutions and glean results from the states of a specific subset of cells.

# 3 Upper bounds for arbitrary order perturbation theory

## 3.1 Structure of the perturbation

In our basic setting we have assumed the perturbation $V$ be block tridiagonalizable with respect to subspaces of $\mathcal{H}^{(i)}$, see Equation 6. Each block $O_{jk}^{(i)}$ by itself has a block structure. Each $O_{jk}^{(i)}$ is a $\dim(\mathcal{P}_j^{(i)}) \times \dim(\mathcal{P}_k^{(i)})$ array of operators $B_{pq,jk}^{(i)}$ that only acts on $\mathcal{B}$. Explicitly,

$$O_{jk}^{(i)} = \begin{pmatrix} B_{11,jk}^{(i)} & B_{12,jk}^{(i)} & \cdots & B_{1K,jk}^{(i)} \\ B_{21,jk}^{(i)} & B_{22,jk}^{(i)} & \cdots & B_{2K,jk}^{(i)} \\ \vdots & \vdots & \ddots & \vdots \\ B_{J1,jk}^{(i)} & B_{J2,jk}^{(i)} & \cdots & B_{JK,jk}^{(i)} \end{pmatrix} \tag{15}$$

where for convenience we define $J = \dim(\mathcal{P}_j^{(i)})$ and $K = \dim(\mathcal{P}_k^{(i)})$. Here $B_{pq,jk}^{(i)}$ describes the action on $\mathcal{B}$ that is coupled with transition from the $p$-th degenerate state in $\mathcal{P}_j^{(i)}$ to the $q$-th degenerate state in $\mathcal{P}_k^{(i)}$.

The following definitions of quantities will become instrumental to our further development in this work.

**Definition 3** (Scalar quantity $\omega$). *Let $\omega$ be an upper bound to the norm of the components in $V$ such that*

$$\omega \geq \|H_\mathcal{B}\|_\infty + \max_{\substack{j=0,\cdots,\ell \\ i=1,\cdots,m}} \|O_{jj}^{(i)}\|_\infty. \tag{16}$$

**Definition 4** (Vector $\boldsymbol{\lambda}$). *Let $\lambda_i$ be an upper bound to the norms of the matrix elements in the off-diagonal blocks $O_{jk}^{(i)}$ (i.e. the blocks with $j$ and $k$ differing by one). In other words,*

$$\lambda_i = \max_{\substack{j,k=0,\cdots,\ell, \\ j \neq k}} \max_{\substack{p=1,\cdots,J, \\ q=1,\cdots,K}} \|B_{pq,jk}^{(i)}\|_\infty. \tag{17}$$

*For convenience we define the vector $\boldsymbol{\lambda} = (\lambda_1, \lambda_2, \cdots, \lambda_m)$.*

**Definition 5** (Matrix $\mathbf{M}$). *For each block $O_{jk}^{(i)}$ as defined in 6 and 15, let $M_{jk}^{(i)}$ be the maximum number of nonzero blocks per row in $O_{jk}^{(i)}$. In precise terms,*

$$M_{jk}^{(i)} = \max_{p=1,\cdots,J} Card\{B_{pq,jk}^{(i)}, q=1,\cdots,K | \|B_{pq,jk}^{(i)}\|_\infty \neq 0\} \tag{18}$$

*where $Card\{\cdot\}$ is the size of a set. Furthermore, let $\mathbf{M}$ be an $\ell \times \ell$ matrix such that $M_{jk} = \max_{i=1,\cdots,m} M_{jk}^{(i)}$.*

Informally, $\lambda_i$ characterizes the "strength" of perturbation $V$ acting on the subsystem $\mathcal{H}_i$ and causing a transition, while $M_{jk}^{(i)}$ characterizes the combinatorial aspect of $V^{(i)}$ inducing transitions between eigenstates in the subspaces $\mathcal{P}_j$ and $\mathcal{P}_k$. Furthermore, $M_{jk}$ represents the maximum possible ways, among all subsystems $i$, in which an unperturbed eigenstate in a subspace $\mathcal{P}_j^{(i)}$ can be transformed into an eigenstate in $\mathcal{P}_k^{(i)}$ via the action of $V^{(i)}$. From a more linear algebraic perspective, it is the maximum row sparsity of the $O_{jk}^{(i)}$ blocks among all subsystems.

## 3.2 Structure of terms at any order

The quantity $T_r = V_{-+}G_+(V_+G_+)^{r-2}V_{+-}$ is a string of matrices multiplied sequentially and we will consider finding upper bounds for the norm of each successively longer substring that starts with the first matrix $V_{-+}$. By definition of block structures introduced in Equations 6 and 15, in the general setting described in Figure 1 of the main text we could express $V_{-+}$ in terms of the finest block division $B_{pq,jk}^{(i)}$ as

$$V_{-+} = \sum_{i=1}^{m} \sum_{j:\mathcal{P}_j^{(i)} \subseteq \mathcal{L}_-^{(i)}} \sum_{k:\mathcal{P}_k^{(i)} \subseteq \mathcal{L}_+^{(i)}} \sum_{p:|\psi_{j,p}^{(i)}\rangle \in \mathcal{P}_j^{(i)}} \sum_{q:|\psi_{k,q}^{(i)}\rangle \in \mathcal{P}_k^{(i)}} B_{pq,jk}^{(i)} \otimes |\psi_{j,p}^{(i)}\rangle\langle\psi_{k,q}^{(i)}| \qquad (19)$$

where we recall that the operators $B_{pq,jk}^{(i)}$ are defined in Equation 15 and the states $|\psi_{j,p}^{(i)}\rangle$ are defined in 4.

Following Equation 19 we could also express $G_+$, $V_+$ and $V_{+-}$ in terms of blocks $B_{pq,jk}^{(i)}$ and unperturbed eigenstates $|\psi_{j,p}^{(i)}\rangle$. Starting from $G_+(z) = \Pi_+(zI - H)^{-1}\Pi_+$, before expanding $G_+$ we introduce the following notions of *energy combination* and *energy configuration* of a given eigenstate of $H$. These notions are also important in our further algorithmic development.

**Definition 6** (Energy configuration). *For an eigenstate $|\psi\rangle$ of $H$ where the energy of each subsystem $H^{(i)}$ is $E^{(i)} = \langle\psi|H^{(i)}|\psi\rangle \in \{E_0, E_1, \cdots, E_\ell\}$. We define the* energy configuration *of the eigenstate $|\psi\rangle$ as a vector $\mathbf{c} \in \{0, 1, \cdots, \ell\}^m$ with each element $\mathbf{c}_i$ be such that $E^{(i)} = E_{c_i}$. We use the notation $\mathbf{c}(|\psi\rangle)$ to refer to the energy configuration of $|\psi\rangle$.*

**Definition 7** (Energy combination). *Given an energy configuration $\mathbf{c}$, for each energy level $j$ ranging from 0 to $\ell - 1$, let $n_j$ be the number of subsystems with energy $j$. In other words $n_j = Card\{i = 1, \cdots, m | c_i = j\}$ where $Card\{\cdot\}$ is the cardinality of a set. Then we define the* energy combination *of the energy configuration $\mathbf{c}$ as $\mathbf{n}(\mathbf{c}) = (n_1, n_2, \cdots, n_\ell) \in \{1, 2, \cdots, m\}^\ell$. Conversely, let $\mathcal{C}(\mathbf{n}) = \{\mathbf{c}|\forall j \in \{0, \cdots, \ell\}, \sum_{i:c_i=j} 1 = n_j\}$ be the set of energy configuration that gives rise to a given energy combination $\mathbf{n}$.*

Informally one could think of $\mathbf{n}$ as representing the eigenstates of $H$ in a "number basis". Then $G_+(z)$ can be expressed as

$$G_+(z) = \sum_{\mathbf{n} \in \mathcal{N}_+} \frac{1}{z - E(\mathbf{n})} \sum_{\mathbf{c} \in \mathcal{C}(\mathbf{n})} P(\mathbf{c}) \qquad (20)$$

where $E(\mathbf{n}) = \sum_{j=1}^{\ell} n_j E_j$ is the total energy of the current energy combination. $\mathcal{N}_+ = \{\mathbf{n}|E(\mathbf{n}) > \Delta/2\}$ is the set of energy combination that correspond to an eigenstate of $H$ in $\mathcal{L}_+$. Similarly we could also define $\mathcal{N}_- = \{\mathbf{n}|E(\mathbf{n}) < \Delta/2\}$. $P(\mathbf{c}) = \bigotimes_{j=1}^{m} P_{c_j}^{(j)}$ is the projector onto the subspace of each subsystem as described by the energy configuration $\mathbf{c}$. Each of the projector $P_{c_j}^{(j)}$ could be further expressed as projectors onto individual eigenstates by (4).

The expression for $V_+$ in terms of blocks $B_{pq,jk}^{(i)}$ and unperturbed eigenstates $|\psi_{j,p}^{(i)}\rangle$ can be obtained by replacing $\mathcal{L}_-^{(i)}$ in the summation over $j$ in (19) by $\mathcal{L}_+^{(i)}$. Similarly, the expression for $V_{+-}$ can be obtained by replacing $\mathcal{L}_-^{(i)}$ in the summation over $j$ in (19) by $\mathcal{L}_+^{(i)}$ and at the same time replacing $\mathcal{L}_+^{(i)}$ in the summation over $k$ in (19) by $\mathcal{L}_-^{(i)}$.

## 3.3 Walk in the space of unperturbed eigenstates

With the notation $P(\mathbf{c})$ introduced in Equation 20 we could express $\Pi_-$ and $\Pi_+$ explicitly as

$$\Pi_- = \sum_{\mathbf{n} \in \mathcal{N}_-} \sum_{\mathbf{c} \in \mathcal{C}(\mathbf{n})} P(\mathbf{c}), \qquad \Pi_+ = \sum_{\mathbf{n} \in \mathcal{N}_+} \sum_{\mathbf{c} \in \mathcal{C}(\mathbf{n})} P(\mathbf{c}). \qquad (21)$$

Combining Equation (21) with the definitions of $P_j^{(i)}$ in Equation (4) we could see that the term

$$T_r = V_{-+}(G_+V_+)^{r-2}G_+V_{+-}$$

for any $r \geq 3$ consists of products of $B_{pq,jk}^{(i)} \otimes |\psi_{j,p}^{(i)}\rangle\langle\psi_{k,q}^{(i)}|$ with each term $|\psi_{j,p}^{(i)}\rangle\langle\psi_{k,q}^{(i)}|$ multiplied together forming a sequence of virtual transitions

$$\underbrace{|\phi^{(0)}\rangle\langle\phi^{(1)}|}_{V_{-+}} \cdot \underbrace{|\phi^{(1)}\rangle\langle\phi^{(1)}|}_{G_+} \cdot \underbrace{|\phi^{(1)}\rangle\langle\phi^{(2)}|}_{V_+} \cdot \underbrace{|\phi^{(2)}\rangle\langle\phi^{(2)}|}_{G_+} \cdots \underbrace{|\phi^{(r-1)}\rangle\langle\phi^{(r)}|}_{V_{+-}} \qquad (22)$$

that corresponds to a walk among the eigenstates of $H$. For convenience in the subsequent discussions we temporarily condense all the subscripts $j, p$ and superscript $(i)$ of the state $|\psi_{j,p}^{(i)}\rangle$ into a single-number superscript. To avoid confusion with the superscript notation in $|\psi_{j,p}^{(i)}\rangle$ we use $\phi$ instead of $\psi$. The superscript for $\phi$ indicates the step of a walk while the superscript for $\psi$ indicates the subsystem. We will only use $|\phi^{(i)}\rangle$ notation when referring to a generic walk among eigenstates of $H$. Here in Equation (22) the operators indicated under the brackets "$\underbrace{\quad}$" are the operators that contributes the respective projector $|\cdot\rangle\langle\cdot|$ in $T_r$. We formally define such walk in the context of bounding $\|T_r\|_2$ as the following.

**Definition 8** (Walk in the space of $H$ eigenstates). *We define an $r$-step walk in the space of $H$ eigenstates as a sequence of unperturbed eigenstates $|\phi^{(0)}\rangle \to |\phi^{(1)}\rangle \to \cdots \to |\phi^{(r)}\rangle$ such that*

$$\begin{aligned} |\phi^{(0)}\rangle \in \mathcal{L}_-, \quad & |\phi^{(r)}\rangle \in \mathcal{L}_- \\ |\phi^{(i)}\rangle \in \mathcal{L}_+, \quad & i = 1, \cdots, r-1. \end{aligned} \tag{23}$$

*In addition, we require that $\|\langle \phi^{(i)}|V|\phi^{(i+1)}\rangle\| \neq 0$ for any $i = 0, 1, \cdots, r-1$. Let $E^{(i)}$ be the energy of $|\phi^{(i)}\rangle$, namely $E^{(i)} = \langle \phi^{(i)}|H|\phi^{(i)}\rangle$.*

Definition 8 is laid out specifically for enumerating terms in $T_r$. The following lemma describes the explicit connection between the $r$-th order perturbative term $T_r$ and the $r$-step walk in Definition 8.

**Lemma 1.** *For an $r$-step walk $|\phi^{(0)}\rangle \to |\phi^{(1)}\rangle \to \cdots \to |\phi^{(r)}\rangle$, let $\mathbf{B}^{(i)}$ be the $B_{pq,jk}^{(i)}$ block in $V$ (Equation 6 and 15) associated with the transition $|\phi^{(i-1)}\rangle \to |\phi^{(i)}\rangle$. In other words[2], $\mathbf{B}^{(i)} \otimes |\phi^{(i-1)}\rangle\langle\phi^{(i)}| = B_{pq,jk}^{(i)} \otimes |\psi_{j,p}^{(i)}\rangle\langle\psi_{k,q}^{(i)}|$. Then*

$$T_r = \sum_{|\phi^{(0)}\rangle \in \mathcal{L}_-} \sum_{|\phi^{(r)}\rangle \in \mathcal{L}_-} \sideset{}{'}\sum \mathbf{B}^{(1)} \cdot \frac{1}{|z - E^{(1)}|} \cdot \mathbf{B}^{(2)} \cdot \frac{1}{|z - E^{(2)}|} \cdots \frac{1}{|z - E^{(r-1)}|} \cdot \mathbf{B}^{(r)} \otimes |\phi^{(0)}\rangle\langle\phi^{(r)}| \tag{24}$$

*where $\sum'$ sums over all $r$-step walks in the space of $H$ eigenstates, as in Definition 8, but restricted to a fixed pair of $|\phi^{(0)}\rangle$ and $|\phi^{(r)}\rangle$.*

*Proof.* In Section 2.3 we interpret powers of block matrices as walks on a weighted directed graph with each edge carrying a "weight" that is a block. Applying this intuition to the block partitioning of the perturbation $V$ introduced in Section 3.1, we could see that $T_r$ is also a block matrix of $\dim(\mathcal{L}_-) \otimes \dim(\mathcal{L}_-)$ blocks with the $(i, j)$ block being the sum over all of the contributions from walks in the space of $H$ eigenstates (Definition 8) that start from the $i$-th low energy eigenstate and end at the $j$-th low energy eigenstate. With $|\phi^{(0)}\rangle$ being the $i$-th low energy eigenstate and $|\phi^{(r)}\rangle$ being the $j$-th, one could see that a term in $T_r$ corresponding to a walk $|\phi^{(0)}\rangle \to |\phi^{(1)}\rangle \to \cdots \to |\phi^{(r)}\rangle$ takes the form

$$\underbrace{(\mathbf{B}^{(1)} \otimes |\phi^{(0)}\rangle\langle\phi^{(1)}|)}_{V_{-+}} \cdot \underbrace{\left(\frac{1}{z - E^{(1)}}|\phi^{(1)}\rangle\langle\phi^{(1)}|\right)}_{G_+} \cdot \underbrace{(\mathbf{B}^{(2)} \otimes |\phi^{(1)}\rangle\langle\phi^{(2)}|)}_{V_+} \cdots \\ \cdots \underbrace{\left(\frac{1}{z - E^{(r-1)}}|\phi^{(r-1)}\rangle\langle\phi^{(r-1)}|\right)}_{G_+} \cdot \underbrace{(\mathbf{B}^{(r)} \otimes |\phi^{(r-1)}\rangle\langle\phi^{(r)}|)}_{V_{+-}}. \tag{25}$$

With the notation introduced in Equation 25 we could build up an expression for $T_r$ term by term. As a start, we could express $V_{-+}$, $V_{+-}$, $V_+$, and $G_+$ as

$$\begin{aligned} V_{-+} &= \sum_{|\phi\rangle \in \mathcal{L}_-} \sum_{|\phi'\rangle \in \mathcal{L}_+} \mathbf{B}_{\phi,\phi'} \otimes |\phi\rangle\langle\phi'|, & V_+ &= \sum_{|\phi\rangle \in \mathcal{L}_+} \sum_{|\phi'\rangle \mathcal{L}_+} \mathbf{B}_{\phi,\phi'} \otimes |\phi\rangle\langle\phi'| \\ V_{+-} &= \sum_{|\phi\rangle \in \mathcal{L}_+} \sum_{|\phi'\rangle \in \mathcal{L}_-} \mathbf{B}_{\phi,\phi'} \otimes |\phi^{(r-1)}\rangle\langle\phi^{(r)}|, & G_+(z) &= \sum_{|\phi\rangle \in \mathcal{L}_+} \frac{1}{z - E_\phi}|\phi\rangle\langle\phi|. \end{aligned} \tag{26}$$

---

[2] To avoid confusion with the $(i)$ superscripts we use $\mathbf{B}$ instead of $B$. Here the superscript $(i)$ of $\mathbf{B}^{(i)}$ stands for the $i$-th step in the walk, while superscript $(i)$ of $B_{pq,jk}^{(i)}$ represents the $i$-th subsystem.

where $\mathbf{B}_{\phi,\phi'}$ is the $B^{(i)}_{pq,jk}$ block in $V$ (Equation 6 and 15) that corresponds to transition from $|\phi\rangle$ to $|\phi'\rangle$, both of which are eigenstates of $H$. $E_\phi = \langle\phi|H|\phi\rangle$. Multiplying with $G_+V_+$ gives

$$\begin{aligned}
V_{-+}G_+V_+ &= \sum_{|\phi^{(0)}\rangle \in \mathcal{L}_-} \sum_{|\phi^{(1)}\rangle \in \mathcal{L}_+} \sum_{|\phi^{(2)}\rangle \in \mathcal{L}_+} \left(\mathbf{B}^{(1)} \otimes |\phi^{(0)}\rangle\langle\phi^{(1)}|\right) \cdot \left(\frac{1}{z-E^{(1)}}|\phi^{(1)}\rangle\langle\phi^{(1)}|\right) \cdot \left(\mathbf{B}^{(2)} \otimes |\phi^{(1)}\rangle\langle\phi^{(2)}|\right) \\
&= \sum_{|\phi^{(0)}\rangle \in \mathcal{L}_-} \sum_{|\phi^{(1)}\rangle \in \mathcal{L}_+} \sum_{\substack{|\phi^{(2)}\rangle \in \mathcal{L}_+ \\ \|\langle\phi^{(1)}|V|\phi^{(2)}\rangle\| \neq 0}} \mathbf{B}^{(1)} \cdot \frac{1}{z-E^{(1)}} \cdot \mathbf{B}^{(2)} \otimes |\phi^{(0)}\rangle\langle\phi^{(2)}|.
\end{aligned} \quad (27)$$

Continue carrying out computations similar in nature to Equation 27 to the $r$-th step $|\phi^{(r)}\rangle$ gives us the full expression of $T_r$ in terms of walks on $H$ eigenstates in Equation 24. $\square$

We are now ready to derive a general upper bound for $\|T_r\|_2$ in a similar spirit to Equation 14. Following Lemma 1 as well as properties of $\infty$-norm mentioned in Section 2.3, the 2-norm of $T_r$ can be bounded from above as

$$\|T_r\|_2 \leq \max_{|\phi^{(0)}\rangle \in \mathcal{L}_-} \sum_{|\phi^{(r)}\rangle \in \mathcal{L}_-} \left\|{\sum}' \mathbf{B}^{(1)} \cdot \frac{1}{|z-E^{(1)}|} \cdot \mathbf{B}^{(2)} \cdot \frac{1}{|z-E^{(2)}|} \cdots \frac{1}{|z-E^{(r-1)}|} \cdot \mathbf{B}^{(r)}\right\|_\infty \quad (28)$$

where the maximum and the first summation are taken over eigenstates of $H$ in $\mathcal{L}_-$.

Equation 28 serves as a starting point for finding tight upper bounds for $\|T_r\|_2$, because each $\|\mathbf{B}^{(i)}\|_\infty$ can be bounded from above by an appropriate choice of element from the vector $\boldsymbol{\lambda}$ (Definition 4). In Appendix B we show a concrete example where the upper bound in Equation 28 is derived explicitly in terms of elements in $\boldsymbol{\lambda}$ and $\mathbf{M}$.

Since the dimension of the Hilbert space $\mathcal{H}$ grows exponentially as $m$ grows, any algorithm that naively computes the right hand side of Equation 28 term by term is likely going to cost $O((D\ell)^{mr})$ where $D = \max_{i=1,\cdots,m} \dim(\mathcal{P}_i)$ is the maximum degeneracy of any subspace. As a first simplification, we could reduce this to $O(\ell^{mr})$ by considering walking in the space of energy configuration (Definition 6) instead of $H$ eigenstates.

### 3.4 Walking in the configuration space

The summation in Equation 28 is over $r$-step walks on the $H$ eigenstates. Note from Equation 20 that we could partition eigenstates of $H$ according to their energy configurations (Definition 6). We could use this partition simplify this summation by first grouping walks that go through the same changes in energy configurations. Let $\mathbf{c}^{(i)}$ be the energy configuration of $|\phi^{(i)}\rangle$ in an $r$-step walk in the space of $H$ eigenstates. Then the type of walks that appear in terms of $T_r$ must consist of $r$ steps and satisfy (refer to Definition 7 for $\mathbf{n}(\mathbf{c})$)

$$\begin{array}{ll} \mathbf{n}(\mathbf{c}^{(0)}) \in \mathcal{N}_-, & \mathbf{n}(\mathbf{c}^{(r)}) \in \mathcal{N}_- \\ \mathbf{n}(\mathbf{c}^{(i)}) \in \mathcal{N}_+, & i = 1, \cdots, r-1. \end{array} \quad (29)$$

In other words, the type of walks, or sequences of transitions, must start and end in the low-energy subspace $\mathcal{L}_-$, but stays in the high energy subspace $\mathcal{L}_+$ in between.

Since each term in $V$ acts on one unperturbed subsystem $\mathcal{H}_i$, at each step $|\psi^{(i)}\rangle\langle\psi^{(i+1)}|$, the energy configurations $\mathbf{c}^{(i)}$ and $\mathbf{c}^{(i+1)}$ must differ in at most one element. Furthermore, because $V$ is block-tridiagonal with respect to any subsystem (Equation 6), the difference between the respective elements in $\mathbf{c}^{(i)}$ and $\mathbf{c}^{(i+1)}$ must be at most 1. Hence the properties of sequences can be summarized as the following definition.

**Definition 9** (Walk in the configuration space). *We define an $r$-step walk in the space of configurations $\mathbf{c}$ (or walk in $\mathbf{c}$ for short) as a sequence of configurations $\mathbf{c}^{(0)} \to \mathbf{c}^{(1)} \to \cdots \to \mathbf{c}^{(r)}$ such that in addition to satisfying Equation 29, $\{\mathbf{c}^{(i)}\}_{i=0}^r$ also satisfies the property that for every step from $\mathbf{c}^{(i)}$ to $\mathbf{c}^{(i+1)}$ with $i = 2, \cdots, r-1$, either one of the following is true:*

1. $\mathbf{c}^{(i)} = \mathbf{c}^{(i+1)}$, *OR*

2. $\mathbf{c}^{(i+1)}$ *is obtained by incrementing or decrementing one element in $\mathbf{c}^{(i)}$ by 1.*

*The initial step $\mathbf{c}^{(0)} \to \mathbf{c}^{(1)}$ and the final step $\mathbf{c}^{(r-1)} \to \mathbf{c}^{(r)}$ only satisfy case 2 above.*

The following lemma relates the set of $r$-step walks in the space of configuration, as defined above, to that in the space of $H$ eigenstates, as in Definition 8.

**Lemma 2.** *For any $r$-step walk $|\phi^{(0)}\rangle \to |\phi^{(1)}\rangle \to \cdots \to |\phi^{(r)}\rangle$ described in Definition 8 there is a walk $\mathbf{c}^{(0)} \to \mathbf{c}^{(1)} \to \cdots \to \mathbf{c}^{(r)}$ as defined in Definition 9 such that $\mathbf{c}(|\phi^{(i)}\rangle) = \mathbf{c}^{(i)}$.*

*Proof.* By Definition 8, $\|\langle \phi^{(i)}|V|\phi^{(i+1)}\rangle\| \neq 0$ for any $i = 0, \cdots, r-1$. Because of the block tridiagonal structure of $V$ as in Equation 6, the energy configurations $\mathbf{c}(|\phi^{(i)}\rangle)$ and $\mathbf{c}(|\phi^{(i+1)}\rangle)$ differ at most one element and the difference is at most 1. In particular, the initial step of the walk from $|\phi^{(0)}\rangle \in \mathcal{L}_-$ to $|\phi^{(1)}\rangle \in \mathcal{L}_+$ and the final step from $|\phi^{(r-1)}\rangle \in \mathcal{L}_+$ to $|\phi^{(r)}\rangle \in \mathcal{L}_-$ satisfies $\mathbf{c}(|\phi^{(i)}\rangle) \neq \mathbf{c}(|\phi^{(i+1)}\rangle)$, which fall into case 2 of Definition 9. Hence if we let $\mathbf{c}^{(i)} = \mathbf{c}(|\phi^{(i)}\rangle)$, the walk $\mathbf{c}^{(0)} \to \mathbf{c}^{(1)} \to \cdots \to \mathbf{c}^{(r)}$ satisfies Definition 9. □

For computing a tight upper bound to the $\infty$-norm of a term in $\|T_r\|_\infty$ that corresponds to a particular walk satisfying the above Definition 9, the definitions of $\lambda_i$ and $M_{jk}$ then come into play. Generally speaking, every step from $\mathbf{c}^{(i)}$ to $\mathbf{c}^{(i+1)}$ contributes a factor. The product of these factors form an upper bound to a term in $T_r$ that corresponds to an entire walk. If a step falls into the case 1 in the above Definition 9, then this step contributes a factor $\omega$ (Definition 3). Otherwise if a step falls in the case 2 in Definition 9 then there must be some element, say the $j$-th element, of $\mathbf{c}_i$ that is changed by 1 to yield the new energy configuration $\mathbf{c}_{i+1}$. The contribution of such a step is $\lambda_j$. In other words, a transition has occurred in the subsystem $\mathcal{H}_j$ under the action of $V$. Further, let $j$ and $k$ be such that the step from $|\psi^{(i)}\rangle$ to $|\psi^{(i+1)}\rangle$ is from the subspace $\mathcal{P}_j$ to $\mathcal{P}_k$ for some subsystem. Then the contributing factor of the step is further multiplied by $M_{jk}$. To make the above intuition precise, we state the following lemma.

**Lemma 3.** *Let $f$ be a function of two energy configurations $\mathbf{c}$ and $\mathbf{c}'$ such that*

$$f(\mathbf{c}, \mathbf{c}') = \begin{cases} \lambda_t M_{ss'} & \mathbf{c} \text{ and } \mathbf{c}' \text{ differ at subsystem } t \text{ where } \mathbf{c}_t = s \text{ and } \mathbf{c}'_t = s' \\ \omega & \mathbf{c} = \mathbf{c}'. \end{cases} \quad (30)$$

*Then for any $r \geq 3$,*

$$\|T_r\|_2 \leq \max_{\mathbf{c}^{(0)}: \mathbf{n}(\mathbf{c}^{(0)}) \in \mathcal{N}_-} \sum_{\mathbf{c}^{(r)}: \mathbf{n}(\mathbf{c}^{(r)}) \in \mathcal{N}_-} \sum{}'' f(\mathbf{c}^{(0)}, \mathbf{c}^{(1)}) \cdot \frac{1}{|z - E^{(1)}|} \cdots f(\mathbf{c}^{(r-2)}, \mathbf{c}^{(r-1)}) \cdot \frac{1}{|z - E^{(r-1)}|} \cdot f(\mathbf{c}^{(r-1)}, \mathbf{c}^{(r)}). \quad (31)$$

*Here the summation $\Sigma''$ is over all $r$-step walks in the space of configurations, as defined in Definition 9, with fixed initial configuration $\mathbf{c}^{(0)}$. $E^{(i)}$ is the energy of the configuration $\mathbf{c}^{(i)}$, namely $\sum_{j=1}^m E_{c_j^{(i)}}$.*

*Proof.* We start from Equation 28 and partition the max and summation operations over $H$ eigenstates according to their energy configurations. Using Lemma 2 we could deduce from Equation 28 that

$$\|T_r\|_2 \leq \max_{\mathbf{c}^{(0)}: \mathbf{n}(\mathbf{c}^{(0)}) \in \mathcal{N}_-} \max_{|\phi^{(0)}\rangle: \mathbf{c}(|\phi^{(0)}\rangle) = \mathbf{c}^{(0)}} \sum_{\mathbf{c}^{(r)}: \mathbf{n}(\mathbf{c}^{(r)}) \in \mathcal{N}_-} \sum{}'' \sum_{\substack{|\phi^{(1)}\rangle \to \cdots \to |\phi^{(r)}\rangle \\ \mathbf{c}(|\phi^{(i)}\rangle) = \mathbf{c}^{(i)}}} \left\| \mathbf{B}^{(1)} \cdot \frac{1}{|z - E^{(1)}|} \cdot \mathbf{B}^{(2)} \cdot \frac{1}{|z - E^{(2)}|} \cdots \frac{1}{|z - E^{(r-1)}|} \cdot \mathbf{B}^{(r)} \right\|_\infty, \quad (32)$$

where the summation $\Sigma''$ is defined in the same way as in Equation 31. The first two max operations are equivalent to the max operation on the right hand side of Equation 28. The three summations essentially sums over the set of all $r$-step walks on $H$ eigenstates that are consistent with $r$-step walks in the space of energy configurations. This set should contain the set of all $r$-step walks on $H$ eigenstates that yield non-zero contributions on the right hand side of Equation 28. Hence the right hand side of Equation 32 is a valid upper bound to that of Equation 28. If we remove the max and summation operations over energy configurations in Equation 32 by considering a *fixed* walk $\mathbf{c}^{(0)} \to \mathbf{c}^{(1)} \to \cdots \to \mathbf{c}^{(r)}$, we are left with a term that is bounded from above by

$$\max_{|\phi^{(0)}\rangle: \mathbf{c}(|\phi^{(0)}\rangle) = \mathbf{c}^{(0)}} \sum_{\substack{|\phi^{(1)}\rangle \to \cdots \to |\phi^{(r)}\rangle \\ \mathbf{c}(|\phi^{(i)}\rangle) = \mathbf{c}^{(i)}}} \left\| \mathbf{B}^{(1)} \right\|_\infty \cdot \left\| \frac{1}{|z - E^{(1)}|} \cdot \mathbf{B}^{(2)} \cdot \frac{1}{|z - E^{(2)}|} \cdots \frac{1}{|z - E^{(r-1)}|} \cdot \mathbf{B}^{(r)} \right\|_\infty. \quad (33)$$

Recall that the operator $\mathbf{B}^{(1)}$ is associated with the transition $|\phi^{(0)}\rangle \to |\phi^{(1)}\rangle$. The corresponding change in energy configuration is $\mathbf{c}^{(0)} \to \mathbf{c}^{(1)}$. It is established in Lemma 1 as well as Definition 9 that $\mathbf{c}^{(0)}$ and $\mathbf{c}^{(1)}$ must differ at one element by 1. Let this be the $t$-th element. In other words, $c_t^{(0)} \neq c_t^{(1)}$. Let $c_t^{(0)} = s$ and $c_t^{(1)} = s'$. We could then interpret $\mathbf{c}^{(0)} \to \mathbf{c}^{(1)}$ as the physical process of a transition in subsystem $t$ from $s$-th energy level to the

11

$s'$-th. Furthermore, $\mathbf{B}^{(1)}$ is the operator associated with transitioning from a *specific* eigenstate $|\phi^{(0)}\rangle$ that satisfies $\langle\phi^{(0)}|H^{(t)}|\phi^{(0)}\rangle = E_s$, to another $H$ eigenstate $|\phi^{(1)}\rangle$ with $\langle\phi^{(1)}|H^{(t)}|\phi^{(1)}\rangle = E_{s'}$. Recall that the superscript $(t)$ for $H^{(t)}$ represents the $t$-th subsystem, while the superscript for $|\phi^{(i)}\rangle$ stands for the $i$-th step during the walk. Now we are considering all such transitions from $|\phi^{(0)}\rangle$ to $|\phi^{(i)}\rangle$, summing over all possible $|\phi^{(1)}\rangle$ eigenstates and maximizing over all possible $|\phi^{(0)}\rangle$ eigenstates that are consistent with the (fixed) walk $\mathbf{c}^{(0)} \to \mathbf{c}^{(1)} \to \cdots$. By Definition 4, $\|\mathbf{B}^{(1)}\|_\infty \le \lambda_t$ for any specific step $|\phi^{(0)}\rangle \to |\phi^{(1)}\rangle$. By Definition 5, there are at most $M_{ss'}$ ways to make a transition from $\mathcal{P}_s$ to $\mathcal{P}_{s'}$ for any subsystem. Hence the contribution of the first step $|\phi^{(0)}\rangle \to |\phi^{(1)}\rangle$ to the right hand side of Expression 33 is bounded from above by $\lambda_t M_{ss'}$. Hence Expression 33 is bounded from above by

$$f(\mathbf{c}^{(0)}, \mathbf{c}^{(1)}) \cdot \frac{1}{|z - E^{(1)}|} \max_{\substack{|\phi^{(0)}\rangle : \mathbf{c}(|\phi^{(0)}\rangle) = \mathbf{c}^{(0)}}} \sum_{\substack{|\phi^{(1)}\rangle \to \cdots \to |\phi^{(r)}\rangle \\ \mathbf{c}(|\phi^{(i)}\rangle) = \mathbf{c}^{(i)}}} \left\| \mathbf{B}^{(2)} \right\|_\infty \cdot \left\| \frac{1}{|z - E^{(2)}|} \cdots \frac{1}{|z - E^{(r-1)}|} \cdot \mathbf{B}^{(r)} \right\|_\infty \quad (34)$$

where $f(\mathbf{c}^{(0)}, \mathbf{c}^{(1)}) = \lambda_t M_{ss'}$ following the definition of $f$ in the statement of the Lemma.

The scalar factors $\frac{1}{z - E^{(i)}}$ are constants for all the walks $|\phi^{(1)}\rangle \to \cdots \to |\phi^{(r)}\rangle$ summed over since the walk in configuration space $\mathbf{c}^{(0)} \to \mathbf{c}^{(1)} \to \cdots \to \mathbf{c}^{(r)}$ is fixed for Expression 33. In other words $E^{(i)} = E(\mathbf{n}(\mathbf{c}^{(i)}))$. The contribution of $\|\mathbf{B}^{(2)}\|_\infty$ could be bounded from above by similar arguments that follow Expression 33 that treat $\|\mathbf{B}^{(1)}\|_\infty$, except that one has to consider an alternative possibility when $\mathbf{c}^{(1)} = \mathbf{c}^{(2)}$, in which case the contribution of $\|\mathbf{B}^{(2)}\|_\infty$ over all possible walks on $H$ eigenstates is bounded from above by $\omega$ (Definition 3). We could thus bound Expression 34 from above by

$$f(\mathbf{c}^{(0)}, \mathbf{c}^{(1)}) \cdot \frac{1}{|z - E^{(1)}|} \cdot f(\mathbf{c}^{(1)}, \mathbf{c}^{(2)}) \cdot \frac{1}{|z - E^{(2)}|} \max_{\substack{|\phi^{(0)}\rangle : \mathbf{c}(|\phi^{(0)}\rangle) = \mathbf{c}^{(0)}}} \sum_{\substack{|\phi^{(1)}\rangle \to \cdots \to |\phi^{(r)}\rangle \\ \mathbf{c}(|\phi^{(i)}\rangle) = \mathbf{c}^{(i)}}} \left\| \mathbf{B}^{(3)} \cdots \frac{1}{|z - E^{(r-1)}|} \cdot \mathbf{B}^{(r)} \right\|_\infty. \quad (35)$$

By repeating the arguments that produced Equation 35 from Equation 34 on $\|\mathbf{B}^{(i)}\|_\infty$ for $i = 3, \cdots, r-1$, one could yield upper bounds that are functions of $\omega$, $\boldsymbol{\lambda}$ and $\mathbf{M}$. Finally, apply the same argument for treating $\|\mathbf{B}^{(1)}\|_\infty$ in Expression 33 for $\|\mathbf{B}^{(r)}\|_\infty$ yields Equation 31. □

With Lemma 3 we in essence have accomplished a reduction of the number of walks that need to be enumerated, from $O((D\ell)^{mr})$ as in the case with walks on $H$ eigenstates in Section 3.3, to $O(\ell^{mr})$. In the next section we show how to use symmetry to reduce the exponential dependence on the number of unperturbed subsystems $m$ to polynomial, assuming that both $\ell$ and $r$ are constant.

### 3.5 Introducing symmetry

In order to further reduce the dimension of the space in which a walk is described, we introduce a symmetric version of the energy configuration. We start by laying down the following definition concerning the status of individual elements in an energy configuration during a walk in the space of $\mathbf{c}$.

**Definition 10** (Active and inactive elements). *Consider an energy configuration $\mathbf{c}^{(i)}$ during a walk in the configuration space $\mathbf{c}^{(0)} \to \cdots \to \mathbf{c}^{(i-1)} \to \mathbf{c}^{(i)}$ with $\mathbf{c}^{(1)} = (0, 0, \cdots, 0)$. For any $k$, if the $k$-th element of $c^{(j)}$, which we denote as $c_k^{(j)}$, is 0 for every $j \le i$, then we call $c_k^{(j)}$ an* inactive *element. Otherwise the $k$-th element is an* active *element.*

In other words, if the $k$-th subsystem is never excited from $\mathcal{P}_0$ during the walk then it is inactive. It is worth noting that an active element of an energy configuration may also be 0. In this case the subsystem was excited from $\mathcal{P}_0$ at some point but returns to $\mathcal{P}_0$.

**Definition 11** (Reduced energy configuration). *For an energy configuration $\mathbf{c}$ (Definition 6) we define* reduced energy configuration $\tilde{\mathbf{c}}$ *as the resulting vector of removing all inactive elements in $\mathbf{c}$ and then sorting the active elements in non-decreasing order. In particular, let $\tilde{\mathbf{c}}(\mathbf{c})$ be the reduced energy configuration that corresponds to a configuration $\mathbf{c}$.*

For example, in a setting with $m = 3$ subsystems, the configuration where the first subsystem has energy $E_3$, the second is inactive and thus has energy $E_0$, the third has $E_1$ and the fourth has $E_0$ but is active would have an energy configuration $\mathbf{c} = (3, 0, 1, 0)$. However, in this case the reduced energy configuration $\tilde{\mathbf{c}} = (0, 1, 3)$. If the second subsystem is active then $\tilde{\mathbf{c}} = (0, 0, 1, 3)$ is the reduced energy configuration.

The advantage of introducing this concept is that the space in which the walks are described can be reduced from exponential in $m$ to polynomial, assuming both $\ell$, the total number of energy levels in each unperturbed subsystem, and $r$, the order of the perturbation or the total number of steps in a walk, are constant. For a fixed set of parameters $m, \ell$, the total possible energy configurations $\mathbf{c}$ is $O(\ell^m)$. However, as we show in the following lemma, the set of a possible reduced energy configuration $\tilde{\mathbf{c}}$ is polynomial in $m$.

**Lemma 4.** *Let $f_{m\ell}$ be the total number of possible reduced energy configurations of length $m$ and maximum possible number of energy levels $\ell$. Then $f_{m\ell} \leq m^\ell$ for any $m \geq 2$ and $\ell \geq 1$.*

*Proof.* The last element of a reduced configuration could take any one of $\ell$ values. Since by Definition 11, the elements of a reduced configuration is non-decreasing, the remaining $m-1$ elements of $\tilde{\mathbf{c}}$ has $f_{m-1,\tilde{\mathbf{c}}_m}$ choices where $\tilde{\mathbf{c}}_m \in \{0, \cdots, \ell-1\}$ is the last element of $\tilde{\mathbf{c}}$. We then have the recursion $f_{m\ell} = f_{m-1,\ell} + f_{m-1,\ell-1} + \cdots + f_{m-1,1}$ with boundary condition $f_{k1} = 1$ for any $k \in \{1, \cdots, m\}$ and $f_{1k} = k$ for any $k \in \{0, 1, \cdots, \ell-1\}$. Hence $f_{m\ell} = f_{m-1,\ell} + f_{m,\ell-1} = 1 + \sum_{i=1}^{m} f_{i,\ell-1}$. Starting from $f_{m1} = 1$, we have $f_{m2} = 1 + f_{11} + f_{21} + \cdots + f_{m1} \leq 1 + mf_{m1} = 1 + m$ and $f_{m3} = 1 + f_{12} + f_{22} + \cdots + f_{m2} \leq 1 + m + m^2$. Applying this to $f_{m\ell}$, we have $f_{m\ell} \leq 1 + f_{m,\ell-1} \leq 1 + m(1 + mf_{m,\ell-2}) \leq \cdots \leq 1 + m + \cdots + m^{\ell-1} \leq m^\ell$. $\square$

We now define the notion of walks in the reduced configuration space as the follows.

**Definition 12** (Walk in the space of reduced configurations). *A sequence of reduced configurations $\tilde{\mathbf{c}}^{(0)} \to \tilde{\mathbf{c}}^{(1)} \to \cdots \to \tilde{\mathbf{c}}^{(r-1)} \to \tilde{\mathbf{c}}^{(r)}$ is an $r$-step walk in the space of reduced configurations $\tilde{\mathbf{c}}$ if*

$$\begin{aligned} \mathbf{n}(\tilde{\mathbf{c}}^{(0)}) &= \mathbf{n}(\tilde{\mathbf{c}}_0) \in \mathcal{N}_-, \quad \mathbf{n}(\tilde{\mathbf{c}}^{(r)}) \in \mathcal{N}_- \\ \mathbf{n}(\tilde{\mathbf{c}}^{(i)}) &\in \mathcal{N}_+, \quad i = 1, \cdots, r-1. \end{aligned} \quad (36)$$

*and either one of the following is true for any $i = 2, \cdots, r-1$:*

1. $\tilde{\mathbf{c}}^{(i)} = \tilde{\mathbf{c}}^{(i+1)}$, OR

2. $\tilde{\mathbf{c}}^{(i)}$ and $\tilde{\mathbf{c}}^{(i+1)}$ differ by 1 at one element, OR

3. $|\tilde{\mathbf{c}}^{(i+1)}| = |\tilde{\mathbf{c}}^{(i)}| + 1$.

*As a consequence, for the initial step $\tilde{\mathbf{c}}^{(0)} \to \tilde{\mathbf{c}}^{(1)}$ only case 3 applies and for the final step $\tilde{\mathbf{c}}^{(r-1)} \to \tilde{\mathbf{c}}^{(r)}$ only case 2 applies.*

The following lemma connects the space of reduced energy configurations $\tilde{\mathbf{c}}$ to that of energy configuration $\mathbf{c}$.

**Lemma 5.** *For every walk $\mathbf{c}^{(0)} \to \mathbf{c}^{(1)} \to \cdots \to \mathbf{c}^{(r)}$ in the space of $\mathbf{c}$ as in Definition 9, there is a corresponding walk $\tilde{\mathbf{c}}^{(0)} \to \tilde{\mathbf{c}}^{(1)} \to \cdots \to \tilde{\mathbf{c}}^{(r)}$ in the space of $\tilde{\mathbf{c}}$ as in Definition 12 such that $\tilde{\mathbf{c}}(\mathbf{c}^{(i)}) = \tilde{\mathbf{c}}^{(i)}$. Furthermore, for any permutation $\pi$ over $m$ elements, the walk $\pi(\mathbf{c}^{(0)}) \to \pi(\mathbf{c}^{(1)}) \to \cdots \to \pi(\mathbf{c}^{(r)})$ also maps to the same walk in $\tilde{\mathbf{c}}$. Conversely, for any walk $\mathbf{c}'^{(0)} \to \mathbf{c}'^{(1)} \to \cdots \to \mathbf{c}'^{(r)}$ that satisfies both Definition 9 and $\tilde{\mathbf{c}}(\mathbf{c}'^{(i)}) = \tilde{\mathbf{c}}^{(i)}$, there must be a permutation $\pi'$ such that $\pi'(\mathbf{c}^{(i)}) = \mathbf{c}'^{(i)}$ for any $i$.*

*Proof.* By definition, $\mathbf{n}(\mathbf{c}^{(0)}) \in \mathcal{N}_-$. Since the definition of energy combination $\mathbf{n}$ (Definition 7) is invariant with respect to permutation of unperturbed subsystems, $\mathbf{n}(\tilde{\mathbf{c}}(\mathbf{c}^{(0)})) = \mathbf{n}(\mathbf{c}^{(0)}) \in \mathcal{N}_-$. For every subsequent step $\mathbf{c}^{(i)} \to \mathbf{c}^{(i+1)}$, $i \in \{0, \cdots, r-2\}$, case 1 in Definition 9 leads to $\tilde{\mathbf{c}}(\mathbf{c}^{(i)}) = \tilde{\mathbf{c}}(\mathbf{c}^{(i+1)})$, which fits case 1 of Definition 12. Case 2 in Definition 9 depends on whether an inactive element in $\mathbf{c}^{(i)}$ becomes active in $\mathbf{c}^{(i+1)}$. If this is not the case, then $\tilde{\mathbf{c}}(\mathbf{c}^{(i)})$ and $\tilde{\mathbf{c}}(\mathbf{c}^{(i+1)})$ differ by 1 at one element, matching case 2 in Definition 12. Otherwise the additional active element in $\mathbf{c}^{(i+1)}$ contributes an additional element in $\tilde{\mathbf{c}}(\mathbf{c}^{(i+1)})$, namely $|\tilde{\mathbf{c}}(\mathbf{c}^{(i+1)})| = |\tilde{\mathbf{c}}(\mathbf{c}^{(i)})| + 1$. Finally from $\mathbf{n}(\mathbf{c}^{(r)}) \in \mathcal{N}_-$ we have $\mathbf{n}(\tilde{\mathbf{c}}(\mathbf{c}^{(r)})) \in \mathcal{N}_-$. Hence if we let $\tilde{\mathbf{c}}^{(i)} = \tilde{\mathbf{c}}(\mathbf{c}^{(i)})$ then the walk $\tilde{\mathbf{c}}^{(0)} \to \tilde{\mathbf{c}}^{(1)} \to \cdots \to \tilde{\mathbf{c}}^{(r)}$ matches the Definition 12. This proves the first part of the lemma.

The second part follows by noting that by Definition 11, the reduced energy configuration of an $H$ eigenstate is invariant with respect to permutation of the subsystems, namely $\tilde{\mathbf{c}}(\mathbf{c}^{(i)}) = \tilde{\mathbf{c}}(\pi(\mathbf{c}^{(i)}))$ for any permutation $\pi$ over $m$ elements.

The last part ("Conversely...") can be proved by starting with the observation that for any walk $\mathbf{c}'^{(0)} \to \mathbf{c}'^{(1)} \to \cdots \to \mathbf{c}'^{(r)}$ that satisfies both Definition 12 and $\tilde{\mathbf{c}}(\mathbf{c}'^{(i)}) = \tilde{\mathbf{c}}^{(i)}$, because $\tilde{\mathbf{c}}(\mathbf{c}^{(i)}) = \tilde{\mathbf{c}}^{(i)}$ and by the permutation invariance of reduced energy configuration there must be a permutation $\pi^{(i)}$ such that $\pi^{(i)}(\mathbf{c}^{(i)}) = \mathbf{c}'^{(i)}$ for every $i \in \{0, \cdots, r\}$. Our goal is thus to show that the permutations $\pi^{(i)}$ are identical to the same permutation $\pi'$. For the sake of contradiction suppose $\pi^{(i)} \neq \pi^{(i+1)}$ for some $i$. Then there must be a (non-trivial) permutation $\Delta\pi$ such that $\pi^{(i+1)} = \Delta\pi \cdot \pi^{(i)}$. Since the walk $\mathbf{c}^{(0)} \to \mathbf{c}^{(1)} \to \cdots \to \mathbf{c}^{(r)}$ conforms to Definition 9, either $\mathbf{c}^{(i)} = \mathbf{c}^{(i+1)}$ or $\mathbf{c}^{(i)}$ and $\mathbf{c}^{(i+1)}$ differ by 1 at one element. We discuss each case individually as the following:

13

- Suppose $\mathbf{c}^{(i)} = \mathbf{c}^{(i+1)}$, then $\mathbf{c}'^{(i)} = \pi^{(i)}(\mathbf{c}^{(i)}) = \pi^{(i)}(\mathbf{c}^{(i+1)})$. Hence $\mathbf{c}'^{(i+1)} = \pi^{(i+1)}(\mathbf{c}^{(i+1)}) = \Delta\pi(\pi^{(i)}(\mathbf{c}^{(i+1)})) = \Delta\pi(\mathbf{c}'^{(i)})$, which is impossible if the walk $\mathbf{c}'^{(0)} \to \mathbf{c}'^{(1)} \to \cdots \to \mathbf{c}'^{(r)}$ conforms to Definition 9 because no step $\mathbf{c}'^{(i)} \to \mathbf{c}'^{(i+1)}$ that conforms to case 1 or 2 in Definition 9 corresponds to a non-trivial permutation of $\mathbf{c}'^{(i)}$. Hence in this case the permutations $\pi^{(i)}$ and $\pi^{(i+1)}$ must be identical.

- Suppose $\mathbf{c}^{(i)}$ and $\mathbf{c}^{(i+1)}$ differ by 1 at one element, namely $\mathbf{c}_j^{(i)} \neq \mathbf{c}_j^{(i+1)}$ for some $j$. Then $\pi^{(i)}(\mathbf{c}^{(i)})$ and $\pi^{(i)}(\mathbf{c}^{(i+1)})$ differ at an element $k \neq j$. Since $\mathbf{c}'^{(i+1)} = \Delta\pi(\pi^{(i)}(\mathbf{c}^{(i+1)}))$ and $\mathbf{c}'^{(i)} = \pi^{(i)}(\mathbf{c}^{(i)})$, we see that the step $\mathbf{c}'^{(i)} \to \mathbf{c}'^{(i+1)}$ is realized by incrementing the $k$-th element of $\mathbf{c}'^{(i)}$ by $\mathbf{c}_j^{(i+1)} - \mathbf{c}_j^{(i)}$ and apply a non-trivial permutation $\Delta\pi$. The latter step contradicts Definition 9 since no permutation is possible in a single step with either case 1 or 2 in Definition 9.

Therefore we have shown that the set of $r$-step walks in $\mathbf{c}$ that is consistent with a particular $r$-step walk in $\tilde{\mathbf{c}}$ are merely the same walk in $\mathbf{c}$ with different permutations of the unperturbed subsystems. $\square$

We could then establish an upper bound for $\|T_r\|_2$ that is based on a walk in the space of $\tilde{\mathbf{c}}$ as in Definition 12, which is stated in the following Lemma.

**Lemma 6.** *For an $r$-step walk $\tilde{\mathbf{c}}^{(0)} \to \tilde{\mathbf{c}}^{(1)} \to \cdots \to \tilde{\mathbf{c}}^{(r)}$ in the space of reduced configuration $\tilde{\mathbf{c}}$ as described in Definition 12, consider any $r$-step walk $\mathbf{c}^{(0)} \to \mathbf{c}^{(1)} \to \cdots \to \mathbf{c}^{(r)}$ such that $\tilde{\mathbf{c}}(\mathbf{c}^{(i)}) = \tilde{\mathbf{c}}^{(i)}$. Define the set $\mathcal{F}_i = \{j = 1, \cdots, r | c_i^{(j-1)} \neq c_i^{(j)}\}$, the vector $\mathbf{f} \in \mathbb{N}^m$ such that $\mathbf{f}_i = |\mathcal{F}_i|$ and an integer $k = r - \sum_{i=1}^{m} \mathbf{f}_i$. Let $\mathbf{b} \in \mathbb{N}^m$ be $\mathbf{f}$ sorted in non-increasing order (to match Definition 2). Then*

$$\|T_r\|_2 \leq \max_{\tilde{\mathbf{c}}^{(0)}:\mathbf{n}(\tilde{\mathbf{c}}^{(0)}) \in \mathcal{N}_-} \sum_{\tilde{\mathbf{c}}^{(r)}:\mathbf{n}(\tilde{\mathbf{c}}^{(r)}) \in \mathcal{N}_-} \sum{}^{*} \left(\prod_{i=1}^{r} \frac{1}{|z - E^{(i)}|}\right) \cdot m_{\mathbf{b}}(\boldsymbol{\lambda}) \cdot \left(\prod_{j:\exists i, j \in \mathcal{F}_i} M_{\mathbf{c}_i^{(j-1)}, \mathbf{c}_i^{(j)}}\right) \cdot \omega^k \quad (37)$$

*where $\omega$, $\boldsymbol{\lambda}$ and $\mathbf{M}$ are defined in Definitions 3, 4 and 5 respectively. The summation $\Sigma^*$ is over all $r$-step walks in the space of reduced configurations, as defined in Definition 12, with* fixed *initial reduced configuration $\tilde{\mathbf{c}}^{(0)}$ and final reduced configuration $\tilde{\mathbf{c}}^{(r)}$.*

*Proof.* Starting from Lemma 3, where we bounded from above contributions of individual $r$-step walks in $\mathbf{c}$ by an expression

$$f(\mathbf{c}^{(0)}, \mathbf{c}^{(1)}) \cdot \frac{1}{|z - E^{(1)}|} \cdot f(\mathbf{c}^{(1)}, \mathbf{c}^{(2)}) \cdots f(\mathbf{c}^{(r-2)}, \mathbf{c}^{(r-1)}) \cdot \frac{1}{|z - E^{(r-1)}|} \cdot f(\mathbf{c}^{(r-1)}, \mathbf{c}^{(r)}). \quad (38)$$

For a specific $r$-step walk in $\mathbf{c}$ space, let $\mathcal{F}_i = \{j = 1, \cdots, r | c_i^{(j-1)} \neq c_i^{(j)}\}$. Then using the definition of $f(\mathbf{c}, \mathbf{c}')$ in Lemma 3, we could rewrite expression 38 as

$$\left(\prod_{i=1}^{m} \frac{1}{|z - E^{(i)}|}\right) \cdot \left(\prod_{i=1}^{m} \lambda_i^{|\mathcal{F}_i|}\right) \cdot \left(\prod_{j:\exists i, j \in \mathcal{F}_i} M_{\mathbf{c}_i^{(j-1)}, \mathbf{c}_i^{(j)}}\right) \cdot \omega^k. \quad (39)$$

For a *fixed* walk $\tilde{\mathbf{c}}^{(0)} \to \tilde{\mathbf{c}}^{(1)} \to \cdots \to \tilde{\mathbf{c}}^{(r)}$, consider the set $\mathcal{W}$ of $r$-step walks in the space of $\mathbf{c}$ such that $\tilde{\mathbf{c}}(\mathbf{c}^{(i)}) = \tilde{\mathbf{c}}^{(i)}$. By Lemma 5, $\mathcal{W}$ consists of permutations of some $r$-step walk in $\mathbf{c}$. If the contribution of a single walk in $\mathcal{W}$ can be bounded from above by Equation 39, then the total contribution from the walks in $\mathcal{W}$ can be bounded from above by summing over *all* possible permutations of the unperturbed subsystems, namely

$$\sum_{\pi:[m] \mapsto [m]} \left(\prod_{i=1}^{m} \frac{1}{|z - E^{(i)}|}\right) \cdot \left(\prod_{i=1}^{m} \lambda_{\pi(i)}^{|\mathcal{F}_i|}\right) \cdot \left(\prod_{j:\exists i, j \in \mathcal{F}_i} M_{\mathbf{c}_i^{(j-1)}, \mathbf{c}_i^{(j)}}\right) \cdot \omega^k. \quad (40)$$

Because the reduced energy configuration $\tilde{\mathbf{c}}$ is invariant with respect to the energy configuration $\mathbf{c}$ that it corresponds to, we have

$$\prod_{j:\exists i, j \in \mathcal{F}_i} M_{\mathbf{c}_i^{(j-1)}, \mathbf{c}_i^{(j)}} = \prod_{j:\exists i, j \in \mathcal{F}'_i} M_{\mathbf{c}'_i^{(j-1)}, \mathbf{c}'_i^{(j)}} \quad (41)$$

for any $\mathbf{c}'^{(0)} \to \mathbf{c}'^{(1)} \to \cdots \to \mathbf{c}'^{(r)}$ such that $\pi(\mathbf{c}^{(i)}) = \mathbf{c}'^{(i)}$ for some permutation $\pi$. Here $\mathcal{F}'_i = \{j = 1, \cdots, r | c_i'^{(j-1)} \neq c_i'^{(j)}\}$. Then by Definition 1 and 2, $\sum_{\pi:[m] \mapsto [m]} \lambda_{\pi(i)}^{\mathbf{b}_i} = m_{\mathbf{b}}(\boldsymbol{\lambda})$ where $\mathbf{b}$ is defined in the statement of the Lemma. Expression 40 serves as an upper bound for a *fixed* walk in $\tilde{\mathbf{c}}$. Summing over all $r$-step walks in $\tilde{\mathbf{c}}$ described in Definition 12, and incorporating Equation 41, we can bound the right hand side of Equation 31 by that of Equation 37. $\square$

In Figure 2 of the main text we have already demonstrated the relationship between a walk in **c** and a walk in **c̃**. Furthermore, we presented Equation (5) in the main text without proof. In Appendix C we illustrate Lemma 6 with a concrete derivation of Equation (5) of the main text, in order to provide more intuitive arguments for understanding the construction of the upper bound in Equation 37.

## 4 Efficient algorithm for computing upper bounds

### 4.1 Constructing cellular automaton

In Definition 7 for energy combination, we define $\mathcal{C}(\mathbf{n})$ as the set of energy configurations that give rise to the energy combination $\mathbf{n}$, while $\mathbf{n}(\mathbf{c})$ is the energy combination corresponding to a given energy configuration. Note that the mapping from an energy combination to an energy configuration is not unique (since for example $\mathbf{c} = (0,1)$ and $\mathbf{c} = (1,0)$ both correspond to $\mathbf{n} = (1,1)$) while the mapping in the reverse direction is unique. To enforce uniqueness in both directions, we define *uniquely reduced energy configuration* as the following.

**Definition 13** (Uniquely reduced energy configuration). *Referring to Definition 6, for an energy configuration* **c** *we define* uniquely reduced energy configuration $\hat{\mathbf{c}}$ *as the resulting vector of removing* all *zero elements in* **c** *and then sorting the active elements in ascending order. For each energy combination* **n** *let* $\hat{\mathbf{c}}(\mathbf{n})$ *be the uniquely reduced energy configuration corresponding to* **n**.

Note that Definition 13 is only minutely different from Definition 11 in terms of which zero elements to remove. With Definition 13 for each energy combination **n** there is a unique $\hat{\mathbf{c}}$ that is consistent with **n**. For example consider $\mathbf{c}_1 = (0,1,0,3)$ and $\mathbf{c}_2 = (0,0,3,1)$, both of which belong in the set $\mathcal{C}(\mathbf{n})$ with $\mathbf{n} = (1,0,1)$, but we have a unique $\hat{\mathbf{c}} = (1,3)$ that corresponds to $\mathbf{n} = (1,0,1)$. In fact it is not hard to see that

$$\hat{\mathbf{c}}(\mathbf{n}) = (\underbrace{1,\cdots,1}_{n_1},\underbrace{2,\cdots,2}_{n_2},\cdots,\underbrace{\ell,\cdots,\ell}_{n_\ell}). \tag{42}$$

Our cellular automaton then consists of cells (graph nodes) connected with directed edges. Each cell is associated with a list of 4-tuples $(\tilde{\mathbf{c}}, \mathbf{b}, \xi, \mu)$. An $n$-tuple is an ordered sequence of $n$ elements. Here in our 4-tuple, $\tilde{\mathbf{c}}$ is a reduced energy configuration (Definition 11) and **b** is a reduced partition vector (Definition 2), $\xi$ is a scalar coefficient and $\mu : \tilde{\mathbf{c}} \mapsto \mathbf{b}$ is a one-one mapping from the reduced energy configuration to the reduced partition. Because of its bijective nature, one could also think of $\mu$ as a permutation map. The reason for introducing the mapping $\mu$ is because the reduced partition does not contain all the information about the current configuration.

We construct the cellular automaton with BUILDCA subroutine as described in Algorithm 1. The algorithm produces a directed graph $G(\mathcal{V}, \mathcal{E})$ that represents the cellular automaton. Each node $v_{\mathbf{n}} \in \mathcal{V}$ corresponds to an energy combination **n**. In each node $v_{\mathbf{n}}$ and each directed edge $e(v_{\mathbf{n}}, v_{\mathbf{n}'}) \in \mathcal{E}$ we store a list of 4-tuples $(\tilde{\mathbf{c}}, \mathbf{b}, \xi, \mu)$ denoted as $\mathcal{S}_{\mathbf{n}}$ and $\mathcal{S}_{\mathbf{n},\mathbf{n}'}$ respectively. For a given energy combination vector $\mathbf{n} = (n_0, n_1, n_2, \cdots, n_{\ell-1})$, we introduce the notation

$$\begin{aligned}
\mathbf{n}_0 &= \underbrace{(m, 0, \cdots, 0)}_{\ell} \\
\mathbf{n}'_i &= (n_1, \cdots, n_i - 1, n_{i+1} + 1, \cdots, n_{\ell-1}), \qquad i = 1, \cdots, \ell-2.
\end{aligned} \tag{43}$$

For an energy combination **n** to be *compatible* with our physical setting (Figure 1 of the main text), it is necessary that

$$\sum_{i=0}^{\ell-1} n_i \le m, \quad \text{and} \quad n_i \ge 0, \quad n_i \in \mathbb{Z}, \quad \forall i \in \{0, \cdots, \ell-1\}. \tag{44}$$

Note that the definition of $\mathbf{n}'_i$ in Equation 43 for a given **n** essentially corresponds to a step $\mathbf{c}^{(j)} \to \mathbf{c}^{(j+1)}$ in the space of configurations **c** where $\mathbf{c}^{(j+1)}$ and $\mathbf{c}^{(j)}$ differ by 1 at one subsystem and going from $\mathbf{c}^{(j)}$ to $\mathbf{c}^{(j+1)}$ the subsystem makes a transition from energy level $i$ to $i+1$. The graph $G(\mathcal{V}, \mathcal{E})$ that Algorithm 1 connects any energy combination **n** with another energy combination **n'** as long as there is a walk in **c** (Definition 9) such that at some step $j$, $\mathbf{n}(\mathbf{c}^{(j)}) = \mathbf{n}$ and $\mathbf{n}(\mathbf{c}^{(j+1)}) = \mathbf{n}'$.

Since the energy combination **n** is a vector of length $\ell$ and each element of **n** takes values from $[m]$, there are in total $O(m^{\ell+1})$ possible energy combinations. The most naive implementation of Algorithm 1 takes $O(m^{2(\ell+1)})$. If we consider $\ell$ to be a constant for the physical system, Algorithm 1 costs computational resource that is polynomial in the system size $m$.

---

**Algorithm 1** Cellular automaton construction algorithm

**Input:**

- The number of subsystems $m$ as shown in Figure 1a of the main text;
- The matrix $\mathbf{M} \in \mathbb{R}^{\ell \times \ell}$ as in Definition 5.

**Output:**

- A weighted directed graph $G(\mathcal{V}, \mathcal{E})$ that serves as a representation of the cellular automaton.

**Procedure** $G(\mathcal{V}, \mathcal{E}) = \text{BuildCA}(m, \mathbf{M})$

1. $\mathcal{V} \leftarrow \{v_{\mathbf{n}_0}\}$, $\mathcal{E} \leftarrow \emptyset$;
2. $\text{BuildCell}(\mathbf{n}_0)$;
3. Return $G(\mathcal{V}, \mathcal{E})$.

**Procedure** $\text{BuildCell}(\mathbf{n})$

1. For each $t = 0, 1, \cdots, \ell - 1$, compute $\mathbf{n}'_t$ and test if it satisfies (44). If so, then
    - If $v_{\mathbf{n}'_i} \notin \mathcal{V}$, $\quad \mathcal{V} \leftarrow \mathcal{V} \cup \{v_{\mathbf{n}'_i}\}$;
    - If $e(v_{\mathbf{n}}, v_{\mathbf{n}'_i}) \notin \mathcal{E}$, $\quad \text{AddEdge}(\mathbf{n}, \mathbf{n}'_t)$;
      If $e(v_{\mathbf{n}'_i}, v_{\mathbf{n}}) \notin \mathcal{E}$, $\quad \text{AddEdge}(\mathbf{n}'_t, \mathbf{n})$;

2. If $\mathbf{n} = (0, \cdots, 0, m)$, return.
   Otherwise for each $t = 0, 1, \cdots, \ell - 1$, call $\text{BuildCell}(\mathbf{n}'_t)$.

**Procedure** $\text{AddEdge}(\mathbf{p}, \mathbf{q})$

1. Find $s$ and $t$ such that $q_s = p_s - 1$ and $q_t = p_t + 1$;
2. Add $e(v_{\mathbf{p}}, v_{\mathbf{q}})$ with weight $M_{st}$.

---

## 4.2 Cell update rules

Recall that we are interested in computing an upper bound for $\|T_r\|_\infty$ for any $r$. The goal of this section is to present the update rules for each individual cells so that in the end the upper bound for $\|T_r\|_\infty$ can be gleaned from all nodes $v_\mathbf{n}$ such that $\mathbf{n} \in \mathcal{N}_-$ after $r$ concurrent updates for all nodes in the cellular automaton.

Let $\mathcal{S}_\mathbf{n}$ be the set of 4-tuples associated with the cell $v_\mathbf{n}$. To aid the presentation we define a scalar multiplication rule for the 4-tuples: $C(\tilde{\mathbf{c}}, \mathbf{b}, \xi, \mu) \equiv (\tilde{\mathbf{c}}, \mathbf{b}, C\xi, \mu)$ where $C$ is a scalar quantity. Naturally we extend the multiplication rule to entire sets of the 4-tuples:

$$C\mathcal{S}_\mathbf{n} \equiv \{(\tilde{\mathbf{c}}, \mathbf{b}, \xi, \mu) \in \mathcal{S}_\mathbf{n} | (\tilde{\mathbf{c}}, \mathbf{b}, C\xi, \mu)\}.$$

Similarly we define $\mathcal{S}_{\mathbf{n},\mathbf{n}'}$ as the set of 4-tuples associated with the edge $e(\mathbf{n}, \mathbf{n}') \in \mathcal{E}$. The rules for updating $\mathcal{S}_\mathbf{n}$ for each cell $v_\mathbf{n}$ and $\mathcal{S}_{\mathbf{n},\mathbf{n}'}$ for any edge $e(\mathbf{n}, \mathbf{n}')$ is outlined in the UPDATECELL subroutine in Algorithm 2.

The procedure UPDATECELL($v_\mathbf{n}$) called on a particular cell $v_\mathbf{n}$ contains two main steps: the first updates the tuple list $\mathcal{S}_\mathbf{n}$ of the current cell by combining $\mathcal{S}_\mathbf{n}$ scaled by $\omega/(z - E(\mathbf{n}))$ with the tuple lists on the incident edges scaled by $1/(z - E(\mathbf{n}))$. See Equation 45. The second step is to generate 4-tuple lists for the outgoing edges from the current cell by the OUT($\mathbf{n}, \mathbf{n}', \mathcal{T}$) subroutine. During the first step, the factor $1/(z - E(\mathbf{n}))$ is to account for the contribution of $G_+$ terms in $T_r$. The $\omega$ factor in the first step is to account for the case where the walk in $\tilde{\mathbf{c}}$ (or $\mathbf{c}$) stays at the same configuration. The second step is to compute the correct list of 4-tuples to deliver to each $\mathbf{n}'$ in the next update. For each $\mathbf{n}'$ that is accessible from the current energy combination $\mathbf{n}$, each 4-tuple in $\mathcal{S}_\mathbf{n}$ will contribute an appropriate set of 4-tuples that are stored in $\mathcal{S}_{\mathbf{n},\mathbf{n}'}$. These new 4-tuples must conform to the transition from $\mathbf{n}$ to $\mathbf{n}'$, in the sense that is demonstrated in Figure 1. We will make these intuition precise in the next section, where we prove Theorems 2 and 3.

## 4.3 Algorithm for computing an upper bound at arbitrary order

Now that we have introduced the major subroutines, we could put them together into an algorithm for finding a tight upper bound to $\|T_r\|_\infty$, see Algorithm 3.

We start by recalling that Definition 12 can be thought of as the reduced configuration $\tilde{\mathbf{c}}$ space counterpart to the description of walks in the space of configuration $\mathbf{c}$ in Definition 9 in Section 3.4. We also define an energy combination $\mathbf{n}$ counterpart as the following.

**Definition 14** (Walk in the space of energy combination $\mathbf{n}$). *A sequence of energy combinations $\mathbf{n}^{(0)} \to \mathbf{n}^{(1)} \to \cdots \to \mathbf{n}^{(r-1)} \to \mathbf{n}^{(r)}$ is an $r$-step walk in the space of energy combination $\mathbf{n}$ (or walk in $\mathbf{n}$ for short) if*

$$\begin{aligned} \mathbf{n}^{(0)} \in \mathcal{N}_-, \quad & \mathbf{n}^{(r)} \in \mathcal{N}_- \\ \mathbf{n}^{(i)} \in \mathcal{N}_+, \quad & i = 1, \cdots, r-1. \end{aligned} \tag{47}$$

*For every step from $\mathbf{n}^{(i)}$ to $\mathbf{n}^{(i+1)}$ with $i = 1, \cdots, r-2$, either one of the following is true:*

1. $\mathbf{n}^{(i)} = \mathbf{n}^{(i+1)}$;

2. $\mathbf{n}^{(i+1)} = (n_0^{(i)}, n_1^{(i)}, \cdots, n_j^{(i)} - 1, n_{j+1}^{(i)} + 1, \cdots, n_{\ell-1}^{(i)})$.

*For the initial step $\mathbf{n}^{(0)} \to \mathbf{n}^{(1)}$ and final step $\mathbf{n}^{(r-1)} \to \mathbf{n}^{(r)}$ only case 2 above applies.*

The following definition concerns the step 2j and 2l of Algorithm 3, where the subroutine UPDATECELL of Algorithm 2 is repeatedly invoked in all the cells of the automaton.

**Definition 15** (Trace of the update algorithm). *Let $\mathcal{S}_\mathbf{n}^{(i)}$ and $\mathcal{S}_{\mathbf{n},\mathbf{n}'}^{(i)}$ be the set of 4-tuples associated with the node $v_\mathbf{n}$ and edge $e(v_\mathbf{n}, v_{\mathbf{n}'})$ respectively at the end of the $i$-th call to UPDATECELL at step 2j of Algorithm 3. A trace of Algorithm 3 is a sequence of 4-tuples $\mathcal{T}^{(0)} \to \mathcal{T}^{(1)} \to \cdots \to \mathcal{T}^{(r)}$ that is associated with an $r$-step walk in the space of energy combinations $\mathbf{n}$ (or equivalently on the vertices of the graph $G$ generated by BUILDCA in Algorithm 1). The 4-tuple $\mathcal{T}^{(i)}$ at each step $i$ is given by*

$$\mathcal{T}^{(i)} = \begin{cases} \mathcal{T}_{-+}, & i = 0 \\ \dfrac{1}{|z - E^{(i)}|} \text{OUT}(\mathbf{n}^{(i-1)}, \mathbf{n}^{(i)}, \mathcal{T}^{(i-1)}), & i = 1, \cdots, r-1 \\ \text{OUT}(\mathbf{n}^{(r-1)}, \mathbf{n}^{(r)}, \mathcal{T}^{(r-1)}), & i = r \end{cases} \tag{48}$$

*where $\mathcal{T}_{-+}$ is computed by the initialization steps 2a through 2h of Algorithm 3.*

17

**Algorithm 2** Updating the cells and their outgoing edges

**Input:**
- The node $v_\mathbf{n} \in \mathcal{V}$ from the graph $G(\mathcal{V}, \mathcal{E})$ with $\mathcal{E} = \mathcal{E}_{\text{dashed}} \cup \mathcal{E}_{\text{non-dashed}}$ generated by Algorithm 1.

**Output:**
- Updated list of 4-tuples $\mathcal{S}_\mathbf{n}$ associated with $v_\mathbf{n}$, and $\mathcal{S}_{\mathbf{n},\mathbf{n}'}$ associated with each outgoing edge $e(v_\mathbf{n}, v_{\mathbf{n}'}) \in \mathcal{E}$.

**Procedure** UPDATECELL($v_\mathbf{n}$)

1. Update the list of 4-tuples at each cell $v_\mathbf{n}$:

$$\mathcal{S}_\mathbf{n} \leftarrow \left( \frac{\omega}{|z - E(\mathbf{n})|} \mathcal{S}_\mathbf{n} \right) \cup \left( \frac{1}{|z - E(\mathbf{n})|} \bigcup_{\mathbf{n}':e(\mathbf{n}',\mathbf{n}) \in \mathcal{E}} \mathcal{S}_{\mathbf{n}',\mathbf{n}} \right). \quad (45)$$

2. For each outgoing edge $e(v_\mathbf{n}, v'_\mathbf{n}) \in \mathcal{E}$ do the following

$$\mathcal{S}_{\mathbf{n},\mathbf{n}'} \leftarrow \bigcup_{\mathcal{T} \in \mathcal{S}_\mathbf{n}} \text{OUT}(\mathbf{n}, \mathbf{n}', \mathcal{T}) \quad (46)$$

where OUT is a subroutine described in the OUT subroutine.

**Procedure** $\mathcal{S}_{\text{new}} = \text{OUT}(\mathbf{n}, \mathbf{n}', \mathcal{T})$

1. $\xi_{\text{new}} \leftarrow M_{\mathbf{n},\mathbf{n}'} \xi$, where $M_{\mathbf{n},\mathbf{n}'}$ is the weight of the edge $e(v_\mathbf{n}, v_{\mathbf{n}'})$.

2. Compute $\tilde{\mathbf{c}} = (\underbrace{0, \cdots, 0}_{|\mathbf{b}| - |\hat{\mathbf{c}}(\mathbf{n})|}, \hat{\mathbf{c}}(\mathbf{n}))$ and $\tilde{\mathbf{c}}' = (\underbrace{0, \cdots, 0}_{|\mathbf{b}| - |\hat{\mathbf{c}}(\mathbf{n})|}, \hat{\mathbf{c}}(\mathbf{n}'))$.

3. If $|\hat{\mathbf{c}}(\mathbf{n})| \geq |\hat{\mathbf{c}}(\mathbf{n}')|$,

    (a) Find $k$ such that $\tilde{c}_k \neq \tilde{c}'_k$ and compute $\Delta c = \tilde{c}'_k - \tilde{c}_k$.

    (b) Mark all $\tilde{c}_j$, $j \in \{1, \cdots, |\mathbf{b}|\}$, such that $\tilde{c}_j = \tilde{c}_k$.

    (c) For every marked $j$:
       i. $\tilde{\mathbf{c}}_{\text{new}} \leftarrow \tilde{\mathbf{c}}'$;
       ii. $(\tilde{c}_{\text{new}})_j \leftarrow (\tilde{c}_{\text{new}})_j + \Delta c$;
       iii. $\mu_{\text{new}} \leftarrow \mu$;
       iv. $\mathbf{b}_{\text{new}} \leftarrow \mathbf{b}$;
       v. $\mu((\tilde{c}_{\text{new}})_j) \leftarrow \mu((\tilde{c}_{\text{new}})_j) + 1$;
       vi. $\mathcal{S}_{\text{new}} \leftarrow \mathcal{S}_{\text{new}} \cup \{(\tilde{\mathbf{c}}_{\text{new}}, \mathbf{b}_{\text{new}}, \xi_{\text{new}}, \mu_{\text{new}})\}$.

4. If $|\hat{\mathbf{c}}(\mathbf{n})| < |\hat{\mathbf{c}}(\mathbf{n}')|$,

    (a) If $|\mathbf{b}| < m$,
       i. $\tilde{\mathbf{c}}_{\text{new}} \leftarrow (1 \quad \tilde{\mathbf{c}})$;
       ii. $\mathbf{b}_{\text{new}} \leftarrow (1 \quad \mathbf{b})$;
       iii. $\mu_{\text{new}} \leftarrow \begin{bmatrix} (1 \quad \tilde{\mathbf{c}}) = \tilde{\mathbf{c}}_{\text{new}} \\ \downarrow \quad \downarrow \mu \\ (1 \quad \mathbf{b}) = \mathbf{b}_{\text{new}} \end{bmatrix}$;
       iv. $\mathcal{S}_{\text{new}} \leftarrow \mathcal{S}_{\text{new}} \cup \{(\tilde{\mathbf{c}}_{\text{new}}, \mathbf{b}_{\text{new}}, \xi_{\text{new}}, \mu_{\text{new}})\}$.

    (b) If $|\mathbf{b}| > |\hat{\mathbf{c}}(\mathbf{n})|$, execute the same steps as 3a, 3b, and 3c.

5. If necessary, rearrange the elements of $\tilde{\mathbf{c}}_{\text{new}}$ (and update $\mu_{\text{new}}$ accordingly) such that $\tilde{\mathbf{c}}_{\text{new}}$ conforms to Definition 11.

6. Return $\mathcal{S}_{\text{new}}$.

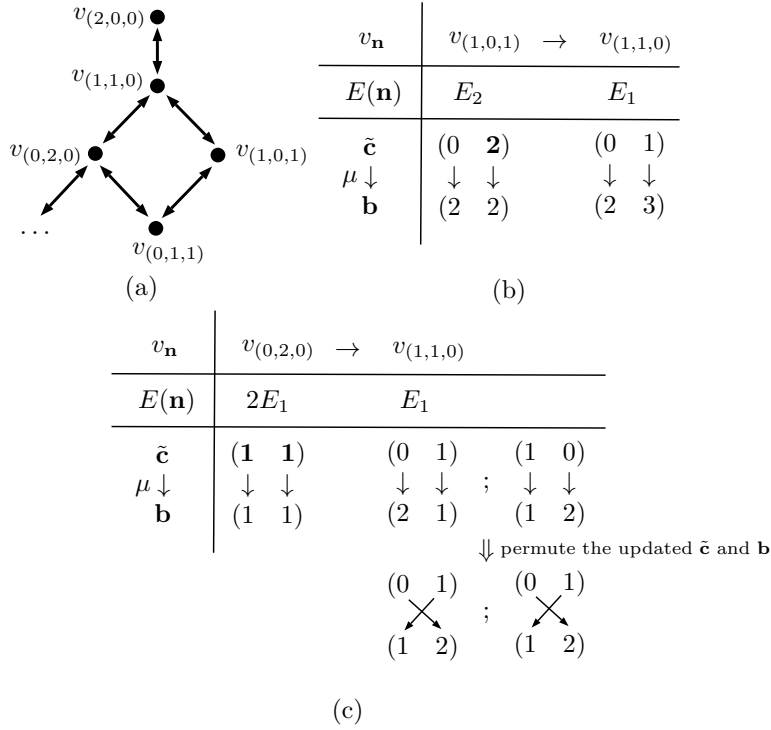

Figure 1: An example illustrating the $\text{Out}(\mathbf{n}, \mathbf{n}', \mathcal{T})$ subroutine in Algorithm 2. Here we let the total number of subsystems be $m = 2$ and each of them has $\ell = 3$ energy levels. Subfigure (a): The graph $G(\mathcal{V}, \mathcal{E})$ generated by Algorithm 1. Here only part of $G$ is shown. Subfigure (b): During a call for $\text{Out}(\mathbf{n}, \mathbf{n}', \mathcal{T})$ with $\mathbf{n} = (0, 1)$ and $\mathbf{n}' = (1, 0)$, the 4-tuple $\mathcal{T} = (\tilde{\mathbf{c}}, \mathbf{b}, \xi, \mu) \in \mathcal{S}_\mathbf{n}$ with $\tilde{\mathbf{c}} = (0, \mathbf{2})$ and $\mathbf{b} = (2, 2)$, which is shown in the left column of (b), is being used for generating a new 4-tuple $(\tilde{\mathbf{c}}_{\text{new}}, \mathbf{b}_{\text{new}}, \xi_{\text{new}}, \mu_{\text{new}}) \in \mathcal{S}_{\mathbf{n}, \mathbf{n}'}$ with $\tilde{\mathbf{c}}_{\text{new}} = (0, 1)$ and $\mathbf{b}_{\text{new}} = (2, 3)$. Here the bold $\mathbf{2}$ in $\tilde{\mathbf{c}}$ represents the "marked" element in step 3b of Algorithm 2. Note that $\mathbf{n}(\mathbf{c}) = \mathbf{n}$ and $\mathbf{n}(\mathbf{c}_{\text{new}}) = \mathbf{n}'$. Subfigure (c): During a call for $\text{Out}(\mathbf{n}, \mathbf{n}', \mathcal{T})$ with $\mathbf{n} = (1, 1)$ and $\mathbf{n}' = (1, 0)$, similar to (b) we use the 4-tuple $\mathcal{S}_\mathbf{n}$ to generate new 4-tuples to be stored in $\mathcal{S}_{\mathbf{n}, \mathbf{n}'}$. However, here both elements of $\tilde{\mathbf{c}} = (\mathbf{1}, \mathbf{1})$ are "marked". Hence step 3c of Algorithm 2 generates two new 4-tuples, each with their $\tilde{\mathbf{c}}_{\text{new}}$ having one distinct element that differs its counterpart in $\tilde{\mathbf{c}}$ by 1. The step with the label "permute the updated $\tilde{\mathbf{c}}$ and $\mathbf{b}$" illustrates the step 6 in $\text{Out}$ in Algorithm 2, where elements of $\mathbf{c}_{\text{new}}$ and $\mathbf{b}_{\text{new}}$ as well as the mapping $\mu_{\text{new}} : \mathbf{c}_{\text{new}} \mapsto \mathbf{b}_{\text{new}}$ are arranged to conform to their respective definitions (Definition 11 for $\tilde{\mathbf{c}}$ and Definition 2 for $\mathbf{b}$).

From Equation 48 we see that $\mathcal{T}^{(i)} \in \mathcal{S}^{(i)}_{\mathbf{n}^{(i)}}$ for any $i = 0, \cdots, r$. Let $\mathcal{P}^{\tilde{\mathbf{c}}}_r$ be the set of $r$-step walks in the reduced configuration space (Definition 9) that starts from the initial reduced configuration $\tilde{\mathbf{c}}_1 = \emptyset$. Let $\mathcal{P}^{\mathcal{T}}_r$ be the set of $r$-step traces (Definition 15) generated by running UPDATECELL procedure $r$ times (Algorithm 2), with the initial input assigned by steps 2a through 2h of Algorithm 3. The following theorem shows that Algorithm 3 captures all the paths in the space of reduced configurations $\tilde{\mathbf{c}}$ that follow Definition 12.

**Theorem 2.** *There is a one-one correspondence (bijective mapping) between the two sets $\mathcal{P}^{\tilde{\mathbf{c}}}_r$ and $\mathcal{P}^{\mathcal{T}}_r$.*

*Proof.* For every $k < r$, let $\mathcal{Q}^{\tilde{\mathbf{c}}}_k$ be the set of $k$-step walks in the space of reduced configuration $\tilde{\mathbf{c}}$ obtained by truncating all $r$-step walks in $\mathcal{P}^{\tilde{\mathbf{c}}}_r$ at step $k$. There could be multiple walks in $\mathcal{P}^{\tilde{\mathbf{c}}}_r$ that share the same first $k$ steps. We count them only once in $\mathcal{Q}^{\tilde{\mathbf{c}}}_k$. Since $k < r$, every step of the $k$-step walks in $\mathcal{Q}^{\tilde{\mathbf{c}}}_k$ is defined using Definition 15 but with all parts concerning $\mathbf{n}^{(r)}$ removed. Similarly, we define $\mathcal{Q}^{\mathcal{T}}_k$ as the set of $k$-step traces of the update algorithm obtained from truncating each trace in $\mathcal{P}^{\mathcal{T}}_r$ at the $k$-th step and counting the redundant elements only once.

To establish the theorem, we first show that for every $k < r$, there is a one-one correspondence between the elements of the two sets $\mathcal{Q}^{\tilde{\mathbf{c}}}_k$ and $\mathcal{Q}^{\mathcal{T}}_k$. Specifically, for any $k$-step walk $q_k \in \mathcal{Q}^{\tilde{\mathbf{c}}}_k$ such that $q_k = \tilde{\mathbf{c}}^{(0)} \to \tilde{\mathbf{c}}^{(1)} \to \cdots \to \tilde{\mathbf{c}}^{(k)}$, there is a trace of Algorithm 3 denoted as $t_k \in \mathcal{Q}^{\mathcal{T}}_K$, that can be described as $t_K = \mathcal{T}^{(0)} \to \mathcal{T}^{(1)} \to \cdots \to \mathcal{T}^{(k)}$ where $\mathcal{T}^{(i)} = (\tilde{\mathbf{c}}^{(i)}, \mathbf{b}^{(i)}, \xi^{(i)}, \mu^{(i)})$ for any $i \in \{1, \cdots, k\}$.

We use induction on $k$. For $k = 1$, $\mathcal{Q}^{\tilde{\mathbf{c}}}_1 = \{\hat{\mathbf{c}}(\mathbf{n}_-)\}$ for some $\mathbf{n}_- \in \mathcal{N}_-$ (step 2 of Algorithm 3), which corresponds to $\mathcal{Q}^{\mathcal{T}}_1 = \{\mathcal{T}_{-+}\}$. For the definition of $\mathcal{T}_{-+}$, refer to step 2g of Algorithm 3 respectively. By inspecting step 2a through 2h it is clear that the reduced energy configuration of $\mathcal{T}_{-+}$ is $\hat{\mathbf{c}}(\mathbf{n}_-)$. Hence the above statement is true for $k = 1$. Suppose the statement is true for all $k \leq K$. Then consider any $K$-step walk $q_K \in \mathcal{Q}^{\tilde{\mathbf{c}}}_K$ such that

$$q_K = \tilde{\mathbf{c}}^{(0)} \to \tilde{\mathbf{c}}^{(1)} \to \cdots \to \tilde{\mathbf{c}}^{(K)}. \tag{49}$$

By induction hypothesis, there must be a $K$-step trace $t_K \in \mathcal{Q}^{\mathcal{T}}_K$ that corresponds to $p_K$. Here the trace $t_K = \mathcal{T}^{(0)} \to \mathcal{T}^{(1)} \to \cdots \to \mathcal{T}^{(K)}$. It then suffices to show that all paths of the form $q'_{K+1} := q_K \to \tilde{\mathbf{c}}'$ has one-one correspondence with traces of the form $t_{K+1} = t_K \to \mathcal{T}_{\text{new}}$ where $\mathcal{T}_{\text{new}} = (\tilde{\mathbf{c}}', \mathbf{b}', \xi', \mu')$ is one of the new 4-tuples generated at either step 1 or 2 of UPDATECELL in Algorithm 2. By Definition 12, $\tilde{\mathbf{c}}'$ has three possibilities:

(i). $\tilde{\mathbf{c}}' = \tilde{\mathbf{c}}^{(K)}$;

(ii). $|\tilde{\mathbf{c}}'| = |\tilde{\mathbf{c}}^{(K)}|$ and $|\tilde{\mathbf{c}}'_j - \tilde{\mathbf{c}}^{(K)}_j| = 1$ for some $j$;

(iii). $|\tilde{\mathbf{c}}'| = |\tilde{\mathbf{c}}^{(K)}| + 1$.

The case (i) is handled by the $\frac{\omega}{z-E(\mathbf{n})}\mathcal{S}_{\mathbf{n}}$ term in step 1 of Algorithm 2, with Equation 45. In other words, in this case $q'_{K+1}$ maps to the trace $t_{K+1} := t_K \to \mathcal{T}_{\text{new}}$ with $\mathcal{T}_{\text{new}} = \frac{\omega}{z-E(\mathbf{n})}\mathcal{T}^{(K)} = (\tilde{\mathbf{c}}^{(K)}, \mathbf{b}^{(K)}, \frac{\omega}{z-E^{(K)}}\xi^{(K)}, \mu^{(K)})$ generated at step 1 of UPDATECELL. Here $E^{(K)} = E(\mathbf{n}(\tilde{\mathbf{c}}^{(K)}))$.

The case (ii) is handled by steps 3 and 4b of OUT in Algorithm 2. By definition, $\mathcal{T}^{(K)} = (\tilde{\mathbf{c}}^{(K)}, \mathbf{b}^{(K)}, \xi^{(K)}, \mu^{(K)})$. Recall $\tilde{\mathbf{c}}^{(K)} = (c^{(K)}_1, c^{(K)}_2, \cdots, c^{(K)}_{|\mathbf{b}^{(K)}|})$. Then $\tilde{\mathbf{c}}'$ is obtained by incrementing or decrementing one of the $\omega_i$ elements by 1. Incrementing or decrementing any $\tilde{c}^{(K)}_i$ element will change the energy combination of $\tilde{\mathbf{c}}^{(K)}$. In particular, if there is a subset of the $\tilde{c}^{(K)}_i$ elements, call them $\tilde{c}^{(K)}_{i_1}, \tilde{c}^{(K)}_{i_2}, \cdots, \tilde{c}^{(K)}_{i_L}$, such that $\tilde{c}^{(K)}_{i_1} = \tilde{c}^{(K)}_{i_2} = \cdots = \tilde{c}^{(K)}_{i_L}$, then incrementing or decrementing any $\tilde{c}^{(K)}_{i_j}$ could yield the same $\mathbf{n}(\tilde{\mathbf{c}}')$. In steps 3 and 4b we mark all such $\tilde{c}^{(K)}_{i_j}$ elements. The case $\tilde{c}^{(K)}_{i_j} = 0$ for any $j = \{1, \cdots, L\}$ is handled in step 3 and the case $\tilde{c}^{(K)}_{i_j} \neq 0$ for any $j = \{1, \cdots, L\}$ is handled in step 4b. In either cases, the new 4-tuple $\mathcal{T}_{\text{new}} = (\tilde{\mathbf{c}}', \mathbf{b}', \xi', \mu')$ generated by OUT is such that we map the path $q'_{K+1} := q_K \to \tilde{\mathbf{c}}'$ to the trace $t_{K+1} := t_K \to \mathcal{T}_{\text{new}}$.

The case (iii) is handled in step 4a of OUT in Algorithm 2. In this case an inactive subsystem is active from $E_0$ to $E_1$. Hence $\tilde{\mathbf{c}}' = (1 \quad \tilde{\mathbf{c}}^{(K)})$. $q'_{K+1}$ then maps to $t_{K+1} := t_K \to \mathcal{T}_{\text{new}}$ where $\mathcal{T}_{\text{new}} = (\tilde{\mathbf{c}}', \mathbf{b}', \xi', \mu')$ is generated at step 4a.

In summary we have shown that for each possible path $q'_{K+1} := q_K \to \tilde{\mathbf{c}}'$ in the reduced configuration space there is a corresponding trace of the algorithm $t_{K+1} := t_k \to \mathcal{T}_{\text{new}}$ where $\mathcal{T}_{\text{new}} = (\tilde{\mathbf{c}}', \mathbf{b}', \xi', \mu')$ is generated at various steps of Algorithm 2. Because these steps are at mutually exclusive branches of IF conditions, no $q'_{K+1}$ maps to two different $t_{K+1}$'s simultaneously and vice versa. We also note that by Definition 12, $q_K$ in Equation 49 must satisfy $\mathbf{n}(\tilde{\mathbf{c}}^{(i)}) \in \mathcal{N}_+$ for all $i = 1, \cdots, K$. This is enforced by step 2i in Algorithm 3, where all edges that goes from $\mathcal{N}_+$ to $\mathcal{N}_-$, namely the "dashed" edges, are removed.

We have thus far shown that for every walk in $\mathcal{Q}^{\tilde{\mathbf{c}}}_k$, there is a corresponding trace in $\mathcal{Q}^{\mathcal{T}}_k$ that maps to it, and this is true for any $k < r$. Conversely, since any new triple $\mathcal{T}^{(i+1)}$ generated by $\mathcal{T}^{(i)}$ comes from either step 1 of

UPDATECELL in Algorithm 2, or step 3 or 4a or 4b of OUT in Algorithm 2, and the cases (i), (ii) and (iii) above has accounted for each of the steps, we conclude that for every trace in $\mathcal{Q}_k^{\mathcal{T}}$ there must be a corresponding walk in $\mathcal{Q}_k^{\tilde{\mathbf{c}}}$. Hence there is a one-one correspondence between the two sets $\mathcal{Q}_k^{\mathcal{T}}$ and $\mathcal{Q}_k^{\tilde{\mathbf{c}}}$ for any $k < r$.

By Definition 12 the final step of any $r$-step walk in the space of reduced configuration has to conform to the case 2 of Definition 12. Similarly, each trace in $\mathcal{P}_r^{\mathcal{T}}$ is associated with an $r$-step walk in the space of energy combination (Definition 14), for which the last step also needs to conform to the case 2 of Definition 14. Hence for any $r$-step walk in $\mathcal{P}_r^{\tilde{\mathbf{c}}}$, if the first $(r-1)$ steps are determined, the final step is also uniquely known. The same goes for any trace in $\mathcal{P}_r^{\mathcal{T}}$. We prove the theorem by using the one-one correspondence between $\mathcal{Q}_{r-1}^{\tilde{\mathbf{c}}}$ and $\mathcal{Q}_{r-1}^{\mathcal{T}}$ established from the previous inductive argument. The condition of returning to $\mathcal{N}_-$ at the last step, namely the restriction $\mathbf{n}(\tilde{\mathbf{c}}^{(r)}) \in \mathcal{N}_-$ in Equation 36 of Definition 12 is enforced in the step 2k of Algorithm 3 by adding back the dashed edges that enable transition from $\mathcal{N}_+$ back to $\mathcal{N}_-$. □

From the above proof we could have a rough upper bound of the complexity of the algorithm. For a walk of $r$ steps where each step has $m$ choices, we have in total $O(m^r)$ possible walks. From the proof of Theorem 2 we have established that at any point during the algorithm, each 4-tuple at a node collects contributions from all possible walks up to the node. Hence at the $r$-th step of the algorithm, there are at most as many 4-tuples stored in all of the nodes as there are $r$-step walks. Each tuple takes $O(m)$ time to update since there are at most $m$ elements of identical values in a reduced configuration $\tilde{\mathbf{c}}$ in case (ii) of OUT in Algorithm 2 while cases (i) and (iii) takes $O(1)$ time to treat. Putting these together, we have that $r$ updates of the algorithm takes $O(rm^r)$. If we fix the order of perturbation $r$, this is polynomial with respect to the system size.

Theorem 2 shows that Algorithm 3 captures all the walks in $\tilde{\mathbf{c}}$ that conform to Definition 12. The theorem below shows that Algorithm 3 indeed computes the right hand side of Equation 37.

**Theorem 3** (Correctness of the CA algorithm). *Given an $r$-step trace $\mathcal{T}^{(0)} \to \mathcal{T}^{(1)} \to \cdots \mathcal{T}^{(r)}$ as described in Definition 15 and (according to Theorem 2) its associated $r$-step walk $\tilde{\mathbf{c}}^{(0)} \to \tilde{\mathbf{c}}^{(1)} \to \cdots \tilde{\mathbf{c}}^{(r)}$ in the space of reduced energy configurations $\tilde{\mathbf{c}}$, let $\mathbf{c}^{(0)} \to \mathbf{c}^{(1)} \to \cdots \to \mathbf{c}^{(r)}$ be an $r$-step walk in $\mathbf{c}$ such that $\tilde{\mathbf{c}}(\mathbf{c}^{(i)}) = \tilde{\mathbf{c}}^{(i)}$. Each step of the trace can be written as $\mathcal{T}^{(i)} = (\tilde{\mathbf{c}}^{(i)}, \mathbf{b}^{(i)}, \xi^{(i)}, \mu^{(i)})$. Then we have*

$$\xi^{(r)} m_{\mathbf{b}^{(r)}}(\boldsymbol{\lambda}) = \left(\prod_{i=1}^{r-1} \frac{1}{|z - E^{(i)}|}\right) \cdot m_{\mathbf{b}}(\boldsymbol{\lambda}) \cdot \left(\prod_{j:\exists i, j \in \mathcal{F}_i} M_{c_i^{(j-1)}, c_i^{(j)}}\right) \cdot \omega^k \tag{50}$$

*where the symbols involved in the right hand side expression Equation 50 are the same as those defined in Equation 37 of Lemma 6.*

*Proof.* The proof of Lemma 3 is based on $r$-step walks that follow Definition 9. In fact from the arguments outlined by Equations 33, 34 and 35 we could see that any such $r$-step walk in the space of configuration $\mathbf{c}$ truncated at step $q$, $\mathbf{c}^{(0)} \to \mathbf{c}^{(1)} \to \cdots \to \mathbf{c}^{(q)}$, contributes a multiplicative factor in one of the terms in the upper bound of $\|T_r\|_2$ (refer to the right hand side of Equation 37) that can be written as

$$f(\mathbf{c}^{(0)}, \mathbf{c}^{(1)}) \cdot \frac{1}{|z - E^{(1)}|} \cdot f(\mathbf{c}^{(1)}, \mathbf{c}^{(2)}) \cdots f(\mathbf{c}^{(q-2)}, \mathbf{c}^{(q-1)}) \cdot \frac{1}{|z - E^{(q-1)}|} \cdot f(\mathbf{c}^{(q-1)}, \mathbf{c}^{(q)}). \tag{51}$$

The first step of the walk in $\tilde{\mathbf{c}}$, $\tilde{\mathbf{c}}^{(0)} \to \tilde{\mathbf{c}}^{(1)}$, falls into either case 2 or 3 of Definition 12. In either case, PERTURBBOUND in Algorithm 3 will produce $\mathcal{T}_{-+}$ (step 2g) with partition $\mathbf{b}^{(1)} = (1)$ and coefficient $\xi^{(1)} = M_{\mathbf{n}_-, \mathbf{n}_+}$, which is correct because by Lemma 5, steps in $\mathbf{c}$ that are consistent with $\tilde{\mathbf{c}}^{(0)} \to \tilde{\mathbf{c}}^{(1)}$ in the sense that $\tilde{\mathbf{c}}(\mathbf{c}^{(0)}) = \tilde{\mathbf{c}}^{(0)}$ and $\tilde{\mathbf{c}}(\mathbf{c}^{(1)}) = \tilde{\mathbf{c}}^{(1)}$ are but the same step $\mathbf{c}^{(0)} \to \mathbf{c}^{(1)}$ with different permutations of the $m$ subsystems (or elements of $\mathbf{c}$). In other words, the multiplicative factor associated with the step in reduced configuration $\tilde{\mathbf{c}}^{(0)} \to \tilde{\mathbf{c}}^{(1)}$ can be written as[3]

$$\sum_{\pi:\mathbb{N}^m \mapsto \mathbb{N}^m} f(\pi(\mathbf{c}^{(0)}), \pi(\mathbf{c}^{(1)})) = \sum_{\pi:[m] \mapsto [m]} \lambda_{\pi(j)} M_{st} = m_{(1)}(\boldsymbol{\lambda}) \cdot M_{st} \tag{52}$$

where we assume that during the step from $\mathbf{c}^{(0)}$ to $\mathbf{c}^{(1)}$, the $j$-th subsystem makes a transition from $\mathcal{P}_s$ to $\mathcal{P}_t$. From Equation 52 we see that the initial partition is indeed (1). Since in this one-step process only the $j$-th subsystem is acted on, $\mathcal{F}_j = \{1\}$ and $\mathcal{F}_i = \emptyset$ for any $i \neq j$ (for the definition of $\mathcal{F}_j$ see Lemma 6). The multiplicative factor $M_{\mathbf{n}_-, \mathbf{n}_+}$ is determined during a call to ADDEDGE in BUILDCELL of Algorithm 1. Since $c_j^{(0)} = s$ and $c_j^{(1)} = t$,

---
[3] Here we abuse the notation $\pi$ to mean a generic permutation over $m$ elements. When $\pi$ acts on an integer it returns another integer that results from the permutation. When $\pi$ is applied on a vector of size $m$ it permutes the $m$ elements.

$n(\tilde{\mathbf{c}}^{(1)})_s = n(\tilde{\mathbf{c}}^{(0)})_s - 1$ and $n(\tilde{\mathbf{c}}^{(1)})_t = n(\tilde{\mathbf{c}}^{(0)})_t + 1$. Hence a call to ADDEDGE($\mathbf{n}(\tilde{\mathbf{c}}^{(0)}), \mathbf{n}(\tilde{\mathbf{c}}^{(1)})$) adds weight $M_{st}$ to the edge between the node for $\mathbf{n}(\tilde{\mathbf{c}}^{(0)})$ and that for $\mathbf{n}(\tilde{\mathbf{c}}^{(0)})$. Because in the context of PERTURBBOUND in Algorithm 3, $\mathbf{n}(\tilde{\mathbf{c}}^{(0)}) = \mathbf{n}_-$ and $\mathbf{n}(\tilde{\mathbf{c}}^{(1)}) = \mathbf{n}_+$, $M_{\mathbf{n}_-, \mathbf{n}_+} = M_{st} = \xi^{(1)}$. We have thus far shown that Equation 52 holds for $r = 1$.

Next we will use induction to show that for any $1 < q < r - 1$,

$$\xi^{(q)} m_{\mathbf{b}^{(q)}}(\boldsymbol{\lambda}) = \left( \prod_{i=1}^{q-1} \frac{1}{|z - E^{(i)}|} \right) \cdot m_{\mathbf{b}^{(q)}}(\boldsymbol{\lambda}) \cdot \left( \prod_{j: \exists i, j \in \mathcal{F}_i^{(q)}} M_{c_i^{(j-1)}, c_i^{(j)}} \right) \cdot \omega^{k_q} \tag{53}$$

where $\mathcal{F}_i^{(q)} = \{j = 1, \cdots, q | c_i^{(j-1)} \neq c_i^{(j)}\}$ and $k_q = q - \sum_{i=1}^{m} |\mathcal{F}_i^{(q)}|$. Let $\mathbf{f}^{(q)}$ be such that $f_i = |\mathcal{F}_i^{(q)}|$, then $\mathbf{b}^{(q)}$ denotes $\mathbf{f}^{(q)}$ with its elements sorted in non-descending order to follow Definition 2 for reduced partitions. With the same rearrangement that leads to Equation 39 from Equation 38, one could see that the right hand side of Equation 53 is equal to Expression 51.

We start the induction by assuming that there is a $Q < r - 1$ such that Equation 53 holds for any $q \leq Q$. Now consider the $Q$-th call to UPDATECELL (Algorithm 2) during the step 2j of PERTURBBOUND in Algorithm 3 on the node associated with the energy combination $\mathbf{n}(\tilde{\mathbf{c}}^{(Q)})$. Depending on the step $\tilde{\mathbf{c}}^{(Q)} \to \tilde{\mathbf{c}}^{(Q+1)}$ there are 3 possible scenarios according to Definition 12:

(i). $\tilde{\mathbf{c}}^{(Q)} = \tilde{\mathbf{c}}^{(Q+1)}$. In this case $\mathcal{T}^{(Q+1)} = \frac{\omega}{|z-E^{(Q)}|} \mathcal{T}^{(Q)}$ from step 1 of UPDATECELL in Algorithm 2. None of the sets $\mathcal{F}_i^{(Q)}$ are changed so $\mathcal{F}_i^{(Q)} = \mathcal{F}_i^{(Q+1)}$ for all $i = 1, \cdots, m$ and $\mathbf{b}^{(Q)} = \mathbf{b}^{(Q+1)}$. Therefore

$$\begin{aligned}
\xi^{(Q+1)} m_{\mathbf{b}^{(Q+1)}}(\boldsymbol{\lambda}) &= \frac{\omega}{|z - E^{(Q)}|} \xi^{(Q)} m_{\mathbf{b}^{(Q)}}(\boldsymbol{\lambda}) \\
&= \frac{\omega}{|z - E^{(Q)}|} \cdot \left( \prod_{i=1}^{Q-1} \frac{1}{|z - E^{(i)}|} \right) \cdot m_{\mathbf{b}^{(Q)}}(\boldsymbol{\lambda}) \cdot \left( \prod_{j: \exists i, j \in \mathcal{F}_i^{(Q)}} M_{c_i^{(j-1)}, c_i^{(j)}} \right) \cdot \omega^{k_Q} \\
&= \left( \prod_{i=1}^{(Q+1)-1} \frac{1}{|z - E^{(i)}|} \right) \cdot m_{\mathbf{b}^{(Q+1)}}(\boldsymbol{\lambda}) \cdot \left( \prod_{j: \exists i, j \in \mathcal{F}_i^{(Q+1)}} M_{c_i^{(j-1)}, c_i^{(j)}} \right) \cdot \omega^{k_{Q+1}}.
\end{aligned} \tag{54}$$

where $k_{Q+1} = Q + 1 - \sum_{i=1}^{m} |\mathcal{F}_i^{(Q+1)}|$. On the second line we used the inductive hypothesis Equation 53 for $q = Q$. By Equation 54 we have established that Equation 53 is also true $q = Q + 1$.

(ii). $\tilde{\mathbf{c}}^{(Q)}$ and $\tilde{\mathbf{c}}^{(Q+1)}$ differ by 1 at one element. Consider a walk in $\mathbf{c}$ with $\tilde{\mathbf{c}}(\mathbf{c}^{(i)}) = \tilde{\mathbf{c}}^{(i)}$. Let $h$ be such that $|\mathbf{c}_h^{(Q+1)} - \mathbf{c}_h^{(Q)}| = 1$. Note that here we are concerned with the walk $\mathbf{c}^{(0)} \to \mathbf{c}^{(1)} \to \cdots \mathbf{c}^{(Q)}$ in $\mathbf{c}$ instead of $\tilde{\mathbf{c}}$, which by similar arguments that lead to Equation 39 from 38 in Lemma 6, contributes a factor

$$\begin{aligned}
f(\mathbf{c}^{(0)}, \mathbf{c}^{(1)}) \cdot \frac{1}{|z - E^{(1)}|} \cdot f(\mathbf{c}^{(1)}, \mathbf{c}^{(2)}) \cdots f(\mathbf{c}^{(Q-2)}, \mathbf{c}^{(Q-1)}) \cdot \frac{1}{|z - E^{(Q-1)}|} \cdot f(\mathbf{c}^{(Q-1)}, \mathbf{c}^{(Q)}) \\
= \left( \prod_{i=1}^{m} \frac{1}{|z - E^{(i)}|} \right) \cdot \left( \prod_{i=1}^{m} \lambda_i^{|\mathcal{F}_i^{(Q)}|} \right) \cdot \left( \prod_{j: \exists i, j \in \mathcal{F}_i^{(Q)}} M_{c_i^{(j-1)}, c_i^{(j)}} \right) \cdot \omega^{k_Q}.
\end{aligned} \tag{55}$$

Applying the inductive hypothesis for $q = Q$, we have that the walk $\mathbf{c}^{(0)} \to \mathbf{c}^{(1)} \to \cdots \to \mathbf{c}^{(Q)} \to \mathbf{c}^{(Q+1)}$ contributes an upper bound

$$\begin{aligned}
f(\mathbf{c}^{(0)}, \mathbf{c}^{(1)}) \cdot \frac{1}{|z - E^{(1)}|} \cdot f(\mathbf{c}^{(1)}, \mathbf{c}^{(2)}) \cdots f(\mathbf{c}^{(Q-1)}, \mathbf{c}^{(Q)}) \cdot \frac{1}{|z - E^{(Q)}|} \cdot \underbrace{f(\mathbf{c}^{(Q)}, \mathbf{c}^{(Q+1)})}_{=\lambda_h M_{c_h^{(Q)}, c_h^{(Q+1)}}} \\
= \left( \prod_{i=1}^{m} \frac{1}{|z - E^{(i)}|} \right) \cdot \left( \prod_{i=1}^{m} \lambda_i^{|\mathcal{F}_i^{(Q)}|} \right) \cdot \lambda_h \cdot \left( \prod_{j: \exists i, j \in \mathcal{F}_i^{(Q)}} M_{c_i^{(j-1)}, c_i^{(j)}} \right) \cdot M_{c_h^{(Q)}, c_h^{(Q+1)}} \cdot \omega^{k_Q}.
\end{aligned} \tag{56}$$

Here in Equation 56 $\lambda_h$ will merge with the $\lambda_h^{\mathcal{F}_h^{(Q)}}$ term in the product, producing $\lambda_h^{|\mathcal{F}_h^{(Q)}|+1}$. Since by definition of $\mathcal{F}_i^{(Q)}$, $\mathcal{F}_h^{(Q+1)} = \mathcal{F}_h^{(Q)} \cup \{Q+1\}$, $|\mathcal{F}_h^{(Q+1)}| = |\mathcal{F}_h^{(Q)}| + 1$. Because $\tilde{\mathbf{c}}^{(Q)} \neq \tilde{\mathbf{c}}^{(Q+1)}$ in this case,

22

$k_Q = k_{Q+1}$. Finally, because $(Q+1) \in \mathcal{F}_h^{(Q+1)}$, $M_{c_h^{(Q)}, c_h^{(Q+1)}}$ merges into the product $\prod_{j:\exists i, j \in \mathcal{F}_i^{(Q)}} M_{c_i^{(j-1)}, c_i^{(j)}}$ and Expression 56 becomes

$$\left( \prod_{i=1}^{(Q+1)-1} \frac{1}{|z - E^{(i)}|} \right) \cdot \left( \prod_{i=1}^{m} \lambda_i^{|\mathcal{F}_i^{(Q+1)}|} \right) \cdot \left( \prod_{j:\exists i, j \in \mathcal{F}_i^{(Q+1)}} M_{c_i^{(j-1)}, c_i^{(j)}} \right) \cdot \omega^{k_{Q+1}}. \tag{57}$$

By Lemma 5, the contribution of the $(Q+1)$-step walk in the reduced configuration $\tilde{\mathbf{c}}$ can be obtained by summing over all permutation $\pi : [m] \mapsto [m]$ of the subsystems, yielding

$$\left( \prod_{i=1}^{(Q+1)-1} \frac{1}{|z - E^{(i)}|} \right) \cdot \underbrace{\sum_{\pi:[m]\mapsto[m]} \left( \prod_{i=1}^{m} \lambda_{\pi(i)}^{|\mathcal{F}_{\pi(i)}^{(Q+1)}|} \right)}_{m_{\mathbf{b}^{(Q+1)}}(\boldsymbol{\lambda})} \cdot \left( \prod_{j:\exists i, j \in \mathcal{F}_i^{(Q+1)}} M_{c_i^{(j-1)}, c_i^{(j)}} \right) \cdot \omega^{k_{Q+1}}. \tag{58}$$

We would like to show that our Algorithm indeed computes expression 58 correctly. First of all, $\mathcal{T}^{(Q+1)}$ could only be generated by first calling OUT($\mathbf{n}^{(Q)}, \mathbf{n}^{(Q+1)}, \mathcal{T}^{(Q)}$) in Algorithm 2. During the OUT call, the algorithm starts out by reconstructing $\tilde{\mathbf{c}}^{(Q)}$ and $\tilde{\mathbf{c}}^{(Q+1)}$ from $\mathbf{b}^{(Q)}$, $\mathbf{n}(\tilde{\mathbf{c}}^{(Q)})$ and $\mathbf{n}(\tilde{\mathbf{c}}^{(Q+1)})$ at step 2 of OUT. Since we assumed that $|c_h^{(Q+1)} - c_h^{(Q)}| = 1$, there must be an $h'$ such that $|\tilde{c}_{h'}^{(Q+1)} - \tilde{c}_{h'}^{(Q)}| = 1$. In other words, $\tilde{c}_{h'}^{(Q+1)} = c_h^{(Q+1)}, \tilde{c}_{h'}^{(Q)} = c_h^{(Q)}$. The algorithm marks all such possible $h'$ indices in $\tilde{\mathbf{c}}^{(Q)}$. With mapping $\mu^{(Q)}$ we are able to locate the element in $\mathbf{b}^{(Q)}$ that stores $|\mathcal{F}_h^{(Q)}|$. The algorithm OUT then correctly increments the element by 1, to generate $\mathbf{b}^{(Q+1)}$. Because of the way ADDEDGE in Algorithm 1 is set up for constructing the cellular automaton which leads to $M_{\tilde{c}_{h'}^{(Q)}, \tilde{c}_{h'}^{(Q+1)}} = M_{\mathbf{n}^{(Q)}, \mathbf{n}^{(Q+1)}}$, the step 1 leads to $\xi^{(Q)} \leftarrow \xi^{(Q)} \cdot M_{\tilde{c}_{h'}^{(Q)}, \tilde{c}_{h'}^{(Q+1)}}$. Putting these together, we can see that OUT($\mathbf{n}^{(Q)}, \mathbf{n}^{(Q+1)}, \mathcal{T}^{(Q)}$) produces a 4-tuple

$$\mathcal{T}^{(Q,Q+1)} = (\tilde{\mathbf{c}}^{(Q+1)}, \mathbf{b}^{(Q+1)}, \xi^{(Q)} \cdot M_{c_h^{(Q)}, c_h^{(Q+1)}}, \mu^{(Q+1)}) \tag{59}$$

where $\mu^{(Q)} = \mu^{(Q+1)}$ because no new element is introduced in $\mathbf{b}^{(Q)}$. Then during step 1 of UPDATECELL in Algorithm 2, $\mathcal{T}^{(Q+1)}$ is finally generated by the operation

$$\mathcal{T}^{(Q+1)} = \frac{1}{|z - E^{(Q)}|} \mathcal{T}^{(Q,Q+1)} = (\tilde{\mathbf{c}}^{(Q+1)}, \mathbf{b}^{(Q+1)}, \xi^{(Q+1)}, \mu^{(Q+1)}) \in \mathcal{S}_{\mathbf{n}(\tilde{\mathbf{c}}^{(Q)}), \mathbf{n}(\tilde{\mathbf{c}}^{(Q+1)})}. \tag{60}$$

It is straightforward to verify that $\xi^{(Q+1)} m_{\mathbf{b}^{(Q+1)}}(\boldsymbol{\lambda})$ equals to expression 58.

(iii). $|\tilde{\mathbf{c}}^{(Q+1)}| = |\tilde{\mathbf{c}}^{(Q)}| + 1$. Consider the same walk $\mathbf{c}^{(0)} \to \mathbf{c}^{(1)} \to \cdots \to \mathbf{c}^{(Q)} \to \mathbf{c}^{(Q+1)}$ as the case ii with $\tilde{\mathbf{c}}(\mathbf{c}^{(i)}) = \tilde{\mathbf{c}}^{(i)}$. In this case there is some $h$ such that $\tilde{c}_h^{(Q)} = 0$ and $\tilde{c}_h^{(Q+1)} = 1$. Also $|\mathcal{F}_h^{(Q)}| = 0$ and $|\mathcal{F}_h^{(Q+1)}| = 1$. In other words a new element is added to $\mathbf{b}^{(Q)}$ to store $|\mathcal{F}_h^{(Q+1)}|$. Hence $\mathbf{b}^{(Q)}$ to store $|\mathcal{F}_h^{(Q+1)}|$. Hence $\mathbf{b}^{(Q+1)} = (\mathbf{b}^{(Q)} \quad 1)$. The algorithm identifies this case by testing if both $|\hat{\mathbf{c}}(\mathbf{n}(\mathbf{c}^{(Q)}))| < |\hat{\mathbf{c}}(\mathbf{n}(\tilde{\mathbf{c}}^{(Q+1)}))|$ and $|\mathbf{b}^{(Q)}| < m$ are true, because if the former is false it implies that $\tilde{\mathbf{c}}^{(Q+1)}$ has one more active element (Definition 10) with energy $E_0$ than $\tilde{\mathbf{c}}^{(Q)}$, which is impossible for any possible step $\tilde{\mathbf{c}}^{(Q)} \to \tilde{\mathbf{c}}^{(Q+1)}$ as stated in Definition 12. If $|\mathbf{b}^{(Q)}| = m$ then there is no $h$ such that $|\mathcal{F}_h^{(Q)}| = 0$, another contradiction. Therefore the algorithm correctly captures the necessary and sufficient condition for this case and once it does, during the $Q$-th call to UPDATECELL on $v_{\mathbf{n}(\tilde{\mathbf{c}}^{(Q)})}$ it generates a new partition $\mathbf{b}^{(Q+1)} = (\mathbf{b}^{(Q)} \quad 1)$, according to step 4(a)ii of OUT in Algorithm 2, and the new element "1" is mapped from $\tilde{c}_h^{(Q+1)}$. The call OUT($\mathbf{n}(\tilde{\mathbf{c}}^{(Q)}), \mathbf{n}(\tilde{\mathbf{c}}^{(Q+1)}), \mathcal{T}^{(Q)}$) returns a new 4-tuple

$$\mathcal{T}^{(Q,Q+1)} = (\tilde{\mathbf{c}}^{(Q+1)}, \mathbf{b}^{(Q+1)}, \xi^{(Q)} \cdot M_{01}, \mu^{(Q+1)}) \in \mathcal{S}_{\mathbf{n}(\tilde{\mathbf{c}}^{(Q)}), \mathbf{n}(\tilde{\mathbf{c}}^{(Q+1)})} \tag{61}$$

The step 1 during the $(Q+1)$-th call to UPDATECELL on $v_{\mathbf{n}(\tilde{\mathbf{c}}^{(Q+1)})}$ generates

$$\mathcal{T}^{(Q+1)} = \frac{1}{|z - E^{(Q)}|} \mathcal{T}^{(Q,Q+1)} \tag{62}$$

with $\xi^{(Q+1)} m_{\mathbf{b}^{(Q+1)}}(\boldsymbol{\lambda})$ being equal to Expression 53 with $q = Q+1$.

At the final step $\mathbf{c}^{(r-1)} \to \mathbf{c}^{(r)}$ only case ii holds. The same arguments carry over here. This concludes our proof. □

**Algorithm 3** Algorithm for computing an upper bound to $\|T_r\|_\infty$

**Input:**

- The order of perturbation, $r$;
- The scalar $\omega \in \mathbb{R}$ as in Definition 3;
- The vector $\boldsymbol{\lambda} \in \mathbb{R}^m$ as in Definition 4;
- The matrix $\mathbf{M} \in \mathbb{R}^{\ell \times \ell}$ as in Definition 5.

**Output:**

- An upper bound for $\|T_r\|_\infty$, which we denote as $\tau_r$.

**Procedure** $\tau_r = \text{PerturbBound}(r, \boldsymbol{\lambda}, \mathbf{M})$

1. Build the graph $G_0$ using Algorithm 1: $G_0(\mathcal{V}_0, \mathcal{E}_0) = \text{BuildCA}(|\boldsymbol{\lambda}|, \mathbf{M})$;

2. For each $\mathbf{n}_- \in \mathcal{N}_-$,

    (a) $G(\mathcal{V}, \mathcal{E}) \leftarrow G_0(\mathcal{V}_0, \mathcal{E}_0)$;

    (b) For all $\mathbf{n}, \mathbf{n}'$, $\mathcal{S}_\mathbf{n} \leftarrow \emptyset$ and $\mathcal{S}_{\mathbf{n},\mathbf{n}'} \leftarrow \emptyset$;

    (c) $\mathcal{E} \leftarrow \mathcal{E}_0 \setminus \bigcup_{\substack{\mathbf{n} \in \mathcal{N}_- \setminus \{\mathbf{n}_-\}, \\ \mathbf{n}' \in \mathcal{N}_-}} e(v_\mathbf{n}, v_{\mathbf{n}'})$;

    (d) $\mathcal{E}_{\text{dashed}} \leftarrow \bigcup_{\substack{\mathbf{n}' \in \mathcal{N}_+ \\ \mathbf{n} \in \mathcal{N}_-}} e(v_{\mathbf{n}'}, v_\mathbf{n})$;

    (e) $\mathcal{T}_- = \{\hat{\mathbf{c}}(\mathbf{n}_-), \quad \mathbf{b} = \underbrace{(0, \cdots, 0)}_{|\hat{\mathbf{c}}(\mathbf{n}_-)|}, \quad \xi = 1, \quad \mu : \hat{\mathbf{c}}(\mathbf{n}_-) \mapsto \mathbf{b}\}$;

    (f) Randomly choose a neighbor $\mathbf{n}_+ \in \mathcal{N}_+$ of $\mathbf{n}_-$;

    (g) Compute $\mathcal{S}_{\mathbf{n}_-, \mathbf{n}_+} = \text{Out}(\mathbf{n}_-, \mathbf{n}_+, \mathcal{T}_-)$ and randomly choose one 4-tuple $\mathcal{T}_{-+} \in \mathcal{S}_{\mathbf{n}_-, \mathbf{n}_+}$;

    (h) $\mathcal{S}_{\mathbf{n}_-, \mathbf{n}_+} \leftarrow \mathcal{T}_{-+}$;

    (i) $\mathcal{E} \leftarrow \mathcal{E} \setminus \mathcal{E}_{\text{dashed}}$;

    (j) Repeat $(r-1)$ times the following: For any $v_\mathbf{n} \in \mathcal{V}$, run $\text{UpdateCell}(v_\mathbf{n})$;

    (k) $\mathcal{E} \leftarrow \mathcal{E} \cup \mathcal{E}_{\text{dashed}}$;

    (l) For any $v_\mathbf{n} \in \mathcal{V}$, run $\text{UpdateCell}(v_\mathbf{n})$;

    (m) $\tau_{r, \mathbf{n}_-} \leftarrow \sum_{\mathbf{n} \in \mathcal{N}_-} \sum_{(\tilde{\mathbf{c}}, \mathbf{b}, \xi, \mu) \in \mathcal{S}_\mathbf{n}} \xi m_\mathbf{b}(\boldsymbol{\lambda})$;

3. Return $\tau_r = \max_{\mathbf{n}_- \in \mathcal{N}_-} \tau_{r, \mathbf{n}_-}$.

## 4.4 Dealing with infinity

Obviously, computing the error exactly requires summing the perturbative series (Equation 9) to infinite order, which is not possible. Hence we make a relaxation by truncating the summation at some finite order $p$ and proving that the norm of the sum from $p+1$ to infinity is bounded from above by some quantity that is easy to calculate. In particular, at $p$-th order, $p \geq 2$, we have the perturbative term $T_p = V_{-+}(G_+V_+)^{p-2}G_+V_{+-}$. Suppose we have found an upper bound $\gamma_p$ such that $\|V_{-+}(G_+V_+)^{p-2}G_+V_{+-}\| \leq \gamma_p$. Then an upper bound for the $p+1$-st order can be established using the inequality $\|AB\| \leq \|A\| \cdot \|B\|$ for submultiplicative norms:

$$\|T_{p+1}\| \leq \|T_p\| \cdot \|G_+V_+\| \leq \gamma_p \|G_+V_+\| \leq \frac{\gamma_p}{\Delta}\|V_+\|. \tag{63}$$

Here in the last inequality we have used the definition of $\Delta$ being the lowest excited state energy in the unperturbed Hamiltonian $H$. Let $r = \|V_+\|/\Delta$. Then we could bound the infinite sum by the triangle inequality:

$$\|\sum_{j=p+1}^{\infty} T_j\| \leq \sum_{j=p+1}^{\infty} \|T_j\| \leq \gamma_p(r + r^2 + r^3 + ....) = \gamma_p \frac{r}{1-r}. \tag{64}$$

To make sure that the series on the right hand side converges, we need $r < 1$, which is true for all the constructions we consider here.

## 5 Numerical example

Here we show an example that demonstrates the effectiveness of our algorithm. Consider the quantum system of 11 spins described in Figure 3a of the main text. The Hamiltonian can be expressed in form of the general setting $\tilde{H} = H + V$ described in Figure 1a of the main text. Here the unperturbed Hamiltonian $H$ and perturbation $V$ are defined as

$$\begin{aligned} H &= H^{(1)} + H^{(2)}, & H^{(1)} &= \frac{\Delta}{4}(Z_{u_1}Z_{u_2} + Z_{u_2}Z_{u_3} + Z_{u_1}Z_{u_3}) \\ & & H^{(2)} &= \frac{\Delta}{4}(Z_{v_1}Z_{v_2} + Z_{v_2}Z_{v_3} + Z_{v_1}Z_{v_3}) \\ V &= V^{(1)} + V^{(2)}, & V^{(1)} &= \mu_1(X_1 X_{u_1} + X_2 X_{u_2} + X_3 X_{u_3}) \\ & & V^{(2)} &= \mu_2(Y_4 X_{v_1} + X_2 X_{v_2} + Z_5 X_{v_3}) \end{aligned} \tag{65}$$

where spins with $u_i$ and $v_i$ labels belong to the two unperturbed subsystems. Here we let $\Delta$ be orders of magnitude larger than $\mu_1$ and $\mu_2$ and keep the coefficients $\mu_1$ and $\mu_2$ as

$$\mu_1 = \left(\frac{\alpha_1 \Delta^2}{6}\right)^{1/3}, \qquad \mu_2 = \left(\frac{\alpha_2 \Delta^2}{6}\right)^{1/3} \tag{66}$$

where $\alpha_1$ and $\alpha_2$ are parameters related to the low energy effective Hamiltonian (see Equation 71). In Figure 3c of the main text we explicitly partition the Hamiltonian in the form of general setting discussed in Section 2.1 (Figure 1a of the main text).

The low-energy subspace of the total Hamiltonian $\tilde{H}$ is then $\mathcal{L}_- = \mathcal{L}_-^{(1)} \otimes \mathcal{L}_-^{(2)}$. Inspecting the expressions $H^{(1)}$ and $H^{(2)}$ gives the low energy subspaces for each subsystem: $\mathcal{L}_-^{(1)} = \text{span}\{|000\rangle_{u_1u_2u_3}, |111\rangle_{u_1u_2u_3}\}$ and $\mathcal{L}_-^{(2)} = \text{span}\{|000\rangle_{v_1v_2v_3}, |111\rangle_{v_1v_2v_3}\}$. For each subsystem $i \in \{1, 2\}$, the subspaces of $H^{(i)}$ and their corresponding energies are

$$\begin{aligned} \mathcal{P}_0 &= \text{span}\{|000\rangle\}, & E_0 &= 0 \\ \mathcal{P}_1 &= \text{span}\{|001\rangle, |010\rangle, |100\rangle\}, & E_1 &= \Delta \\ \mathcal{P}_2 &= \text{span}\{|011\rangle, |101\rangle, |110\rangle\}, & E_2 &= \Delta \\ \mathcal{P}_3 &= \text{span}\{|111\rangle\}, & E_3 &= 0. \end{aligned} \tag{67}$$

In Figure 3d of the main text we show the spectrum of each subsystem. The vector $\boldsymbol{\lambda} = (\lambda_1, \lambda_2)$, which characterizes the "magnitudes" of perturbations onto each subsystem (Definition 4) can be determined based on Equation 65 as

$$\lambda_1 = \mu_1, \qquad \lambda_2 = \mu_2. \tag{68}$$

From the diagram in Figure 3d of the main text we could also determine the matrix $\mathbf{M}$ (see Definition 5) for this system. One could compute the matrix elements $M_{ij}$ from the figure, where $M_{ij}$ is the maximum, over all eigenstates of $H$ in $\mathcal{P}_i$, number of possible transitions from a particular $|u\rangle \in \mathcal{P}_i$ to an eigenstate in $\mathcal{P}_j$. Precisely,

$$M_{ij} = \max_{|u\rangle \in \mathcal{P}_i} Card\{|v\rangle \in \mathcal{P}_j | \|\langle v|V|u\rangle\| \neq 0\} \tag{69}$$

where $Card\{\cdot\}$ stands for cardinality (number of distinct elements) of a set. We could then determine that

$$\mathbf{M} = \begin{array}{c} \\ \mathcal{P}_0 \\ \mathcal{P}_1 \\ \mathcal{P}_2 \\ \mathcal{P}_3 \end{array} \begin{array}{cccc} \mathcal{P}_0 & \mathcal{P}_1 & \mathcal{P}_2 & \mathcal{P}_3 \end{array} \left[ \begin{array}{cccc} & 3 & & \\ 1 & & 2 & \\ & 2 & & 1 \\ & & 3 & \end{array} \right] \tag{70}$$

where the row and column indices start from 0 because the subspaces $\mathcal{P}_0, \mathcal{P}_1, \cdots$, have indices that start from 0.

From Figure 3a and 3c of the main text we can see that the unperturbed system $H$ essentially consists of two identical 4-level systems with energy levels $E_0$, $E_1$, $E_2$ and $E_3$. This gives rise to in total 9 possible energy combinations (Definition 7). Starting from the all-zero energy combination $\mathbf{n}_0 = [2, 0, 0, 0]$ and running Algorithm 1, we could construct a cellular automaton as shown in Figure 4d of the main text. We tabulate all the cells and their relevant information as in Figure 4c of the main text.

With the vector $\boldsymbol{\lambda}$ and the matrix $\mathbf{M}$ worked out as in Equations (68) and (70), we could use Algorithm 3 to find a tight upper bound for $\|T_r\|_\infty$ at any order $r$. After a certain order $p$, when the upper bound becomes sufficiently small (assuming $\|T_r\|_\infty \to 0$ as $r \to \infty$), we use Equation 64 to bound the terms from $p+1$ to infinity.

Using the perturbation series in Equation (9) we could show that if we truncate the series at the 3rd order, namely $\Sigma_-(z) = H_{\text{eff}} + T_4 + T_5 + \cdots$, we have the effective 3-body Hamiltonian

$$H_{\text{eff}} = \alpha_1 X_1 X_2 X_3 + \alpha_2 X_2 Y_4 Z_5 + \gamma \mathbf{1} \tag{71}$$

for some $\gamma$ that signifies the magnitude of the spectral shift. Here we let $\alpha_1 = 0.1$ and $\alpha = 0.2$. Then the entire Hamiltonian $\tilde{H} = H + V$ in Equation 65 is only dependent on a free parameter $\Delta$. In order to test our algorithm for bounding perturbative terms, we treat terms from 4th order onward as errors in the perturbation series. This amounts to estimating $\|\Sigma_-(z) - H_{\text{eff}}\|$. We could compute this value by explicitly computing $\Sigma_-(z)$ by its definition $z\mathbf{1} - (\tilde{G}_-(z))^{-1}$ and then evaluating $\|\Sigma_-(z) - H_{\text{eff}}\|$. This method is inefficient but yields an accurate estimation for the error $\|\Sigma_-(z) - H_{\text{eff}}\|$. We will use it as a benchmark for comparison with the upper bound computed by the new algorithm developed here. As shown Figure 5 of the main text, the upper bounds computed by the cellular automaton algorithms are tight with respect to the exact calculation. For the purpose of comparison we also compute the error bound due to triangle inequality (see Equation 12). We explicitly computed $\|V\|_2$ and bounded $\|G_+\|$ from above by $1/E_1$. Hence the simple bound based on Equation 12 becomes $\sum_{r=4}^\infty \|V\|_2^r / E_1^{r-1} = \|V\|_2^4 / (E_1^2 (E_1 - \|V\|_2))$.

Note from Figure 5 of the main text that our upper bound based on the output of the CA algorithm only differs from the simple bound by a constant factor. This provides empirical justification for the method to treat infinity described in Section 4.4. When implementing the CA algorithm for the numerical example concerned in this section, we compute $\tau_r = \text{PerturbBound}(r, \boldsymbol{\lambda}, \mathbf{M})$ for $r$ from 4 to a value $p$ such that $\tau_p \leq 10^{-20}$. Then we resort to Equation 64 for computing an upper bound to $\|T_{p+1} + \cdots\|_2$.

# 6 Discussions

- Our algorithms are constructed based on a physical setting that is not without assumptions. The first major assumption concerns the structure of $V$ as described in Equation 5 and 6. The block tridiagonal structure of $V^{(i)}$ has a direct consequence on what transitions are possible during one step of a walk, be it in $H$ eigenstates (Definition 8), energy configuration $\mathbf{c}$ (Definition 9), reduced energy configuration $\tilde{\mathbf{c}}$ (Definition 12) or energy combination $\mathbf{n}$ (Definition 14). In case one would like to relax the assumption of $V^{(i)}$ being tridiagonal and would like to instead treat $V^{(i)}$'s that are band diagonal with the band width being greater than 3, the definitions of the walks will need to be modified to account for $V$ being able to change an element of $\mathbf{c}$ by more than 1 during a single step $\mathbf{c}^{(i)} \to \mathbf{c}^{(i+1)}$. The algorithms will also need to be adjusted accordingly.

  A second assumption concerns the magnitude of $V$. Here in order to guarantee the convergence of perturbation series $\Sigma_-(z)$ in the regime of $z$ specified by Theorem 1, we assume that $\|V\|_2 \leq \Delta/2$. In general this assumption could be weakened [3] to a statement that ultimately is not dependent on any global property of $V$, such as $\|V\|_2$, and the series in $\Sigma_-(z)$ still converges and Theorem 1 could still hold.

- We derive the upper bound using symmetric polynomials, as one could see from Lemma 3 and Lemma 6. An implicit assumption on using symmetric polynomials is that the terms in $V$ commute with each other. Otherwise for example if $V$ contains terms that are proportional to $\lambda_1 X_i$ and $\lambda_2 Z_i$ operating on the same spin $i$, at high orders one may expect terms such as $\lambda_1\lambda_2 X_i Z_i + \lambda_2\lambda_1 Z_i X_i$, which is vanishing but the symmetric polynomial would include such terms as $\lambda_1\lambda_2 + \lambda_2\lambda_1 = 2\lambda_1\lambda_2$, which is non-zero. This unawareness of non-commutativity will cause the upper bound computed by the algorithm to be less tight than the case shown in Section 5, where all terms in $V$ commute.

- Perhaps one of the areas where our algorithm could find direct application is adiabatic quantum computation, where one often works with quantum systems with simple, restricted forms of interaction but wishes to realize some effective interactions $H_{\text{eff}}$ that are more complicated. A common idea is to construct a Hamiltonian $\tilde{H}$ for which perturbation theory gives rise to $H_{\text{eff}}$ at the first few orders. Then it becomes instrumental to have accurate estimation of how large the higher order error terms are. In fact a seemingly minor improvement in error estimation could lead to significant reduction in the resource required for producing $H_{\text{eff}}$ using constructions of $\tilde{H}$, see for example [2]. Our algorithm certainly will enable improvement on a broader class of constructions of $\tilde{H}$ for adiabatic quantum computing than prior works by providing accurate error estimates that are not available with simple techniques (such as those that lead to Equation 12).

- The parallel nature of the update rules in cellular automata could facilitate parallelism in the software implementation of our algorithms, which will further speed up the computation. For example, with $O(m)$ processors each storing the information of one cell and its out going edges, the algorithm takes $O(rh(r))$ time. Here $h(r)$ is the maximum number of 4-tuples stored in any cell or edge during the algorithm.

# A  Glossary of notations

As a general guideline, throughout this Supplementary Material we use lower case Greek letters for scalar quantities, lower case bold English letters for representing vectors and capital case English letters for representing matrices and operators. Calligraphic fonts (such as $\mathcal{H}$ for the letter 'H') are reserved for representing vector spaces and sets of vertices (as in $\mathcal{E}$). For a vector $\mathbf{v}$, the subscript in the notation $v_i$ represents the $i$-th element of $\mathbf{v}$. Superscripts in parentheses have two possible meanings: depending on the context, they could mean either the subsystem that

the operator acts on (as in Figure 1a of the main text) or the step in a walk. Tables 1 and 2 contain the main recurring notations introduced in this Supplementary Material.

## B   An example for illustrating walks in unperturbed eigenspaces

Consider the setting described in Figure 2 with $m = 2$ and $\ell = 2$. This means that there are in total 2 copies of identical unperturbed systems. Let $\mathcal{H}_1$ and $\mathcal{H}_2$ be their respective Hilbert spaces. $\ell = 2$ means that each of the unperturbed systems are 3-level systems with energy levels $E_0^{(1)}$, $E_1^{(1)}$ and $E_2^{(1)}$ for system 1 and similarly for system 2, with the superscript '(1)' replaced with '(2)'. We assume the subspace $\mathcal{P}_1$ for both unperturbed systems is 2-fold degenerate with eigenstates $|\psi_{1,1}\rangle$ and $|\psi_{1,2}\rangle$, as shown in Figure 2b. Under the basis of the unperturbed eigenstates with ordering $|\psi_{0,1}\rangle, |\psi_{1,1}\rangle, |\psi_{1,2}\rangle, |\psi_{2,1}\rangle$, the unperturbed Hamiltonian for each subsystems $H^{(1)}$ and $H^{(2)}$ can be written as

$$H^{(1)} = \begin{pmatrix} E_0^{(1)} & & & \\ & E_1^{(1)} & & \\ & & E_1^{(1)} & \\ & & & E_2^{(1)} \end{pmatrix} \otimes \mathbf{I}_{\mathcal{H}_2}, \tag{72}$$

$$H^{(2)} = \mathbf{I}_{\mathcal{H}_1} \otimes \begin{pmatrix} E_0^{(2)} & & & \\ & E_1^{(2)} & & \\ & & E_1^{(2)} & \\ & & & E_2^{(2)} \end{pmatrix} \tag{73}$$

where $\mathbf{I}$ is the identity operator of appropriate dimension. We assume that there is a (large) gap $\Delta$ between $E_0$ and $E_1$ of each subsystem and $E_* = \frac{E_0+E_1}{2}$ is the cutoff. Let the low energy subspace $\mathcal{L}_- = \mathcal{P}_0^{(1)} \otimes \mathcal{P}_0^{(2)}$. This is illustrated in Figure 2a.

We let the components $V^{(1)}$ and $V^{(2)}$ of the perturbation $V = V^{(1)} + V^{(2)}$ be such that

$$\begin{aligned} V^{(1)} &= B_{11,01}^{(1)} \otimes (|\psi_{0,1}^{(1)}\rangle\langle\psi_{1,1}^{(1)}| + |\psi_{1,1}^{(1)}\rangle\langle\psi_{0,1}^{(1)}|) \\ &+ B_{12,01}^{(1)} \otimes (|\psi_{0,1}^{(1)}\rangle\langle\psi_{1,2}^{(1)}| + |\psi_{1,2}^{(1)}\rangle\langle\psi_{0,1}^{(1)}|) \\ &+ B_{11,12}^{(1)} \otimes (|\psi_{1,1}^{(1)}\rangle\langle\psi_{2,1}^{(1)}| + |\psi_{2,1}^{(1)}\rangle\langle\psi_{1,1}^{(1)}|) \\ &+ B_{21,12}^{(1)} \otimes (|\psi_{1,2}^{(1)}\rangle\langle\psi_{2,1}^{(1)}| + |\psi_{2,1}^{(1)}\rangle\langle\psi_{1,2}^{(1)}|) \end{aligned} \tag{74}$$

and $V^{(2)}$ is the same as $V^{(1)}$ but with all superscripts replaced with '(2)'. In matrix forms,

$$V^{(1)} = \begin{pmatrix} & B_{11,01}^{(1)} & B_{12,01}^{(1)} & \\ B_{11,10}^{(1)} & & & B_{11,12}^{(1)} \\ B_{21,10}^{(1)} & & & B_{21,12}^{(1)} \\ & B_{11,21}^{(1)} & B_{12,21}^{(1)} & \end{pmatrix} \otimes \mathbf{I}, \tag{75}$$

$$V^{(2)} = \mathbf{I} \otimes \begin{pmatrix} & B_{11,01}^{(2)} & B_{12,01}^{(2)} & \\ B_{11,10}^{(2)} & & & B_{11,12}^{(2)} \\ B_{21,10}^{(2)} & & & B_{21,12}^{(2)} \\ & B_{11,21}^{(2)} & B_{12,21}^{(2)} & \end{pmatrix}. \tag{76}$$

As shown in Figure 2b, we can represent the component of $V^{(i)}$ acting on $\mathcal{H}_i$ as a graph with the operator $B_{mn,jk}^{(i)}$ as the "weight" of the edge that corresponds to the transition $|\psi_{j,m}^{(i)}\rangle\langle\psi_{k,n}^{(i)}|$. The factors $\lambda_i$ in this case are

$$\begin{aligned} \lambda_1 &= \max\{\|B_{11,01}^{(1)}\|_\infty, \|B_{12,01}^{(1)}\|_\infty, \|B_{11,12}^{(1)}\|_\infty, \|B_{21,12}^{(1)}\|_\infty\}, \\ \lambda_2 &= \max\{\|B_{11,01}^{(2)}\|_\infty, \|B_{12,01}^{(2)}\|_\infty, \|B_{11,12}^{(2)}\|_\infty, \|B_{21,12}^{(2)}\|_\infty\}. \end{aligned} \tag{77}$$

From the diagram we could see that to excite the eigenstate $|\psi_{0,1}\rangle$ of $\mathcal{P}_0$ into $\mathcal{P}_1$, there are in total 2 ways: $|\psi_{0,1}\rangle \to |\psi_{1,1}\rangle$ and $|\psi_{0,1}\rangle \to |\psi_{1,2}\rangle$. Hence $M_{01} = 2$. Following a similar line of argument we can see that $M_{10} = 1$, $M_{12} = 1$, and $M_{21} = 2$. Because we assume that $V$ is block tridiagonalizable with respect to any subsystem $i$, there will not be any transition from $\mathcal{P}_0$ to $\mathcal{P}_2$.

| Symbol | Meaning and first appearance |
|---|---|
| $\mathbf{a}$ | Partition of a symmetric polynomial $m_\mathbf{a}$, see Definition 1 in Section 2.4 |
| $\mathbf{b}$ | Reduced partition of a monomial symmetric polynomial. See Definition 2. |
| $\mathcal{B}$ | Hilbert space for the "bath" in the basic setting in Figure 1a of the main text. |
| $B^{(i)}_{pq,jk}$ | The $pq$-th block of $O^{(i)}_{jk}$ (Eq. 15). It contributes a term $B^{(i)}_{pq,jk} \otimes |\psi^{(i)}_{j,p}\rangle\langle\psi^{(i)}_{k,q}|$ to $V$. See Eq. 19. |
| $\mathbf{c}, \mathbf{c}(|\psi\rangle)$ | Energy configuration of an eigenstate $|\psi\rangle$ of $H$. See Def. 6. |
| $\tilde{\mathbf{c}}, \tilde{\mathbf{c}}(\mathbf{c})$ | Reduced energy configuration of a set of $H$ eigenstates with energy configuration $\mathbf{c}$. See Def. 11. |
| $\hat{\mathbf{c}}(\mathbf{n})$ | Uniquely reduced configuration associated with an energy combination $\mathbf{n}$. See Def. 13. |
| $E^{(j)}_i$ | The $i$-th energy level of the subsystem $H^{(j)}$ (Fig. 2b of the main text). Also written as $E_i$. |
| $E^{(i)}$ | The energy of the $i$-th step during a walk in $H$ eigenstates, $\mathbf{c}$, $\tilde{\mathbf{c}}$ or $\mathbf{n}$. See Def. 8. |
| $E(\mathbf{n})$ | The energy of an energy combination $\mathbf{n}$. See Equation 20. |
| $G(z)$ | Operator-valued resolvent, or Green's function. See Section 2.1 after Equation 2. |
| $G(\mathcal{V},\mathcal{E})$ | The graph generated by Algorithm 1. $\mathcal{V}$ and $\mathcal{E}$ are the sets of nodes and edges respectively. |
| $\mathcal{H}^{(i)}$ | Hilbert space of the $i$-th subsystem, see text after Equation 3. |
| $H$ | Unperturbed Hamiltonian for all subsystems (Figure 1a of the main text) |
| $H^{(i)}$ | The Hamiltonian for the $i$-th unperturbed subsystem. See Equation 5. |
| $H_\mathcal{B}$ | The part of $\tilde{H}$ that only acts on $\mathcal{B}$. See Equation 2. |
| $\tilde{H}$ | Perturbed Hamiltonian that equals to $H + V$. See Section 2.1 after Equation 2. |
| $\ell$ | Total number of energy levels in each subsystem $H^{(i)}$. See Section 2.1 after Equation 2. |
| $\mathcal{L}_-, \mathcal{L}_+$ | Low- and high- energy subspaces of $H$. See Section 2.1 after Equation 2. |
| $\mathcal{L}^{(i)}_-, \mathcal{L}^{(i)}_+$ | The low- and high- energy subspace of $H^{(i)}$. |
| $m$ | Total number of subsystems. See Figure 1a of the main text and Equation 5. |
| $m_\mathbf{b}(\mathbf{x})$ | Symmetric polynomial over variables $\mathbf{x} \in \mathbb{C}^n$ with reduced partition $\mathbf{b}$. See Section 2.4. |
| $\mathbf{M}, M_{jk}$ | Basic quantity for constructing an upper bound to $\|T_r\|_2$. See Definition 5. |
| $\mathcal{N}_-, \mathcal{N}_+$ | The set of energy combinations that corresponds to $\mathcal{L}_-$ and $\mathcal{L}_+$ respectively. See after Eq. 20. |
| $\mathbf{n}, \mathbf{n}(\mathbf{c}), \mathbf{n}(\tilde{\mathbf{c}})$ | Energy combination an $H$ eigenstate with energy configuration $\mathbf{c}$. Same for $\mathbf{n}(\tilde{\mathbf{c}})$. See Def. 7. |
| $O^{(i)}_{jk}$ | The $jk$-th block of the perturbation $V^{(i)}$ corresponding to transition from $\mathcal{P}^{(i)}_j$ to $\mathcal{P}^{(i)}_k$, see Eq. 6 |
| $\mathcal{P}^{(j)}_i$ | The $i$-th subspace of the $j$-th subsystem $H^{(j)}$. Sometimes also written as $\mathcal{P}_i$ if context permits. |
| $P^{(j)}_i$ | Projector onto $\mathcal{P}^{(j)}_i$. Defined in Equation 4. |
| $P(\mathbf{c})$ | Projector onto the subspace of each subsystem as described by energy configuration $\mathbf{c}$. See Eq. 20. |
| $\mathcal{S}_\mathbf{n}, \mathcal{S}_{\mathbf{n},\mathbf{n}'}$ | Set of 4-tuples stored in the node $v_\mathbf{n}$ or edge $e(v_\mathbf{n}, v_{\mathbf{n}'})$ in $G(\mathcal{V},\mathcal{E})$ generated in Alg. 1. See Sec. 4.2. |
| $T_r$ | The $r$-th order term in the self energy expansion $\Sigma_-(z)$. See Equations 9 and 11. |
| $V^{(i)}$ | Perturbation that acts on the Hilbert space $\mathcal{H}^{(i)} \otimes \mathcal{B}$. See Figure 1a of the main text and Eq. 5. |
| $V$ | Total perturbation $H_\mathcal{B} + V^{(1)} + ... + V^{(m)}$, see Equation 5 |
| $z$ | Expansion parameter for perturbation series. See Section 2.1 after Equation 2. |

Table 1: Table of notations (English alphabet) that have recurring appearance in the Supplementary Material.

| Symbol | Meaning and first appearance |
|---|---|
| $\Delta$ | The spectral gap between the ground and the first excited state of $H$. See Section 2.1 opening. |
| $\boldsymbol{\lambda}, \lambda_i$ | Basic quantity for constructing an upper bound to $\|T_r\|_2$. See Definition 4. |
| $\Pi_-, \Pi_+$ | Projectors onto $\mathcal{L}_-$ and $\mathcal{L}_+$ respectively. See text before Equation 7 and also Equation 21. |
| $|\psi_{j,p}^{(i)}\rangle$ | The $p$-th degenerate eigenvector of $\mathcal{P}_j^{(i)}$. See Equation 4. |
| $\omega$ | Basic quantity for constructing an upper bound to $\|T_r\|_2$. See Definition 3. |

Table 2: Table of notations (Greek alphabet) that have recurring appearance in the Supplementary Material.

The projections of $V_+$ then can be determined by taking the subgraphs in Figure 2b on the eigenstates that belong to $\mathcal{L}_+$:

$$
\begin{aligned}
V_+ &= B_{11,12}^{(1)} \otimes (|\psi_{1,1}^{(1)}\rangle\langle\psi_{2,1}^{(1)}| + |\psi_{2,1}^{(1)}\rangle\langle\psi_{0,1}^{(1)}|) \\
&+ B_{21,12}^{(1)} \otimes (|\psi_{1,2}^{(1)}\rangle\langle\psi_{2,1}^{(1)}| + |\psi_{2,1}^{(1)}\rangle\langle\psi_{1,2}^{(1)}|) \\
&+ B_{11,12}^{(2)} \otimes (|\psi_{1,1}^{(2)}\rangle\langle\psi_{2,1}^{(2)}| + |\psi_{2,1}^{(2)}\rangle\langle\psi_{0,1}^{(2)}|) \\
&+ B_{21,12}^{(2)} \otimes (|\psi_{1,2}^{(2)}\rangle\langle\psi_{2,1}^{(2)}| + |\psi_{2,1}^{(2)}\rangle\langle\psi_{1,2}^{(2)}|).
\end{aligned}
\tag{78}
$$

The projections $V_{-+}$ (resp. $V_{+-}$) are respectively cuts of edges that go from $\mathcal{L}_-$ to $\mathcal{L}_+$ (resp. $\mathcal{L}_+$ to $\mathcal{L}_-$):

$$
\begin{aligned}
V_{-+} &= B_{11,01}^{(1)} \otimes |\psi_{0,1}^{(1)}\rangle\langle\psi_{1,1}^{(1)}| + B_{12,01}^{(1)} \otimes |\psi_{0,1}^{(1)}\rangle\langle\psi_{1,2}^{(1)}| \\
&+ B_{11,01}^{(2)} \otimes |\psi_{0,1}^{(2)}\rangle\langle\psi_{1,1}^{(2)}| + B_{12,01}^{(2)} \otimes |\psi_{0,1}^{(2)}\rangle\langle\psi_{1,2}^{(2)}| \\
V_{+-} &= B_{11,10}^{(1)} \otimes |\psi_{1,1}^{(1)}\rangle\langle\psi_{0,1}^{(1)}| + B_{21,10}^{(1)} \otimes |\psi_{0,1}^{(1)}\rangle\langle\psi_{1,2}^{(1)}| \\
&+ B_{11,10}^{(2)} \otimes |\psi_{0,1}^{(2)}\rangle\langle\psi_{1,1}^{(2)}| + B_{21,10}^{(2)} \otimes |\psi_{0,1}^{(2)}\rangle\langle\psi_{1,2}^{(2)}|.
\end{aligned}
\tag{79}
$$

The operator valued resolvent $G_+(z) = (z\mathbf{I} - H)^{-1}$ could then be written as

$$
\begin{aligned}
G_+(z) &= \frac{1}{z - E_1}(|\psi_{1,1}^{(1)}\rangle\langle\psi_{1,1}^{(1)}| + |\psi_{1,2}^{(1)}\rangle\langle\psi_{1,2}^{(1)}|) \otimes |\psi_{0,1}^{(2)}\rangle\langle\psi_{0,1}^{(2)}| \\
&+ \frac{1}{z - E_1}|\psi_{0,1}^{(1)}\rangle\langle\psi_{0,1}^{(1)}| \otimes (|\psi_{1,1}^{(2)}\rangle\langle\psi_{1,1}^{(2)}| + |\psi_{1,2}^{(2)}\rangle\langle\psi_{1,2}^{(2)}|) \\
&+ \frac{1}{z - 2E_1}(|\psi_{1,1}^{(1)}\rangle\langle\psi_{1,1}^{(1)}| + |\psi_{1,2}^{(1)}\rangle\langle\psi_{1,2}^{(1)}|) \otimes (|\psi_{1,1}^{(2)}\rangle\langle\psi_{1,1}^{(2)}| + |\psi_{1,2}^{(2)}\rangle\langle\psi_{1,2}^{(2)}|) \\
&+ \frac{1}{z - E_2}(|\psi_{2,1}^{(1)}\rangle\langle\psi_{2,1}^{(1)}| \otimes |\psi_{0,1}^{(2)}\rangle\langle\psi_{0,1}^{(2)}| + |\psi_{0,1}^{(1)}\rangle\langle\psi_{0,1}^{(1)}| \otimes |\psi_{2,1}^{(2)}\rangle\langle\psi_{2,1}^{(2)}|)
\end{aligned}
\tag{80}
$$

In our projector notations, we could rewrite $G_+$ as

$$
\begin{aligned}
G_+(z) &= \frac{1}{z - E_1}(P_1^{(1)} \otimes P_0^{(2)} + P_0^{(1)} \otimes P_1^{(2)}) + \frac{1}{z - 2E_1} P_1^{(1)} \otimes P_1^{(2)} \\
&+ \frac{1}{z - 2E_1}(P_2^{(1)} \otimes P_0^{(2)} + P_0^{(1)} \otimes P_2^{(2)}) \\
&= \frac{1}{z - E_1}(P([1,0]) + P([0,1])) + \frac{1}{z - 2E_1} P([1,1]) \\
&+ \frac{1}{z - E_2}(P([2,0]) + P([0,2])).
\end{aligned}
\tag{81}
$$

With definitions in eqs. (78) to (80) we could express any $r$-th order term $T_r = V_{-+}(G_+V_+)^{r-2} G_+V_{+-}$ as a sum of terms involving $B_{mn,jk}^{(i)}$ operators. For example,

$$
T_2 = V_{-+} G_+ V_{+-} = \frac{1}{z - E_1}(B_{11,01}^{(1)} B_{11,10}^{(1)} + B_{12,01}^{(1)} B_{21,10}^{(1)} + B_{11,01}^{(2)} B_{11,10}^{(2)} + B_{12,01}^{(2)} B_{21,10}^{(2)}) \otimes \Pi_-. \tag{82}
$$

Note in (82) that there are in total four terms, two for each subsystem. The fact that there are two terms for each subsystem is due to the fact that for each subsystem there are at most two ways to transform, through

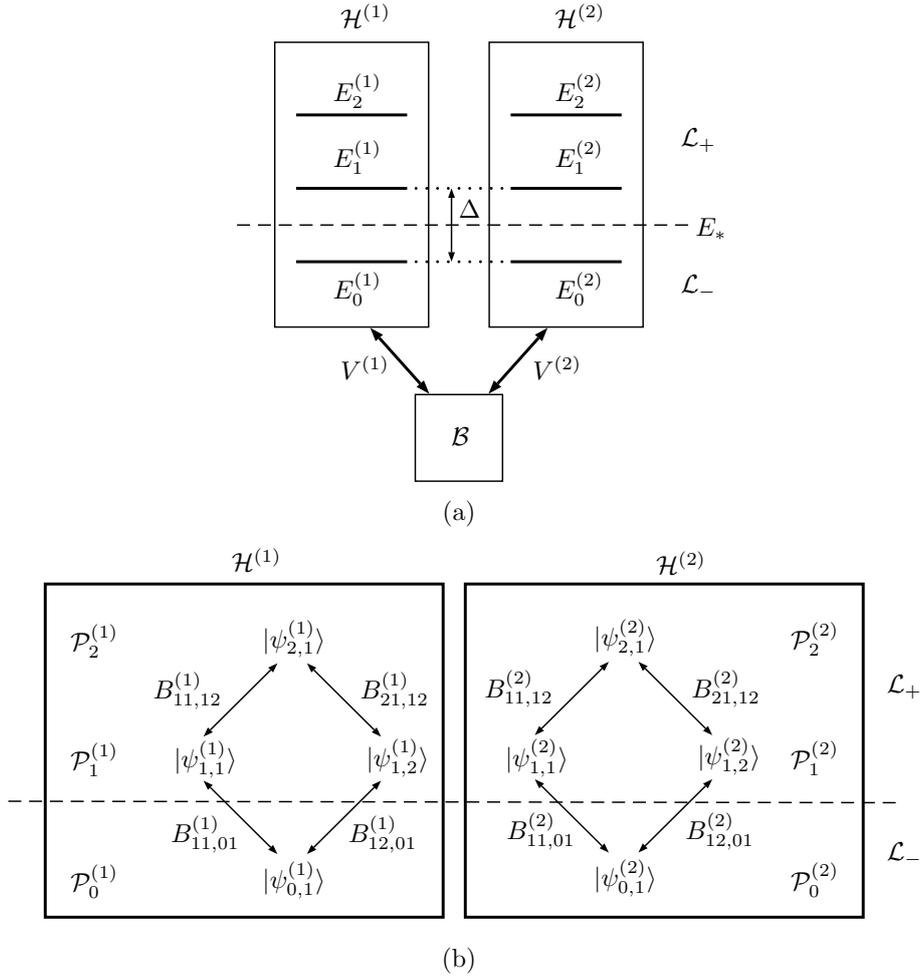

Figure 2: An example for illustrating the setting of perturbation theory that is concerned in this work.

perturbation $V$, an eigenstate (of $H$) in $\mathcal{P}_0$ to one in $\mathcal{P}_1$ (Figure 2b). In other words, $M_{01} = 2$. For an eigenstate in $\mathcal{P}_1$, there are at most one way to be transformed into $\mathcal{P}_0$ or $\mathcal{P}_2$ (or in other words, $M_{10} = 1$ and $M_{12} = 1$). Applying the definitions of $\lambda_i$, we have an upper bound to the $\infty$-norm of $T_2$ as

$$\|T_2\|_\infty = \|V_{-+}G_+V_{+-}\|_\infty \leq \frac{1}{z-E_1}2(\lambda_1^2 + \lambda_2^2) = \frac{1}{z-E_1}M_{01}M_{10}m_{(2)}. \tag{83}$$

The upper bound in the above equation can be interpreted diagrammatically as in Figure 3. The diagram shows how the upper bound to the $\infty$-norm "evolve" as we compute the upper bounds to $\|V_{-+}\|_\infty$, $\|V_{-+}G_+\|_\infty$, and $\|V_{-+}G_+V_{+-}\|_\infty$:

$$\begin{aligned}\|V_{-+}\|_\infty &\leq 2(\lambda_1 + \lambda_2) = M_{01}m_{(1)} \\ \|V_{-+}G_+\|_\infty &\leq \frac{1}{z-E_1} \cdot 2(\lambda_1 + \lambda_2) = \frac{1}{z-E_1}M_{01}m_{(1)}\end{aligned} \tag{84}$$

and an upper bound to $\|T_2\|_\infty$ is computed in (83).

## C  An example for illustrating walks in reduced configurations

Lemma 3 has established the basic idea that $T_r$ is essentially a sum of operator products associated with specific types of walks in the space of energy configuration **c**. For each particular walk, we could bound the $\infty$-norm of its corresponding operator product using a product of scalar quantities $\lambda_i$, $M_{jk}$ introduced in Definition 4 and 5 and $\frac{1}{z-E}$ where $E$ is taken from the set described in Equation (3). For a setting with $m$ unperturbed subsystems, $T_r$ is

31

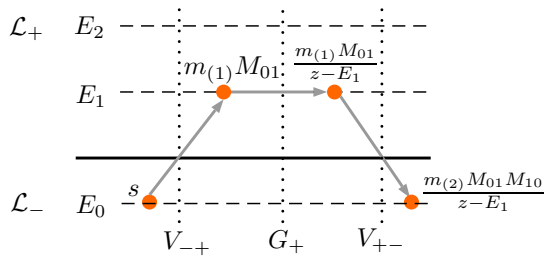

Figure 3: Diagram illustrating the virtual transitions associated with $T_2$. Here each horizontal line represents an (unperturbed) energy level. Each vertical line represents an operator in $T_r$ (here we show the diagram for $r = 2$). Each edge is associated both horizontally with an energy level and vertically with the operator corresponding to the vertical line that the edge crosses. Each node is associated with an upper bound to $\|Q_{e_1} Q_{e_2} \cdots Q_{e_k}\|_\infty$ with $e_1, \cdots, e_k$ forming a path from the starting node $s$ to the current node and $Q_e$ being the operator associated with edge $e$.

a summation of contributions from $O(m^r)$ walks. For example in $T_r = V_{-+}(G_+ V_+)^{r-2} G_+ V_{+-}$ for any $r$, the first factor $V_{-+}$ corresponds to the first step in the walk that departs from $\mathcal{L}_-$ into $\mathcal{L}_+$. To accomplish such departure one could excite any of the $m$ subsystems to raise the total energy into the high energy subspace $\mathcal{L}_+$, which gives a sum

$$\lambda_1 M_{01} + \cdots + \lambda_m M_{01} \tag{85}$$

as shown in Equation (84). Each term in the sum corresponds to a distinct walk. If we consider the lowest order term $T_2$, which sums over contributions from 2-step walks that first enters $\mathcal{L}_+$ and immediately return to $\mathcal{L}_-$, each walk that contributes to $T_2$ must first excite a subsystem and subsequently de-excite it so that the total state returns to $\mathcal{L}_-$. Hence an upper bound to $\|T_2\|_\infty$ can be computed as

$$\frac{1}{z - E_1}[(\lambda_1 M_{01})(\lambda_1 M_{10}) + (\lambda_2 M_{01})(\lambda_2 M_{10}) + \cdots + (\lambda_m M_{01})(\lambda_m M_{10})] \tag{86}$$

where $E_1$ is the first energy level above the cutoff $\lambda_*$. Expression 86 is identical to the right hand side of Equation 83 in the Appendix B, where a far more detailed derivation is presented. Expression 86 is written in a way that highlights the structure of a summation over contributions from 2-step walks. The term in each pair of parenthesis $(\cdot)$ corresponds to the factor contributed from a single step. For general $T_r$ we have $O(r^m)$ products of such $(\cdot)$ terms to sum over, which could quickly become computationally infeasible for large systems. Using symmetric polynomials to represent the summation, as can be seen in Equations 83 and 84, alleviates this concern by turning the problem of managing expressions such as Equations 85 and 86 into the problem managing the reduced partitions (Definition 2) of symmetric polynomials. The process of summing over walks in $\mathbf{c}$ hence becomes summing over walks in the space of reduced configurations $\tilde{\mathbf{c}}$.

We now consider 4-th order perturbation theory *i.e.* $r = 4$. Figure 4 illustrates the process of finding an upper bound to $\|T_4\|_\infty$ according to Lemma 6. There are in total 3 distinct walks in $\tilde{\mathbf{c}}$ and indeed the upper bound of $\|T_r\|_\infty$, denoted as $\triangle$ in Figure 4, consists of 3 terms of symmetric polynomials with distinct partitions. Each step of the walk is driven by an operator in $T_r$. Each node that the walk passes through corresponds to both a specific energy configuration and a particular position in the walk. Each node is also associated with a scalar number that serves as an upper bound to the $\infty$-norm of the product of operators so far.

An analogous diagram for $T_2$ is shown in Figure 3 in Appendix B. The upper bounds associated with the nodes passed through by the walk undergo a certain kind of "evolution" as the walk progresses, as can be observed both Figures 3 and 4. Informally the "evolution" can be described as the following: we start from an upper bound for $\|V_{-+}\|_\infty$. By modifying the upper bound according to some fixed rules, we arrive at an upper bound for $\|V_{-+}G_+\|_\infty$. Then by further modifying the upper bound for $\|V_{-+}G_+\|_\infty$ we get an upper bound for $\|V_{-+}G_+V_+\|_\infty$ etc.

The goal of the algorithms presented in the Sections 4.1 and 4.2 is to efficiently automate this "evolution" of walks using cellular automaton as the basic data structure. In the context of Algorithm 3, each horizontal line in Figure 4 corresponds to a cell (or a node) of the graph $G(\mathcal{V}, \mathcal{E})$ generated by BUILDCA in Algorithm 1 and each vertical column of nodes corresponds to a snapshot of the cell states at a given repetition of cell updates during step 2j of PERTURBBOUND in Algorithm 3. An upper bound for $\|T_r\|_\infty$ is computed by evolving the cellular automaton $r$ times in total ($r - 1$) times during step 2j and once during step 2l). Each path in Figure 4 corresponds to a walks in $\tilde{\mathbf{c}}$, which by Theorem 2, also corresponds to a trace of the algorithm.

32

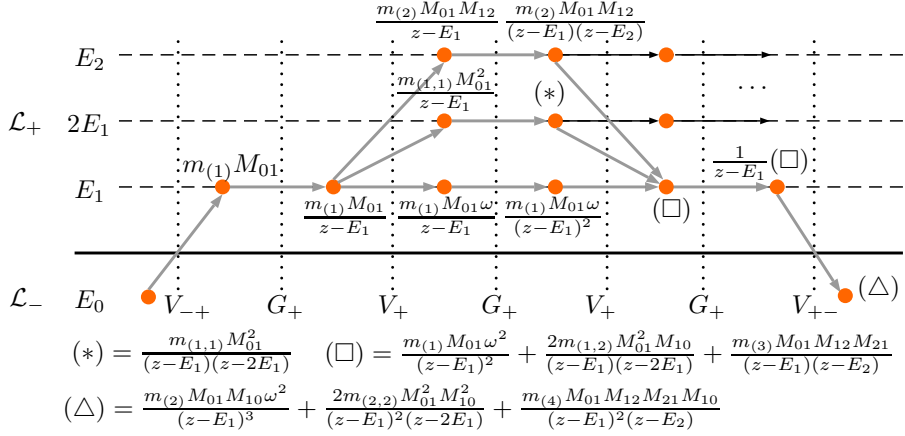

Figure 4: An example of enumerating 4-step walks in $\tilde{\mathbf{c}}$. Each path marked with bold edges corresponds to a walk in $\tilde{\mathbf{c}}$ with $\tilde{\mathbf{c}}^{(0)} = \tilde{\mathbf{c}}^{(4)} = (0, \cdots, 0)$. Due to limited space we replace some of the longer expressions with symbols $(*)$, $(\square)$ and $(\triangle)$ in the diagram and provide their full expressions below the diagram. Here we assume that $\mathcal{L}_-^{(i)} = \mathcal{P}_0^{(i)}$ for any $i$. Each horizontal line represents an energy level of the total unperturbed system $\mathcal{H}^{(1)} \otimes \mathcal{H}^{(2)} \otimes \cdots \otimes \mathcal{H}^{(m)}$, or equivalently an energy combination $\mathbf{n}$. Each vertical line represents an operator in $T_r$. Each edge is associated both horizontally with an energy level and vertically with the operator corresponding to the vertical line that the edge crosses.

Another observation concerns the property of monomial symmetric polynomials. Note first that $m_{\mathbf{b}}(\boldsymbol{\lambda})$ contains terms that have one-one correspondence with walks that consists of $b_1$ transitions on one subsystem, $b_2$ transitions on another system, $b_3$ transitions on another system etc. For example, if we have $m = 3$ subsystems, then $\boldsymbol{\lambda} = (\lambda_1, \lambda_2, \lambda_3)$ and the symmetric polynomial $m_{(1,3)}(\boldsymbol{\lambda}) = \lambda_1 \lambda_2^3 + \lambda_1 \lambda_3^3 + \lambda_2 \lambda_1^3 + \lambda_2 \lambda_3^3 + \lambda_3 \lambda_1^3 + \lambda_3 \lambda_2^3$ represents a collection of 4-step walks (because the sum of elements in the reduced partition is 4). Each term in $m_{(1,3)}(\boldsymbol{\lambda})$ corresponds to a type of 4-step walk. If we consider 5-step walks that are continuation of 4-step walks included in $m_{(1,3)}(\boldsymbol{\lambda})$, naturally we could choose any subsystem to act on for the 5-th step. An algebraic way of describing this freedom of choice is to use the sum $\lambda_1 + \lambda_2 + \lambda_3 = m_{(1)}(\boldsymbol{\lambda})$. Hence the collection of 5-step walks with the first 4 steps being any walk contained $m_{(1,3)}(\boldsymbol{\lambda})$ can be represented as [1, Lemma 1]

$$m_{(1,3)}(\boldsymbol{\lambda}) m_{(1)}(\boldsymbol{\lambda}) = m_{(2,3)}(\boldsymbol{\lambda}) + m_{(1,4)}(\boldsymbol{\lambda}) + m_{(1,1,3)}(\boldsymbol{\lambda}). \tag{87}$$

The above equation shows an example of generating terms for $(t+1)$-step walks from terms for $t$-step walks. As can be noticed from Figure 4, such "generation" mechanism of high-order symmetric polynomials from lower-order ones as exemplified in Equation 87 plays an important role in the "evolution" of upper bounds mentioned in the previous paragraph.

If one runs PERTURBBOUND$(4, \boldsymbol{\lambda}, \mathbf{M})$ as described in Algorithm 3 with the initial assignment of cell state being $\mathbf{n}_- = \mathbf{n}_0$ and $\mathbf{n}_+ = (m-1, 1, 0, \cdots, 0)$ during step 2a through 2h, the returned value $\tau_{4,\mathbf{n}_-}$ at step 2m should be the total value of the list of 4-tuples shown in Table 3, which is

$$\tau_{4,\mathbf{n}_-} = \sum_{\mathcal{T}=(\tilde{\mathbf{c}},\mathbf{b},\xi,\mu) \in \mathcal{S}_{\mathbf{n}_0}} m_{\mathbf{b}}(\boldsymbol{\lambda}) = \frac{M_{01} M_{10} \omega^2}{(z-E_1)^3} m_{(2)} + \frac{2 M_{01}^2 M_{10}^2}{(z-E_1)^2 (z-2E_1)} m_{(2,2)} + \frac{M_{01} M_{12} M_{21} M_{10}}{(z-E_1)^2 (z-E_2)} m_{(4)} = (\triangle) \tag{88}$$

where by $(\triangle)$ we refer to Figure 4. Equation 88 is also one of the terms on the right hand side of Equation 37 in Lemma 6 with $\tilde{\mathbf{c}}^{(0)} = \tilde{\mathbf{c}}^{(4)} = (0, \cdots, 0)$. See Figure 4.

| $\tilde{\mathbf{c}}$ | $\mathbf{b}$ | $\xi$ | $\mu$ |
|---|---|---|---|
| $(0)$ | $(2)$ | $\dfrac{M_{01}M_{10}\omega^2}{(z-E_1)^3}$ | $\tilde{\mathbf{c}} = (0)$ <br> $\downarrow$ <br> $\mathbf{b} = (2)$ |
| $(0,0)$ | $(2,2)$ | $\dfrac{2M_{01}^2 M_{10}^2}{(z-E_1)^2(z-2E_1)}$ | $\tilde{\mathbf{c}} = (0\ \ 0)$ <br> $\downarrow\ \ \downarrow$ <br> $\mathbf{b} = (2\ \ 2)$ |
| $(0)$ | $(4)$ | $\dfrac{M_{01}M_{12}M_{21}M_{10}}{(z-E_1)^2(z-E_2)}$ | $\tilde{\mathbf{c}} = (0)$ <br> $\downarrow$ <br> $\mathbf{b} = (4)$ |

Table 3: 4-tuple list associated with the cell $\mathcal{S}_{\mathbf{n}_0}$, representing the the expression ($\triangle$) in Figure 4, which is the final upper bound computed for $\|T_4\|_\infty$.



\end{document}